\documentclass[]{aastex701}

\usepackage{graphicx}
\usepackage{subcaption}
\usepackage{natbib}
\usepackage{here}
\usepackage{amsmath} 
\usepackage{threeparttable}
\usepackage{multirow}
\usepackage{subcaption}
\usepackage{comment}
\usepackage{textgreek}
\usepackage{CJKutf8}

\newcommand{\exomolco}{\mathrm{ExoMol\,\,CO}}
\newcommand{\hitempairco}{\mathrm{HITEMP_{Air}\,\,CO}}
\newcommand{\hitemphco}{\mathrm{HITEMP_{H_2}\,\,CO}}
\newcommand{\CO}{\mathrm{C/O}}
\graphicspath{{./}{fig_revise_ver1/}}

\begin{document}

\title{ExoJAX Retrievals of VLT/CRIRES Spectra of Luhman 16AB: C/O Ratios and Systematic Uncertainties}

\correspondingauthor{Hibiki Yama}
\email{yama@iral.ess.sci.osaka-u.ac.jp}

\author[0000-0001-7692-0581]{Hibiki Yama}
\affiliation{Department of Earth and Space Science, Graduate School of Science, Osaka University, 1-1 Machikaneyama-cho, Toyonaka, Osaka 560-0043, Japan}
\email{yama@iral.ess.sci.osaka-u.ac.jp}

\author[0000-0003-1298-9699]{Kento Masuda}
\affiliation{Department of Earth and Space Science, Graduate School of Science, Osaka University, 1-1 Machikaneyama-cho, Toyonaka, Osaka 560-0043, Japan}
\email{kmasuda@ess.sci.osaka-u.ac.jp}

\author[0000-0003-3800-7518]{Yui Kawashima}
\affiliation{Department of Astronomy, Graduate School of Science, Kyoto University, Kitashirakawa Oiwake-cho, Sakyo-ku, Kyoto 606-8502, Japan}
\email{ykawashima@kusastro.kyoto-u.ac.jp}

\author[0000-0003-3309-9134]{Hajime Kawahara}
\affiliation{Institute of Space and Astronautical Science, Japan Aerospace Exploration Agency, 3-1-1 Yoshinodai, Chuo-ku, Sagamihara, Kanagawa 252-5210, Japan}
\affiliation{Department of Astronomy, Graduate School of Science, The University of Tokyo, 7-3-1 Hongo, Bunkyo-ku, Tokyo 113-0033, Japan}
\email{kawahara@ir.isas.jaxa.jp}



\begin{abstract}

We present atmospheric retrievals of the benchmark brown dwarf binary Luhman 16AB using high-resolution VLT/CRIRES spectra and the differentiable framework ExoJAX. We derive elemental abundances and temperature--pressure ($T$--$P$) profiles while explicitly testing the robustness of the results against major sources of systematic uncertainty. We first perform retrievals with a power-law $T$--$P$ profile and assess the sensitivity of inferred molecular abundances and C/O ratios to different CO line lists (ExoMol, HITEMP with air- and H2-broadening). We then introduce a flexible Gaussian process-based $T$--$P$ profile, allowing a non-parametric characterization of the thermal structure and a more conservative treatment of uncertainties. For both components, we infer C/O ratios of about 0.67, slightly above solar, with line list systematics at the 7 percent level emerging as the dominant source of uncertainty, whereas assumptions about $T$--$P$ parameterization or photometric variability play a lesser role. The retrieved $T$--$P$ profiles and molecular abundances are broadly consistent with atmospheric models and equilibrium chemistry. Our results establish Luhman 16AB as a key anchor for substellar C/O measurements, demonstrate the utility of flexible $T$--$P$ modeling in high-resolution retrievals, and highlight the importance of systematic tests --- particularly line list uncertainties --- for robust comparisons between brown dwarfs and giant exoplanets.

\end{abstract}

\keywords{Brown dwarfs (185) --- Exoplanet atmospheres (487) --- High resolution spectroscopy (2096) --- Molecular spectroscopy (2095) --- T dwarfs (1679)}


\section{Introduction}

Detailed characterization of brown dwarfs and exoplanets is becoming possible with high-resolution spectroscopy. In addition to molecular detections with the cross-correlation function, retrieval analyses that directly model spectra to derive molecular abundances and temperature-pressure ($T$--$P$) structures are now commonly applied \citep[e.g.,][]{Birkby2018,Snellen2025}.
Atmospheric properties derived from such retrieval analyses provide valuable insights into the formation pathways of giant planets and brown dwarfs. For example, the relationship between a gas giant's C/O ratio and its formation environment has long been discussed as an important diagnostic: generally, star-like formation is expected to inherit the elemental ratios of the natal cloud, whereas core accretion within a disk --- where solids and gas evolve separately --- can alter the initial composition depending on the formation location \citep[e.g.,][]{Oberg2011}. Empirically, C/O ratios measured for wide-orbit gas giants/super-Jupiters from emission spectroscopy have tended to be roughly solar, whereas those inferred for hot Jupiters from transmission spectroscopy, including with JWST, exhibit a wider dispersion, potentially reflecting differences in formation processes \citep[e.g.,][]{Walsh2025}.
While this study focuses on the C/O ratio, elemental abundances such as [C/H] can also provide complementary diagnostics for distinguishing between planetary and brown dwarf formation pathways \citep[e.g.,][]{Xuan2024a, Wang2025}.

\begin{figure*}[h!]
    \centering
    \includegraphics[width=0.9\textwidth]{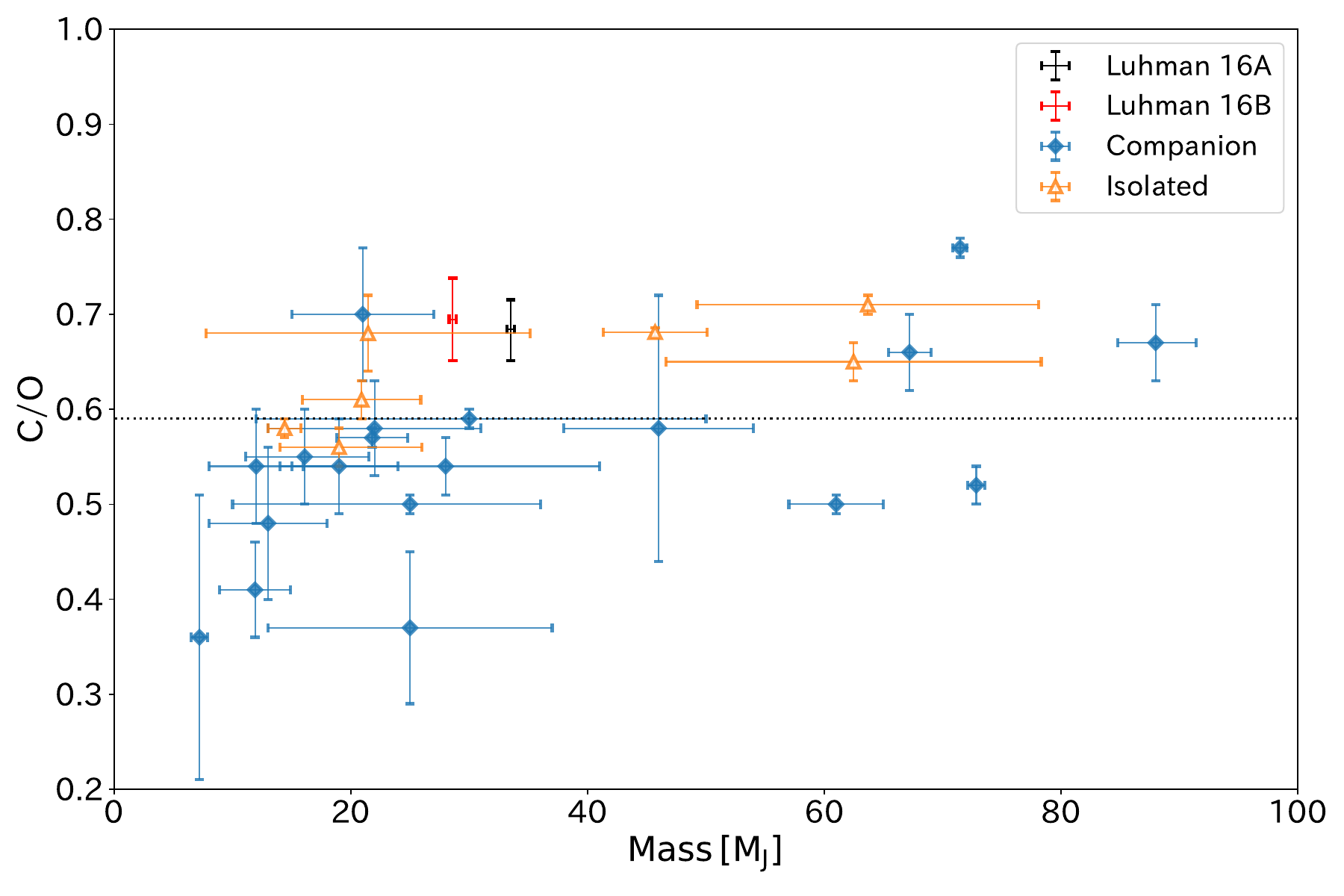}
    \caption{Comparison of C/O ratios versus mass for brown dwarfs and directly-imaged super Jupiters based on high-resolution spectra. Literature values and masses are summarized in Table~\ref{tab:co_lit}; the measurements for Luhman 16A and 16B are from this work. Blue symbols denote objects that are wide-orbit companions, and orange symbols denote isolated objects. For Luhman 16AB, the vertical error bars include the systematic uncertainty from the choice of CO line list quantified in this study.}
    \label{fig:co_mass}
\end{figure*}

To advance these discussions, brown dwarfs are important comparison targets, providing anchor points in addition to gas giants: while the formation of cold Jupiters remains debated, brown dwarfs likely form in a star-like manner and thus can provide a baseline for C/O. However, the number of available high-resolution measurements remains limited (Figure~\ref{fig:co_mass}). Since brown dwarfs are often observable at higher S/N than planetary-mass objects, they also offer valuable testbeds for assessing retrieval systematics. Despite rapid progress of atmospheric retrieval, high-resolution retrievals still carry systematic uncertainties from cloud modeling, molecular line lists, $T$--$P$ parameterizations, and chemical–equilibrium assumptions, and a community standard has yet to emerge. For example, \citet{Picos2024} compared spectral models built on two distinct gradient assumptions of $T$--$P$ profiles and showed that the inferred surface gravity can shift appreciably with the chosen $T$--$P$ model, while \citet{deRegt2024} demonstrated that using alternative $\mathrm{H_2O}$ line lists leads to different inferences for the isotopologue ratio $^{12}\mathrm{CO}/^{13}\mathrm{CO}$. Testing their impact with high-quality data is important both for comparative planetology and for understanding these processes themselves.

In this work, we perform retrievals on VLT/CRIRES spectra of the benchmark brown dwarf binary Luhman 16AB. Owing to its proximity and brightness, high-S/N data are available \citep{Crossfield2014}, and extensive external constraints exist, making the system not only a key anchor for C/O but also a valuable testbed for evaluating model systematics and trialing new methodologies, as pursued here. 
In particular, astrometric monitoring yields dynamical masses for both components \citep{lazorenko2018, Bedin2024}, and recent kinematic analyses identify Luhman 16AB as a member of the Oceanus association, providing an age estimate \citep{gagne2023}. Combining these with brown dwarf evolutionary model \citep{Baraffe2003} furnishes informative priors on the surface gravity. Imposing such priors helps to break the well-known metallicity–$\log g$ degeneracy in high-resolution spectral retrievals, thereby enabling more robust inferences on elemental abundances. 
Furthermore, such external information, together with the binary nature, is beneficial for assessing the validity of the retrievals.

The paper is organized as follows. Section~\ref{sec:Data} describes the data, and Section~\ref{sec:Framework} outlines the modeling framework used for the retrievals. In Section~\ref{sec:PLTP}, we analyze the line-list dependence within a power-law $T$--$P$ framework. Section~\ref{sec:GPTP} then introduces a more flexible Gaussian-process $T$--$P$ profile, demonstrates its utility, and discusses how assumptions about the $T$--$P$ structure propagate into other parameters and their interpretation. Section~\ref{sec:Summary} summarizes our findings, discusses implications, and outlines directions for future work.

\begin{figure*}[h!]
    \includegraphics[width=0.5\textwidth]{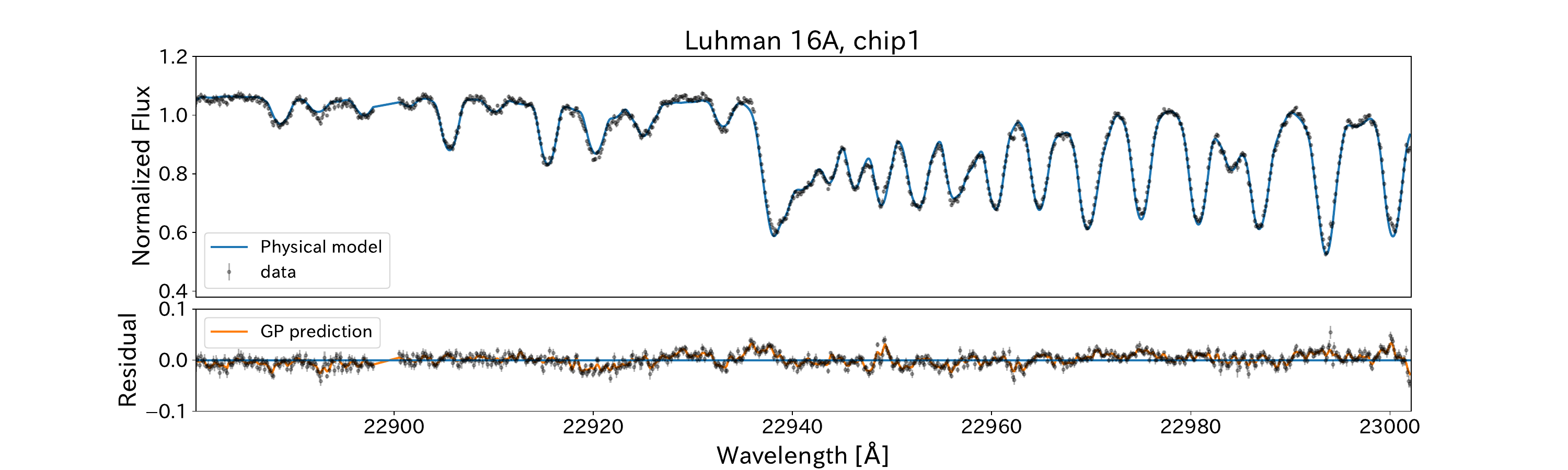}
    \includegraphics[width=0.5\textwidth]{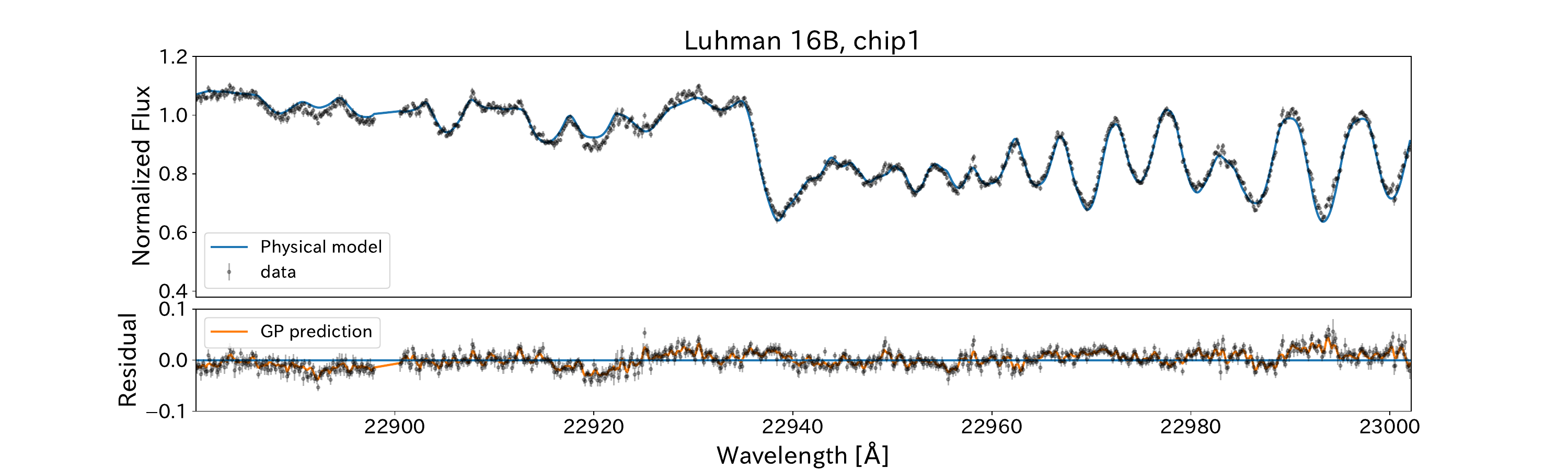}

    \includegraphics[width=0.5\textwidth]{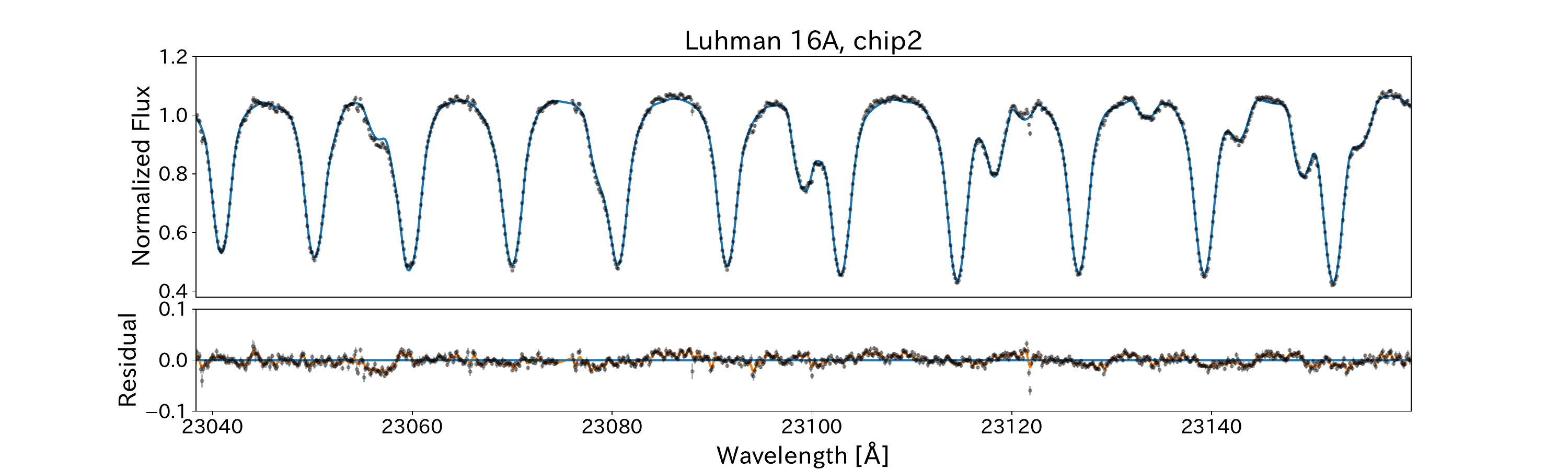}
    \includegraphics[width=0.5\textwidth]{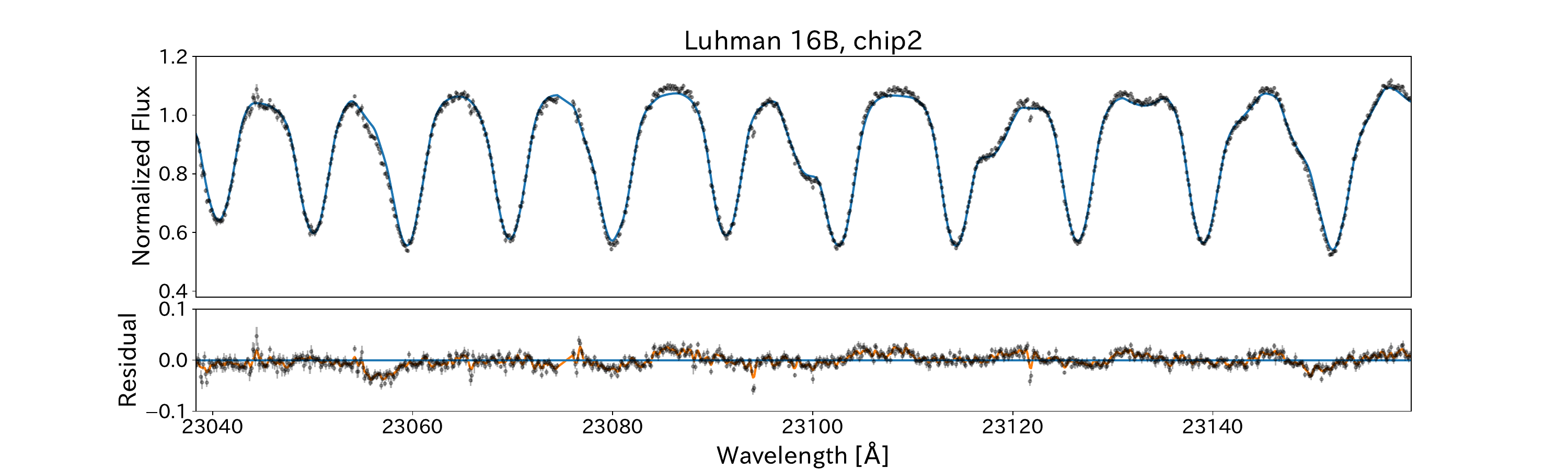}

    \includegraphics[width=0.5\textwidth]{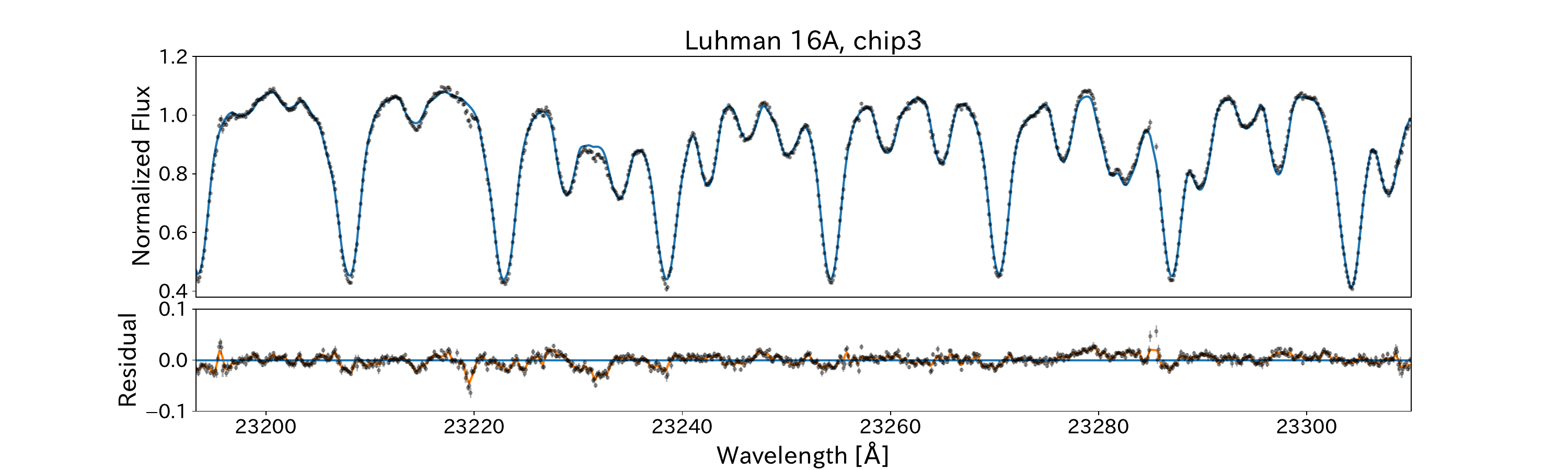}
    \includegraphics[width=0.5\textwidth]{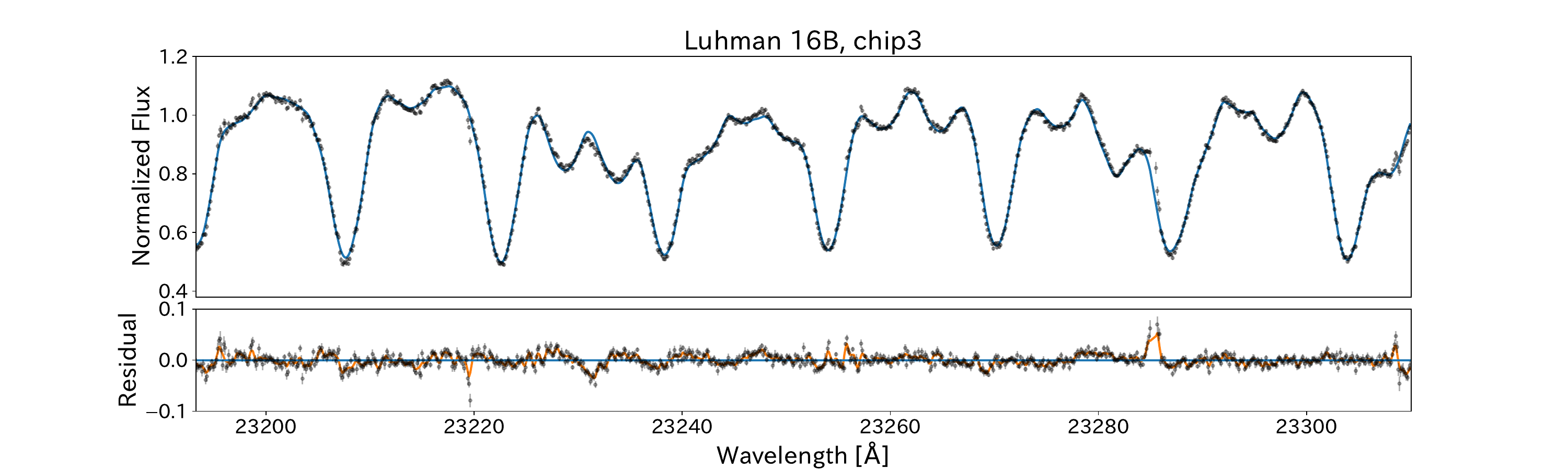}

    \includegraphics[width=0.5\textwidth]{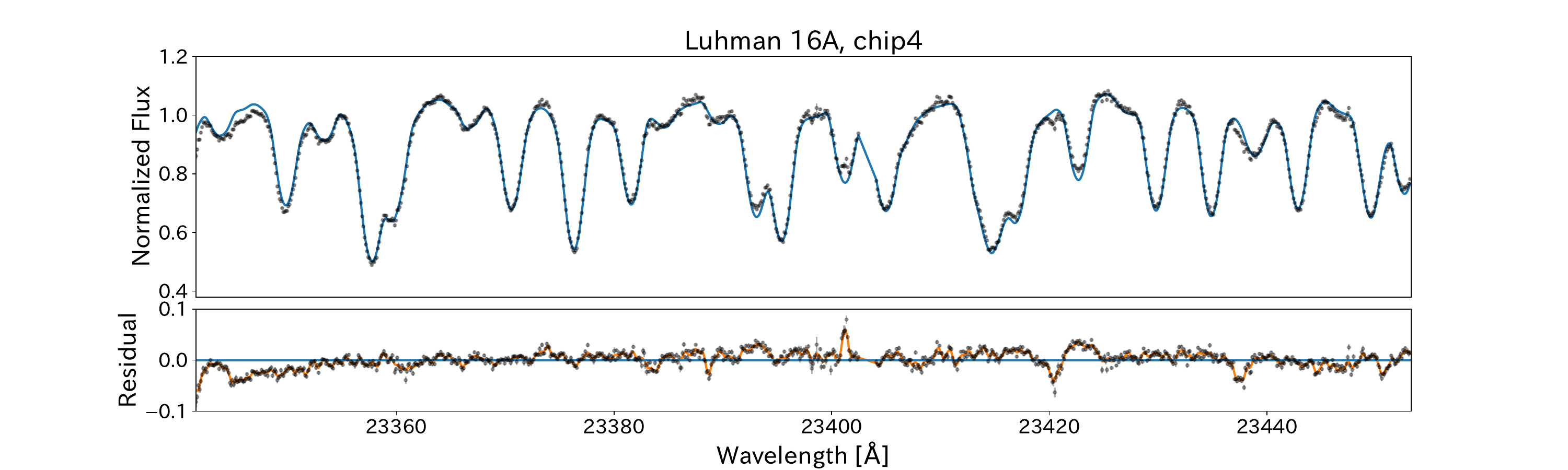}
    \includegraphics[width=0.5\textwidth]{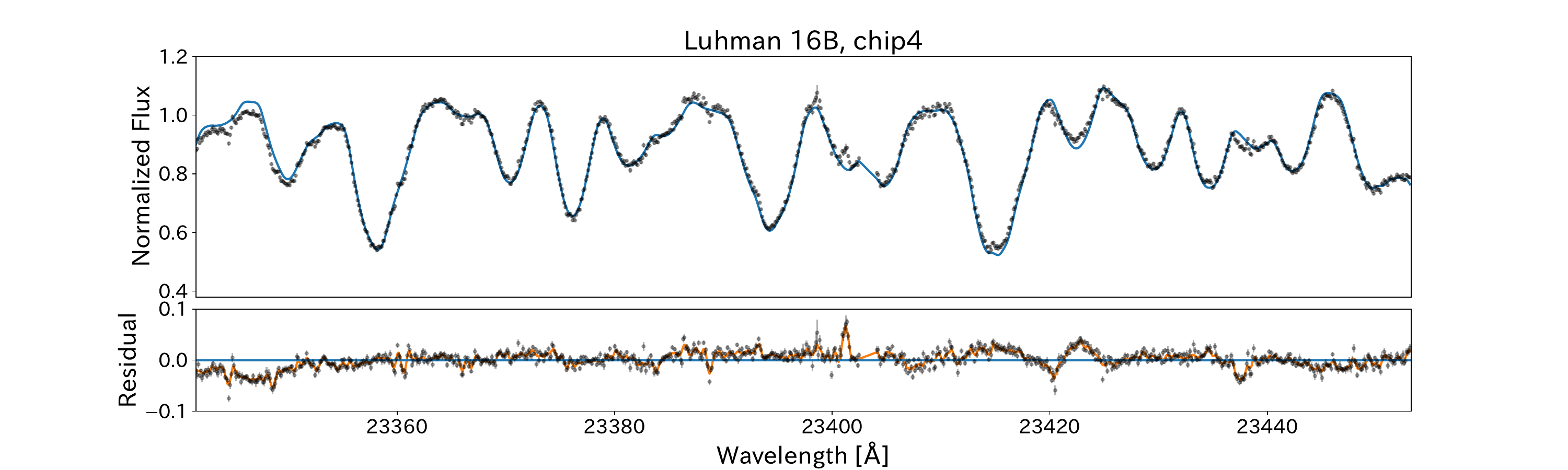}
    
    \caption{Retrieval results for the VLT/CRIRES high-resolution spectra of Luhman 16A (left) and Luhman 16B (right), modeled with the $\hitemphco$ line list and a power-law $T$--$P$ profile. The wavelength coverage is split into four detector segments (chips 1–4). For each chip, the top panel shows the normalized flux data (black points) and the best-fit physical model (blue line). The bottom panel shows the residuals (data minus the best-fit model); orange lines indicate the correlated noise predicted by the GP. Each chip is normalized independently.}
    \label{fig:fit_PLTP_HITEMPH2}
\end{figure*}

\section{Data}
\label{sec:Data}

In this work, we analyze the $K$-band ($2.288$–$2.345\,\mu\mathrm{m}$) high-resolution spectra of Luhman 16AB obtained with the CRyogenic high-resolution InfraRed Echelle Spectrograph \citep[CRIRES;][]{CRIRES} mounted on the Very Large Telescope (VLT), i.e., the same data used for Doppler imaging by \citet{Crossfield2014}. The observations were carried out for 5 hr on UT 2013-05-05. The stacked spectrum achieves a resolving power of $R\sim10^{5}$ with $\mathrm{S/N}\sim180$. Because CRIRES records spectra on four detectors aligned in the dispersion direction, the analyzed data are further divided into four wavelength segments within the observed bandpass; in what follows we refer to these as chip 1–4 from shorter to longer wavelengths. Each segment is normalized independently (Figure~\ref{fig:fit_PLTP_HITEMPH2}).

\section{Retrieval Framework}
\label{sec:Framework}

\subsection{Spectrum Model}

\begin{deluxetable*}{l c c l}
\tablecaption{Parameters and Priors for spectral model\label{tab:prior_Luhman16AB}}
\tabletypesize{\footnotesize}     
\tablewidth{\columnwidth}         
\tablehead{
\colhead{Parameter} & \colhead{Unit} & \colhead{Prior} & \colhead{Description}
}
\startdata
$\log g$ & cgs & $\mathcal{N}_{[4.66, 5.20]}(4.93, 0.09)$ (A), $\mathcal{N}_{[4.59, 5.13]}(4.86, 0.09)$ (B)  & Surface gravity \\
$RV$ & $\mathrm{km} / \mathrm{s}$ & $\mathcal{U}(20, 30)$ & Radial velocity \\
$\log_{10}\mathrm{VMR_{\mathrm{CO}}}$ & - & $\mathcal{U}(-6.0, -1.0)$ & Volume mixing ratio of $\mathrm{^{12}C^{16}O}$ \\ 
$\log_{10}\mathrm{VMR_{\mathrm{H_2O}}}$ & - & $\mathcal{U}(-6.0, -1.0)$  & Volume mixing ratio of $\mathrm{^{1}H_{2}^{16}O}$ \\ 
$\log_{10}\mathrm{VMR_{\mathrm{CH_4}}}$ & - & $\mathcal{U}(-6.0, -1.0)$ & Volume mixing ratio of $\mathrm{^{12}C^{1}H_4}$ \\ 
$\log_{10}\mathrm{VMR_{\mathrm{HF}}}$ & - & $\mathcal{U}(-10.0, -5.0)$ &  Volume mixing ratio of $\mathrm{^{1}H^{19}F}$ \\ 
$\log_{10} P_{\mathrm{c}}$ & bar & $\mathcal{U}(-2.0, 2.0)$ & Pressure of cloud center height \\
$V \sin i$ & $\mathrm{~km} / \mathrm{s}$ & $\mathcal{U}(10, 20)$ (A), $\mathcal{U}(20, 30)$ (B) & Projected rotational velocity \\ 
$q_1$ & - & $\mathcal{U}(0.0, 1.0)$ &  Limb darkening parameter \\ 
$q_2$ & - & $\mathcal{U}(0.0, 1.0)$ &  Limb darkening parameter \\ 
$\log_{10} a_{\mathrm{GP}}$ & - & $\mathcal{U}(-4.5, 2.0)$  & Amplitude of GP \\ 
$\log_{10} \tau_{\mathrm{GP}}$ & \AA & $\mathcal{U}(-1.5, 0.5)$  & Length scale of GP \\ 
$A_1$ & - & $\mathcal{U}(1.0, 1.2)$ & Normalize factor (chip1) \\ 
$A_2$ & - & $\mathcal{U}(1.0, 1.2)$ & Normalize factor (chip2) \\ 
$A_3$ & - & $\mathcal{U}(1.0, 1.2)$ & Normalize factor (chip3) \\ 
$A_4$ & - & $\mathcal{U}(1.0, 1.2)$ & Normalize factor (chip4) \\
\cutinhead{\itshape (Power-law $T$--$P$)}
$T_0$ & K & $\mathcal{U}(1000, 1700)$  & Temperature at $1\,\mathrm{bar}$ \\ 
$\alpha$ & - & $\mathcal{U}(0.05, 0.20)$  & Power-law index \\
\cutinhead{\itshape (GP $T$--$P$)}
$T_{0,\,m}$ & K & $\mathcal{U}\left(1000, 1700\right)$ & Temperature at $1\,\mathrm{bar}$ in the GP mean model \\ 
$\alpha_{m}$ & - & $\mathcal{U}(0.05, 0.20)$  & Power-law index in the GP mean model \\
$\log_{10}a_{u}$ & - & $\mathcal{U}\left(-1.0, 0.5\right)$ & Amplitude of GP $T$--$P$ \\
$\log_{10}\tau_{u}$ & - & $\mathcal{U}\left(-0.5, 1.4\right)$ & Length scale for GP $T$--$P$ \\
\enddata
\end{deluxetable*}

We modeled the high-resolution spectra of Luhman 16AB with the ExoJAX spectral framework \citep{Kawahara2022, Kawahara2024}. ExoJAX solves the radiative transfer by computing, for each pressure interval $\Delta P$ between adjacent levels, the optical–depth increment as the sum of the molecular contribution in a given layer $\sum_i\Delta\tau_{\mathrm{mol},i}$, collision–induced absorption (CIA) $\Delta\tau_{\mathrm{CIA}}$, and cloud opacity $\Delta\tau_\mathrm{c}$:
\begin{equation}
\Delta\tau(\nu)=\sum_i \Delta\tau_{\mathrm{mol},i}(\nu)+\Delta\tau_{\mathrm{CIA}}(\nu)+\Delta\tau_{\mathrm{c}},
\label{eq:dtau}
\end{equation}
where $\nu$ is the wavenumber. We discretize the atmosphere into 101 pressure levels, logarithmically spaced from $10^{-4}$--$10^{2}$ bar.
Here, the molecular contribution in a given layer $\Delta\tau_{\mathrm{mol},i}$ is
\begin{equation}
    \label{eq:dtau_mol}
    \Delta\tau_{\mathrm{mol},i}\left(\nu\right)=\frac{\mathrm{VMR}_{i}}{\mu\,m_{u}\,g}\,\sigma_{i}\left(\nu\right)\,\Delta P,
\end{equation}
where $\mathrm{VMR}_i$ denotes the volume mixing ratio of absorbing species $i$, $m_u$ is the atomic mass unit, and $g$ is the surface gravity.
The mean molecular weight $\mu$ is calculated self-consistently from the adopted VMRs of all atmospheric components, accounting for both the absorbing species and the bulk gases. The molecular cross section $\sigma_i$ is computed from line strengths $S\left(T\right)$ and Voigt profiles $V\left(\nu,\,\beta,\,\gamma_{L}\right)$ as
\begin{equation}
    \sigma_{i}\left(\nu\right)=\sum_{l} S_{l}\left(T\right)\,V\left(\nu-\hat{\nu},\,\beta_{l}\left(T\right),\,\gamma_{L,\,l}\left(T,\,P\right)\right).
\end{equation}
Here, $l$ indexes the absorption lines ($l=1,\ldots,N_l$) and $\hat{\nu}$ is the line center.
ExoJAX evaluates the Voigt profiles only for the lines located within the specified wavelength range, and for each line the Voigt function is computed over that interval.
The Gaussian (thermal) width is
\begin{equation}
    \label{eq:sigma_thermal_broadening}
    \beta=\sqrt{\frac{k_{\mathrm{B}}T}{M_im_{u}}}\,\frac{\hat{\nu}}{c}, 
\end{equation}
where $M_i$ is molecular mass of species $i$ and $c$ is the speed of light. The pressure broadening parameter $\gamma_{L}$ and the line strengths $S_l(T)$ are computed from the line lists HITRAN/HITEMP \citep{HiTRAN, HiTEMP} and ExoMol \citep{ExoMol2016,ExoMol2024} in combination with the assumed temperature–pressure structure $T(P)$ (see Sections~\ref{sec:PLTP}, \ref{subsec:GPTP}). 
We explicitly include the trace absorbers $i=\mathrm{CO},\,\mathrm{H_2O},\,\mathrm{CH_4},\,\mathrm{HF}$, with free, vertically uniform VMRs; inclusion of HF is motivated by its reported $K$-band detection in Luhman 16AB from high-resolution IGRINS/Gemini South spectroscopy \citep{Ishikawa2025}. The remainder of the atmosphere is assigned to the bulk gas, $\mathrm{H_2}$ and $\mathrm{He}$. After specifying the four trace-species VMRs, the residual fraction $1-\sum_i \mathrm{VMR}_i$ is distributed between $\mathrm{H_2}$ and $\mathrm{He}$ according to the solar mass ratio 0.74:0.25 \citep{Asplund2021}. Because the retrieval is performed in VMR, we convert this mass ratio into a volume ratio by weighting with the molecular masses, i.e., in proportion to $(0.74/M_{\mathrm{H_2}}):(0.25/M_{\mathrm{He}})$, which ensures that the total across every included species---$\mathrm{CO},\,\mathrm{H_2O},\,\mathrm{CH_4},\,\mathrm{HF},\,\mathrm{H_2},$ and $\mathrm{He}$---sums to unity. 
The collision-induced absorption (CIA) contribution $\Delta\tau_{\mathrm{CIA}}$ is
\begin{equation}
    \label{eq:dtau_CIA}
    \Delta\tau_{\mathrm{CIA}}\left(\nu\right)=\left(n_{\mathrm{H_2}}^2\,\alpha_{\mathrm{H_2-H_2}}\left(T,\nu\right)+n_{\mathrm{H_2}}\,n_{\mathrm{He}}\,\alpha_{\mathrm{H_2-He}}\left(T,\nu\right)\right)\,\frac{k_{\mathrm{B}}T}{\mu m_{u}g}\,\frac{\Delta P}{P},
\end{equation}
where $n_{\mathrm{H_2}}$ and $n_{\mathrm{He}}$ are number densities, $\mu$ is the mean molecular weight, and $\alpha_{\mathrm{CIA}}(T,\nu)$ are CIA coefficients taken from the HITRAN database. 
The cloud optical depth per layer, $\Delta\tau_{\mathrm c}$, is modeled as a Gaussian in $\log_{10}P$ with width $w_{\mathrm c}$ and central pressure $P_{\mathrm c}$:
\begin{equation}
    \label{eq:dtau_cloud}
    \Delta\tau_\mathrm{c}=\frac{C}{\sqrt{2\pi w_\mathrm{c}^2}}\,\exp{\left[-\frac{(\log_{10}P-\log_{10}P_{\mathrm{c
    }})^2}{2w_\mathrm{c}^2}\right]},
\end{equation}
where $C$ is the (dimensionless) column optical depth of the cloud. This functional form is motivated by the vertical cloud opacity profiles in the Sonora diamondback models \citep{Morley2024}.
Given the narrow wavelength range analyzed here, we neglect any wavelength dependence of the cloud optical depth. Because $w_{\mathrm c}$ and $P_{\mathrm c}$ are strongly degenerate, we fix $w_{\mathrm c}=0.3$.
To emulate an effectively optically thick gray deck, we fix the column optical depth to a large arbitrary constant, $C=500$.

Using the resulting $\Delta\tau(\nu)$, we solve the radiative transfer with an eight-stream approximation to obtain the local one-dimensional emergent spectrum $S_{\mathrm{raw}}(\nu)$. We then convolve with a rotational broadening kernel computed from the limb-darkening law and the projected rotational velocity $V\sin i$ to obtain the disk-integrated spectrum $F_{\mathrm{raw}}(\nu)$. For limb darkening, we adopt the quadratic law with the $(q_1, q_2)$ parametrization as proposed by \citet{Kipping2013}.

To model the Earth-observed, per-chip normalized spectra, we first apply the Doppler shift due to the brown dwarf radial velocity $RV$. For an intrinsic wavenumber $\nu_0$, the observed wavenumber is
\begin{equation}
    \label{eq:RV_nu}
    \nu_{\mathrm{obs}} = \nu_0\,\frac{c}{c + RV}.
\end{equation}
Because the observed spectra are normalized per detector, the model is likewise normalized in each wavelength segment. Denoting the model for the four detector segments by $\boldsymbol{F}_i$ ($i=1,2,3,4$) and their mean values by $\boldsymbol{\bar{F}}_i$, the normalized spectra are
\begin{equation}
    \label{eq:normalize_nu}
    \boldsymbol{f}_{i} = \frac{\boldsymbol{F}_i}{A_i \boldsymbol{\bar{F}}_i},
\end{equation}
with $A_i$ a per-segment normalization factor. The final model used in the analysis is the concatenation
\begin{equation}
    \label{eq:final_model_flux}
    \boldsymbol{f} = \left[\boldsymbol{f}_{1},\, \boldsymbol{f}_{2},\, \boldsymbol{f}_{3},\, \boldsymbol{f}_{4}\right].
\end{equation}

\subsection{Inference}

ExoJAX is implemented with JAX \citep{jax2018github}, a high-performance Python library that supports GPU acceleration and automatic differentiation. Consequently, our spectral model can be evaluated on GPUs and differentiated with respect to arbitrary parameters, enabling gradient-based inference.

Leveraging these features, we infer posterior distributions of the model parameters with Hamiltonian Monte Carlo \citep[HMC;][]{HMC}, specifically the No-U-Turn Sampler \citep[NUTS;][]{NUTS} implemented in NumPyro. With prior $p(\boldsymbol{\theta})$, the posterior is given by Bayes' theorem:
\begin{equation}
    \label{eq:Bayes}
    p(\boldsymbol{\theta}\mid\boldsymbol{d})\propto\mathcal{L}(\boldsymbol{\theta})\,p(\boldsymbol{\theta}).
\end{equation}
To account for correlated noise, we use a Gaussian-process (GP) likelihood \citep{Kawahara2022},
\begin{equation}
    \label{eq:liklihood_model}
    \mathcal{L}(\boldsymbol{\theta})=\frac{1}{\sqrt{(2\pi)^n\,|\boldsymbol{\Sigma}|}}\,\exp{\left[-\frac{1}{2}\,(\boldsymbol{d}-\boldsymbol{f(\boldsymbol{\theta})})^{\mathrm{T}}\,\boldsymbol{\Sigma}^{-1}\,(\boldsymbol{d}-\boldsymbol{f(\boldsymbol{\theta})})\right]},
\end{equation}
where $\boldsymbol{d}$ is the $n$-point observed spectrum and $\boldsymbol{f}(\boldsymbol{\theta})$ is the model. The covariance matrix is
\begin{equation}
    \label{eq:cov_GP}
    \Sigma_{ij}=K_{ij}+\sigma_{e,\,i}^2\,\delta_{ij},
\end{equation}
with $\sigma_{e,i}$ the $i$-th data uncertainty and $\delta_{ij}$ the Kronecker delta. We adopt an RBF kernel,
\begin{equation}
    \label{eq:ker_GP}
    K_{ij}=a_{\mathrm{GP}}\,\exp{\left[-\frac{(\lambda_i-\lambda_j)^2}{2\tau_{\mathrm{GP}}^2}\right]},
\end{equation}
where $\lambda_i,\,\lambda_j$ are wavelength of $i$-th and $j$-th data, $a_{\mathrm{GP}}$ is the GP amplitude, and $\tau_{\mathrm{GP}}$ (in $\mathrm{\AA}$) is the correlation length. 

The set of inferred parameters and their priors is listed in Table~\ref{tab:prior_Luhman16AB}.
For $\log g$, we imposed Normal priors informed by external constraints. Using the system age of $510\pm95\,\mathrm{Myr}$ from \citet{gagne2023} together with the dynamical masses $M_A=33.5\pm0.3\,\mathrm{M_J}$ and $M_B=28.6\pm0.3\,\mathrm{M_J}$ from \citet{lazorenko2018}, we interpolated the brown dwarf evolutionary model of \citet{Baraffe2003} to obtain surface gravities. Since the surface gravity is given by $g = GM/R^{2}$ with brown dwarf mass $M$, radius $R$, and the gravitational constant $G$, we propagated the quoted mass uncertainties ($\sim1\%$) and assumed a 10\% fractional uncertainty on the radius to set the prior widths, yielding Normal priors $\log g_A = 4.93 \pm 0.09$ and $\log g_B = 4.86 \pm 0.09$, each truncated at $3\sigma$.
Sampling used HMC–NUTS with a dense mass matrix. For numerical stability and mixing, parameters were affinely re-scaled so that the prior support maps to $\mathcal{O}(1)$ in the sampler's internal coordinates. Sampling/adaptation were performed in the transformed space and results are reported after back-transformation.

\section{Retrievals Using a Power-Law $T$--$P$ Profile}
\label{sec:PLTP}

In this section, we adopt a simple power-law $T$--$P$ profile given by
\begin{equation}
    \label{eq:PLTP}
    T\left(P\right)=T_{0}\,\left(\frac{P}{1\,\mathrm{bar}}\right)^{\alpha}.
\end{equation}
In high-resolution retrievals of brown dwarf and exoplanet atmospheres, line shapes depend on the pressure broadening parameters provided by the molecular line lists, which can in turn impart systematic effects on the inferred molecular abundances and atmospheric structure. To assess the robustness of the results, it is therefore essential to analyze a given target with multiple line lists. Motivated by this, we focus on the CO line list and compare three models based on ExoMol and HITEMP.

\subsection{Line Lists}
\label{subsec:line_list}

High-resolution retrievals for brown dwarfs commonly use line lists from ExoMol and HITEMP. ExoMol provides theoretically computed line data tailored to brown dwarf and exoplanet atmospheres, assuming an $\mathrm{H_2}$–$\mathrm{He}$ background. HITEMP, by contrast, is based largely on high-temperature laboratory measurements and typically adopts an Earth background (air) for pressure broadening. More recently, an $\mathrm{H_2}$–$\mathrm{He}$ broadened variant of HITEMP has become available, obtained by converting the air-broadening parameters; the conversion employs semi-empirical models anchored to validated laboratory measurements and theoretical predictions \citep{tan2022}.

\subsubsection{ExoMol}
\label{subsubsec:ExoMol}
According to \citet{ExoMol2016}, ExoMol supplies transition wavenumbers $\hat{\nu}$, Einstein coefficients $A$, state energies $E$, and statistical weights $\mathsf{g}$, from which the line strength $S(T)$ can be computed as
\begin{equation}
    S\left(T\right)=\frac{\mathsf{g_{up}}}{Q\left(T\right)}\,\frac{A}{8\pi c\hat{\nu}^2}\,e^{-hcE_{\mathrm{low}}/k_{B}T}\,\left(1-e^{-hc\hat{\nu}/k_{B}T}\right),
\end{equation}
where $E_{\mathrm{low}}$ is the lower-state energy and $\mathsf{g_{up}}$ is the upper-state statistical weight. For pressure broadening in an $\mathrm{H_2}$ background, ExoMol provides the reference broadening coefficient $\gamma_{\mathrm{ref}}(J_{\mathrm{low}})$ and temperature exponent $\mathsf{n}(J_{\mathrm{low}})$ as functions of the lower-state rotational quantum number. The pressure-broadened Lorentz width is then
\begin{equation}
    \label{eq:gamma_p_L_exomol}
    \gamma^{p}_{L}=\gamma_{\mathrm{ref}}\,\left(\frac{T}{T_{\mathrm{ref}}}\right)^{-\mathsf{n}}\,\left(\frac{P}{P_{\mathrm{ref}}}\right),
\end{equation}
with $T_{\mathrm{ref}}=296\,\mathrm{K}$ and $P_{\mathrm{ref}}=1\,\mathrm{bar}$.

\subsubsection{HITEMP}
According to \citet{HiTEMP}, HITEMP provides $\hat{\nu}$, $S(T_{\mathrm{ref}})$ at $T_{\mathrm{ref}}=296\,\mathrm{K}$, and $E_{\mathrm{low}}$, from which
\begin{equation}
    S\left(T\right)=S\left(T_{\mathrm{ref}}\right)\,\frac{Q\left(T_{\mathrm{ref}}\right)}{Q\left(T\right)}\,\frac{e^{-hcE_{\mathrm{low}}/k_{\mathrm{B}}T}}{e^{-hcE_{\mathrm{low}}/k_{\mathrm{B}}T_{\mathrm{ref}}}}\,\frac{1-e^{-hc\hat{\nu}/k_{\mathrm{B}}T}}{1-e^{-hc\hat{\nu}/k_{\mathrm{B}}T_{\mathrm{ref}}}} 
\end{equation}
is obtained. HITEMP further provides air-broadened reference widths $\gamma_{\mathrm{ref,\,air}}$ and temperature exponents $\mathsf{n}_{\mathrm{air}}$ for each line, yielding
\begin{equation}
    \label{eq:gamma_p_L_hitemp}
    \gamma^{p}_{L}=\gamma_{\mathrm{ref,\,air}}\,\left(\frac{T}{T_{\mathrm{ref}}}\right)^{-\mathsf{n}_{\mathrm{air}}}\,\left(\frac{P}{P_{\mathrm{ref}}}\right)
\end{equation}
with $P_{\mathrm{ref}}=1\,\mathrm{atm}$. For some molecules, $\mathrm{H_2}$-broadened parameters $\gamma_{\mathrm{ref,,H_2}}$ and $\mathsf{n}_{\mathrm{H_2}}$ are available, allowing $\mathrm{H_2}$-background pressure broadening to be computed via Eq.~(\ref{eq:gamma_p_L_hitemp}) \citep{tan2022}. In what follows, we refer to air-broadened HITEMP as $\mathrm{HITEMP_{Air}}$ and $\mathrm{H_2}$-broadened HITEMP as $\mathrm{HITEMP_{H_2}}$.
Nominally, air broadening parameters are not appropriate for brown dwarf atmospheres. However, the laboratory experiment by \citet{2025ApJ...984...92H} suggests that these parameters are not necessarily incorrect, so we include them in the comparison.

\subsubsection{Adopted Line Lists in This Study}

Given that brown dwarf atmospheres are dominated by $\mathrm{H_2}$, we adopt the ExoMol POKAZATEL for $\mathrm{H_2O}$ \citep{POKAZATEL}, ExoMol MM for $\mathrm{CH_4}$ \citep{MM}, and the ExoMol Coxon–Hajig line list for $\mathrm{HF}$ \citep{Coxon-Hajig}. To assess the impact of the CO line list --- which exhibits prominent absorption in the $K$ band --- on the retrieval results, we consider the following three CO line list configurations:
\begin{description}
\item[$\exomolco$] ExoMol CO line list Li2015 \citep{Li2015}, assuming an $\mathrm{H_2}$ background.
\item[$\hitempairco$] HITEMP CO line list \citep{Li2015} with air-broadening parameters.
\item[$\hitemphco$] An $\mathrm{H_2}$-broadened version of HITEMP CO based on the semi-empirical models constructed on available and validated experimental measurements and theoretical predictions \citep{tan2022}.
\end{description}
For all three CO line lists, the same Voigt profile treatment is applied, so that the differences in the retrieval results originate only from the broadening parameters.
The selection of molecular species was referenced from the $K$-band high-resolution spectral analysis of Luhman 16AB with IGRINS/Gemini South reported by \citet{Ishikawa2025}.
We also searched for the isotopologue $^{13}\mathrm{CO}$, but it was not detected in the present analysis. This is consistent with \citet{Ishikawa2025}, who did not detect $^{13}\mathrm{CO}$ in the same wavelength region. Therefore, we did not include $^{13}\mathrm{CO}$ opacity in the spectral model.

\subsection{Results}

\begin{figure*}[h!]
    \includegraphics[width=0.5\textwidth]{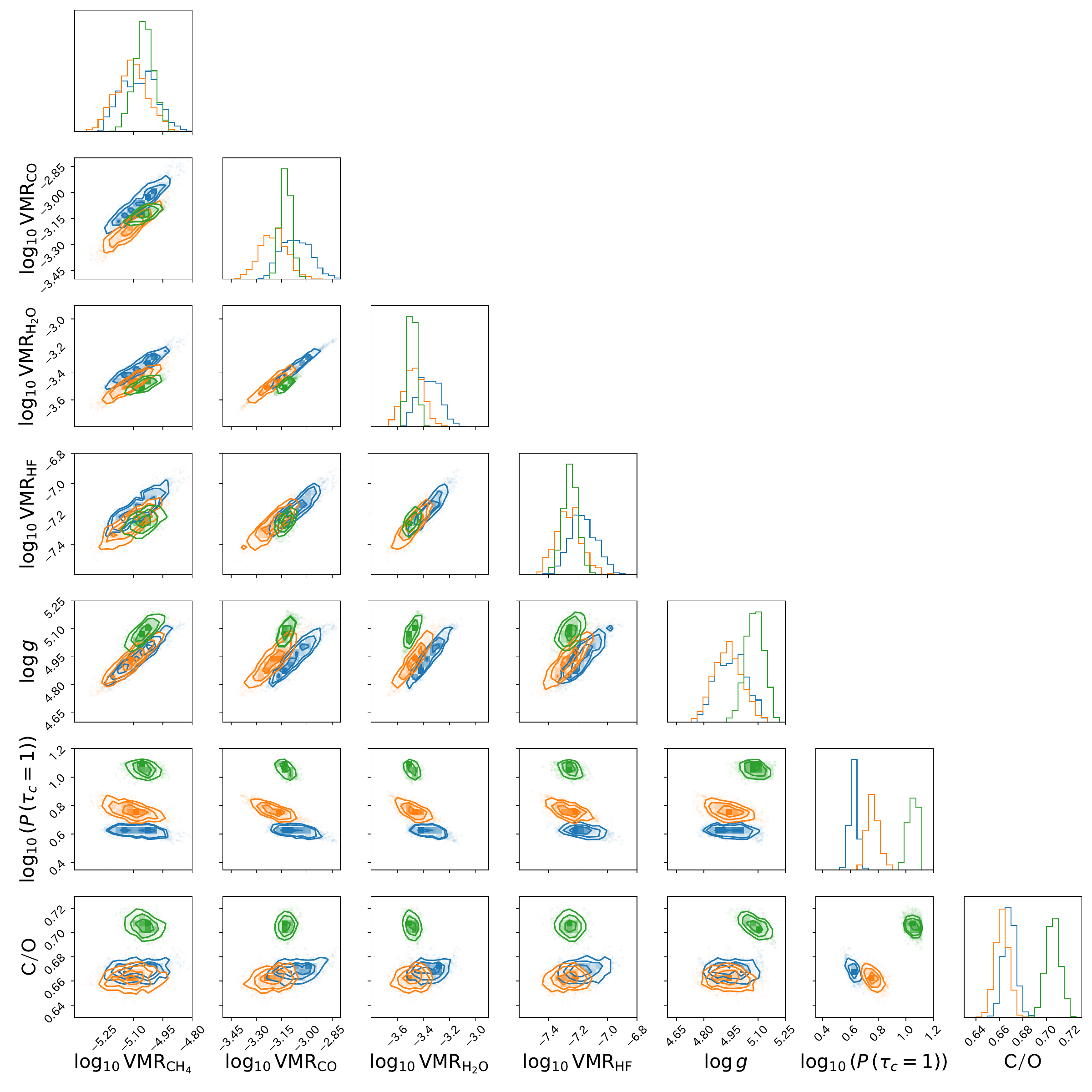}
    \includegraphics[width=0.5\textwidth]{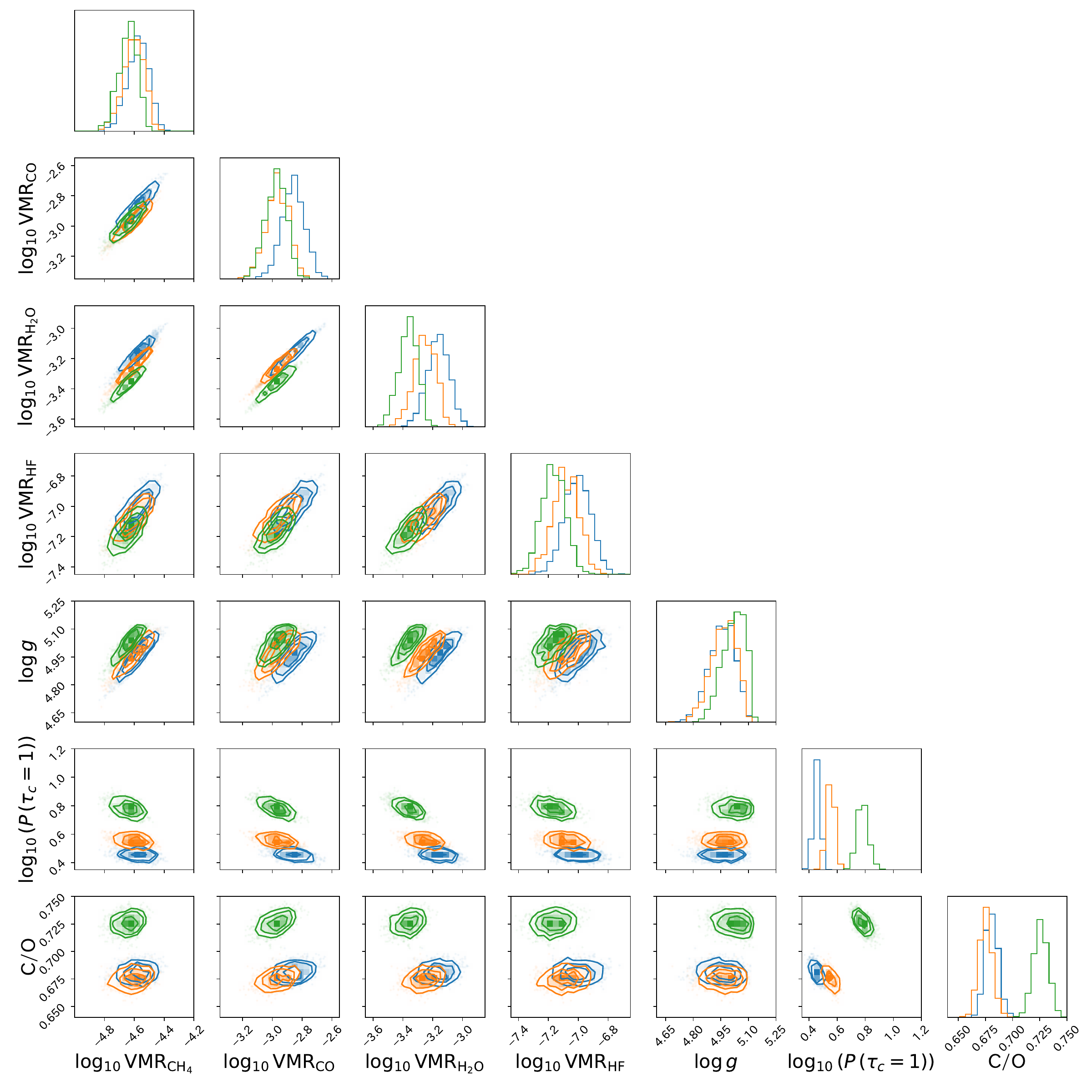}
    \caption{Corner plots for the posterior samples obtained by fitting the spectra with models that use CO line lists from $\exomolco$ (orange), $\hitempairco$ (green), and $\hitemphco$ (blue). (Left) Luhman 16A. (Right) Luhman 16B.}
    \label{fig:corner_plot_line_lists}
\end{figure*}

\begin{figure*}[h!]
    \includegraphics[width=0.5\textwidth]{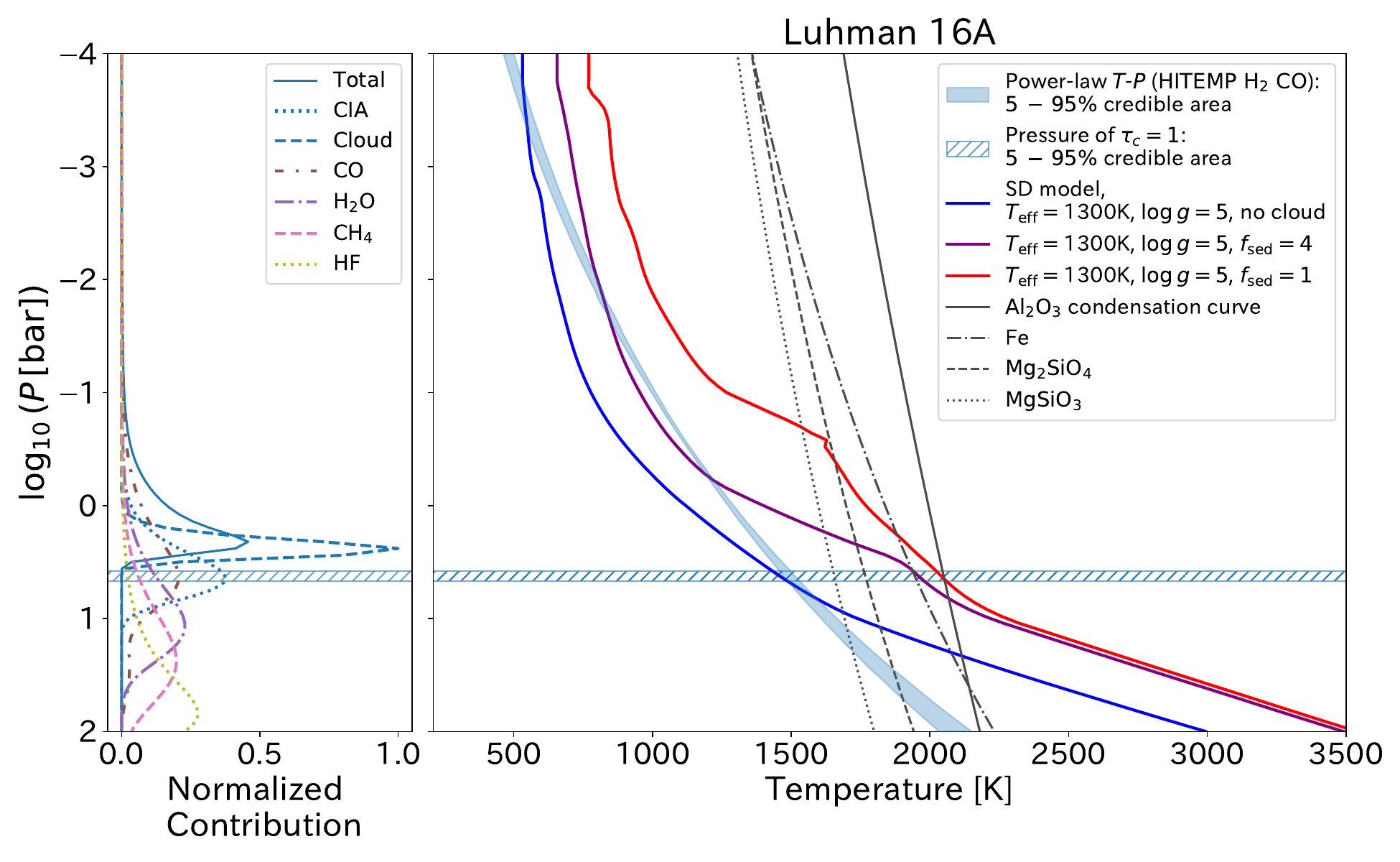}
    \includegraphics[width=0.5\textwidth]{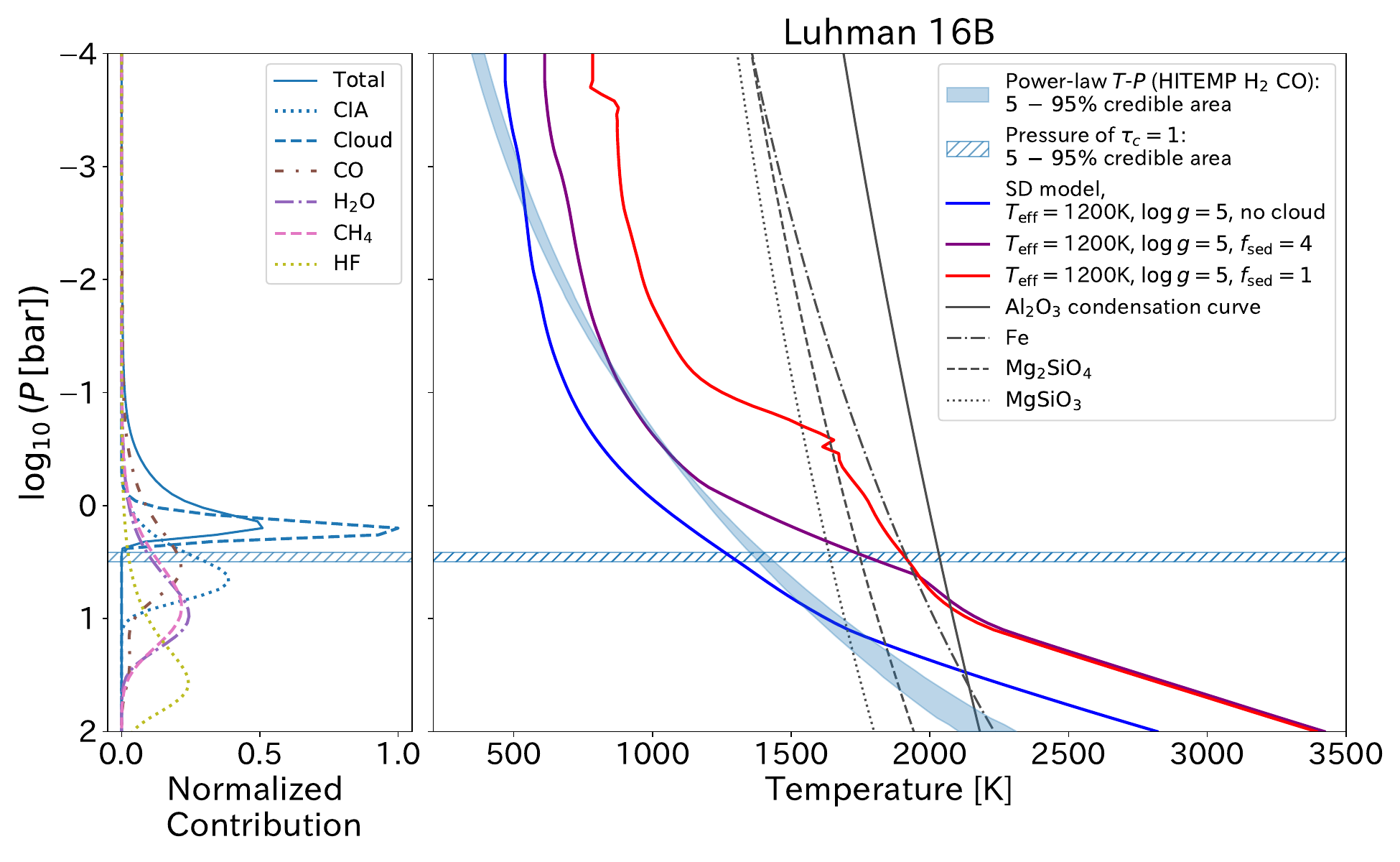}
    \caption{Retrieved $T$--$P$ profiles and normalized contribution fuctions for Luhman 16A (left) and Luhman 16B (right) under the $\hitemphco$ configuration. For each object, the left panel shows the contribution function, and the right panel shows the posterior $T$--$P$ profile. The horizontal shaded band marks the inferred cloud deck pressure, $\log_{10}P\left(\tau_c=1\right)$. Note that $\tau_c$ denotes the cumulative optical depth due only to cloud opacity. Blue, purple, and red lines are Sonora Diamondback (SD) $T$--$P$ models \citep{Morley2024}. Black curves in the right panels indicate condensation curves for the principal brown dwarf cloud species identified by \citet{Morley2024} (condensation curve formulas from \citealt{Visscher2010,Wakeford2017}).}
    \label{fig:est_PLTP_cloud_HITEMPH2}
\end{figure*}

\begin{deluxetable*}{l c c c c c}
\tablecaption{Parameter estimates for Luhman 16A under each modeling configuration.\label{table:result_Luhman16A}}
\tabletypesize{\footnotesize}
\tablewidth{\columnwidth}
\tablehead{
\colhead{Parameter} & \colhead{Unit} &
\colhead{\shortstack{\rule{0pt}{1.2em}PL $T$--$P$\\$\exomolco$}} &
\colhead{\shortstack{\rule{0pt}{1.2em}PL $T$--$P$\\$\hitempairco$}} &
\colhead{\shortstack{\rule{0pt}{1.2em}PL $T$--$P$\\$\hitemphco$}} &
\colhead{\shortstack{\rule{0pt}{1.2em}GP $T$--$P$\\$\hitemphco$}}
}
\startdata
$\log g$ & cgs & $4.93^{+0.07}_{-0.09}$ & $5.08^{+0.04}_{-0.06}$ & $4.95^{+0.08}_{-0.09}$ & $5.06^{+0.08}_{-0.07}$ \\
$RV$ & $\mathrm{km\,s^{-1}}$ & $28.13^{+0.03}_{-0.04}$ & $28.19^{+0.05}_{-0.05}$ & $28.17^{+0.03}_{-0.04}$ & $28.16^{+0.03}_{-0.04}$ \\
$\log_{10}\mathrm{VMR_{CO}}$ & -- & $-3.20^{+0.06}_{-0.10}$ & $-3.12^{+0.03}_{-0.04}$ & $-3.07^{+0.09}_{-0.09}$ & $-3.04^{+0.07}_{-0.08}$ \\
$\log_{10}\mathrm{VMR_{H_2O}}$ & -- & $-3.47^{+0.08}_{-0.07}$ & $-3.49^{+0.03}_{-0.03}$ & $-3.36^{+0.09}_{-0.08}$ & $-3.31^{+0.08}_{-0.07}$ \\
$\log_{10}\mathrm{VMR_{CH_4}}$ & -- & $-5.17^{+0.10}_{-0.07}$ & $-5.10^{+0.05}_{-0.05}$ & $-5.11^{+0.08}_{-0.09}$ & $-5.04^{+0.08}_{-0.05}$ \\
$\log_{10}\mathrm{VMR_{HF}}$ & -- & $-7.27^{+0.08}_{-0.09}$ & $-7.25^{+0.05}_{-0.05}$ & $-7.17^{+0.07}_{-0.11}$ & $-7.15^{+0.1}_{-0.07}$ \\
$\mathrm{[C/H]}$ & -- & $0.11^{+0.09}_{-0.07}$ & $0.19^{+0.03}_{-0.03}$ & $0.26^{+0.07}_{-0.09}$ & $0.28^{+0.07}_{-0.06}$ \\
$\mathrm{[O/H]}$ & -- & $0.06^{+0.08}_{-0.08}$ & $0.11^{+0.04}_{-0.03}$ & $0.21^{+0.06}_{-0.09}$ & $0.23^{+0.07}_{-0.05}$ \\
$\mathrm{[F/H]}$ & -- & $-3.96^{+0.09}_{-0.08}$ & $-3.95^{+0.05}_{-0.05}$ & $-3.85^{+0.09}_{-0.09}$ & $-3.84^{+0.07}_{-0.07}$ \\
$\mathrm{[(C+O)/H]}$ & -- & $0.08^{+0.08}_{-0.08}$ & $0.14^{+0.04}_{-0.03}$ & $0.23^{+0.06}_{-0.10}$ & $0.25^{+0.07}_{-0.05}$ \\
$\mathrm{C/O}$ & -- & $0.662^{+0.005}_{-0.006}$ & $0.704^{+0.007}_{-0.006}$ & $0.668^{+0.005}_{-0.006}$ & $0.660^{+0.005}_{-0.007}$ \\
$\log_{10} P_{\mathrm{c}}$ & bar & $1.66^{+0.03}_{-0.06}$ & $1.95^{+0.05}_{-0.02}$ & $1.52^{+0.02}_{-0.03}$ & $1.90^{+0.09}_{-0.05}$ \\
$\log_{10}P\left(\tau_c=1\right)$ & bar & $0.76^{+0.05}_{-0.03}$ & $1.05^{+0.05}_{-0.02}$ & $0.62^{+0.02}_{-0.03}$ & $1.00^{+0.10}_{-0.04}$ \\
$V\sin i$ & $\mathrm{km\,s^{-1}}$ & $16.21^{+0.37}_{-0.51}$ & $15.89^{+0.22}_{-0.29}$ & $17.09^{+0.27}_{-0.22}$ & $15.62^{+0.19}_{-0.23}$ \\
$q_1$ & -- & $0.82^{+0.17}_{-0.12}$ & $0.73^{+0.26}_{-0.11}$ & $0.90^{+0.10}_{-0.05}$ & $0.77^{+0.23}_{-0.13}$ \\
$q_2$ & -- & $0.43^{+0.17}_{-0.15}$ & $0.53^{+0.14}_{-0.14}$ & $0.29^{+0.09}_{-0.10}$ & $0.63^{+0.15}_{-0.12}$ \\
$\log_{10}a_{\mathrm{GP}}$ & -- & $-3.96^{+0.02}_{-0.02}$ & $-3.87^{+0.02}_{-0.02}$ & $-3.95^{+0.01}_{-0.02}$ & $-4.03^{+0.02}_{-0.02}$ \\
$\log_{10}\tau_{\mathrm{GP}}$ & \AA & $-1.27^{+0.01}_{-0.01}$ & $-1.22^{+0.02}_{-0.01}$ & $-1.27^{+0.01}_{-0.01}$ & $-1.30^{+0.01}_{-0.02}$ \\
$A_1$ & -- & $1.140^{+0.001}_{-0.001}$ & $1.140^{+0.001}_{-0.001}$ & $1.141^{+0.001}_{-0.001}$ & $1.140^{+0.001}_{-0.001}$ \\
$A_2$ & -- & $1.134^{+0.001}_{-0.001}$ & $1.134^{+0.001}_{-0.001}$ & $1.134^{+0.001}_{-0.001}$ & $1.133^{+0.001}_{-0.001}$ \\
$A_3$ & -- & $1.138^{+0.001}_{-0.001}$ & $1.138^{+0.001}_{-0.001}$ & $1.139^{+0.001}_{-0.001}$ & $1.139^{+0.001}_{-0.001}$ \\
$A_4$ & -- & $1.109^{+0.001}_{-0.001}$ & $1.108^{+0.001}_{-0.001}$ & $1.109^{+0.001}_{-0.001}$ & $1.108^{+0.001}_{-0.001}$ \\
\cutinhead{\itshape (Power-law $T$--$P$)}
$T_0$ & K & $1293^{+6}_{-6}$ & $1258^{+5}_{-5}$ & $1283^{+6}_{-6}$ & -- \\
$\alpha$ & -- & $0.109^{+0.003}_{-0.003}$ & $0.108^{+0.002}_{-0.003}$ & $0.107^{+0.003}_{-0.003}$ & -- \\
\cutinhead{\itshape (GP $T$--$P$)}
$T_{0,\,m}$ & K & -- & -- & -- & $1285^{+158}_{-143}$ \\
$\alpha_{m}$ & -- & -- & -- & -- & $0.110^{+0.027}_{-0.041}$ \\
$\log_{10}a_{u}$ & -- & -- & -- & -- & $-0.34^{+0.10}_{-0.18}$\\
$\log_{10}\tau_{u}$ & -- & -- & -- & -- & $-0.40^{+0.06}_{-0.10}$ \\
\cutinhead{\itshape (Negative log-likelihood)}
Normalized high-resolution spectra & -- & $-13177$ & $-13090$ & $-13167$ & $-13257$ \\
JWST medium-resolution flux density & -- & 13 & $-29$ & $-24$ & $-1$ \\
Total & -- & $-13164$ & $-13119$ & $-13190$ & $-13258$ \\
\enddata
\end{deluxetable*}

\begin{deluxetable*}{l c c c c c}
\tablecaption{Parameter estimates for Luhman 16B under each modeling configuration.\label{table:result_Luhman16B}}
\tabletypesize{\footnotesize}
\tablewidth{\columnwidth}
\tablehead{
\colhead{Parameter} & \colhead{Unit} &
\colhead{\shortstack{\rule{0pt}{1.2em}PL $T$--$P$\\$\exomolco$}} &
\colhead{\shortstack{\rule{0pt}{1.2em}PL $T$--$P$\\$\hitempairco$}} &
\colhead{\shortstack{\rule{0pt}{1.2em}PL $T$--$P$\\$\hitemphco$}} &
\colhead{\shortstack{\rule{0pt}{1.2em}GP $T$--$P$\\$\hitemphco$}}
}
\startdata
$\log g$ & cgs & $4.97^{+0.08}_{-0.06}$ & $5.03^{+0.08}_{-0.05}$ & $4.96^{+0.07}_{-0.07}$ & $5.03^{+0.07}_{-0.05}$ \\
$RV$ & $\mathrm{km\,s^{-1}}$ & $25.66^{+0.07}_{-0.06}$ & $25.59^{+0.07}_{-0.07}$ & $25.54^{+0.06}_{-0.08}$ & $25.52^{+0.06}_{-0.06}$ \\
$\log_{10}\mathrm{VMR_{CO}}$ & -- & $-2.96^{+0.09}_{-0.07}$ & $-2.97^{+0.09}_{-0.05}$ & $-2.86^{+0.07}_{-0.09}$ & $-2.95^{+0.06}_{-0.08}$ \\
$\log_{10}\mathrm{VMR_{H_2O}}$ & -- & $-3.25^{+0.09}_{-0.05}$ & $-3.36^{+0.08}_{-0.05}$ & $-3.16^{+0.08}_{-0.07}$ & $-3.21^{+0.06}_{-0.07}$ \\
$\log_{10}\mathrm{VMR_{CH_4}}$ & -- & $-4.65^{+0.08}_{-0.07}$ & $-4.69^{+0.06}_{-0.06}$ & $-4.61^{+0.10}_{-0.06}$ & $-4.61^{+0.07}_{-0.05}$ \\
$\log_{10}\mathrm{VMR_{HF}}$ & -- & $-7.08^{+0.10}_{-0.07}$ & $-7.17^{+0.08}_{-0.07}$ & $-7.00^{+0.10}_{-0.07}$ & $-7.10^{+0.09}_{-0.07}$ \\
$\mathrm{[C/H]}$ & -- & $0.36^{+0.09}_{-0.07}$ & $0.35^{+0.06}_{-0.06}$ & $0.47^{+0.09}_{-0.08}$ & $0.36^{+0.05}_{-0.07}$ \\
$\mathrm{[O/H]}$ & -- & $0.30^{+0.09}_{-0.06}$ & $0.25^{+0.07}_{-0.05}$ & $0.41^{+0.10}_{-0.07}$ & $0.31^{+0.06}_{-0.06}$ \\
$\mathrm{[F/H]}$ & -- & $-3.76^{+0.08}_{-0.09}$ & $-3.85^{+0.07}_{-0.08}$ & $-3.69^{+0.07}_{-0.10}$ & $-3.82^{+0.10}_{-0.07}$ \\
$\mathrm{[(C+O)/H]}$ & -- & $0.33^{+0.09}_{-0.06}$ & $0.29^{+0.06}_{-0.07}$ & $0.43^{+0.09}_{-0.07}$ & $0.33^{+0.06}_{-0.07}$ \\
$\mathrm{C/O}$ & -- & $0.674^{+0.005}_{-0.008}$ & $0.725^{+0.007}_{-0.006}$ & $0.679^{+0.006}_{-0.006}$ & $0.660^{+0.007}_{-0.007}$ \\
$\log_{10} P_{\mathrm{c}}$ & bar & $1.45^{+0.03}_{-0.03}$ & $1.68^{+0.03}_{-0.04}$ & $1.35^{+0.03}_{-0.02}$ & $1.93^{+0.07}_{-0.03}$ \\
$\log_{10}P\left(\tau_c=1\right)$ & bar & $0.55^{+0.02}_{-0.03}$& $0.78^{+0.03}_{-0.04}$ & $0.45^{+0.03}_{-0.02}$ & $1.04^{+0.07}_{-0.03}$ \\
$V\sin i$ & $\mathrm{km\,s^{-1}}$ & $27.65^{+0.28}_{-0.53}$ & $27.30^{+0.43}_{-0.37}$ & $27.79^{+0.46}_{-0.48}$ & $26.93^{+0.56}_{-0.47}$ \\
$q_1$ & -- & $0.81^{+0.19}_{-0.08}$ & $0.81^{+0.18}_{-0.09}$ & $0.80^{+0.19}_{-0.09}$ & $0.79^{+0.21}_{-0.11}$ \\
$q_2$ & -- & $0.36^{+0.12}_{-0.11}$ & $0.40^{+0.12}_{-0.12}$ & $0.34^{+0.11}_{-0.12}$ & $0.48^{+0.14}_{-0.14}$ \\
$\log_{10}a_{\mathrm{GP}}$ & -- & $-3.83^{+0.02}_{-0.02}$ & $-3.75^{+0.02}_{-0.02}$ & $-3.83^{+0.02}_{-0.02}$ & $-3.88^{+0.02}_{-0.02}$ \\
$\log_{10}\tau_{\mathrm{GP}}$ & \AA & $-1.32^{+0.01}_{-0.02}$ & $-1.28^{+0.02}_{-0.02}$ & $-1.32^{+0.01}_{-0.02}$ & $-1.35^{+0.01}_{-0.01}$ \\
$A_1$ & -- & $1.141^{+0.001}_{-0.001}$ & $1.141^{+0.001}_{-0.001}$ & $1.142^{+0.001}_{-0.001}$ & $1.140^{+0.001}_{-0.001}$ \\
$A_2$ & -- & $1.141^{+0.001}_{-0.001}$ & $1.141^{+0.001}_{-0.001}$ & $1.141^{+0.001}_{-0.001}$ & $1.140^{+0.001}_{-0.001}$ \\
$A_3$ & -- & $1.141^{+0.001}_{-0.001}$ & $1.140^{+0.001}_{-0.001}$ & $1.142^{+0.001}_{-0.001}$ & $1.142^{+0.001}_{-0.001}$ \\
$A_4$ & -- & $1.108^{+0.001}_{-0.001}$ & $1.107^{+0.001}_{-0.001}$ & $1.108^{+0.001}_{-0.001}$ & $1.108^{+0.001}_{-0.001}$ \\
\cutinhead{\itshape (Power-law $T$--$P$)}
$T_0$ & K & $1219^{+7}_{-9}$ & $1177^{+6}_{-6}$ & $1215^{+8}_{-9}$ & -- \\
$\alpha$ & -- & $0.129^{+0.003}_{-0.006}$ & $0.123^{+0.005}_{-0.004}$ & $0.128^{+0.004}_{-0.005}$ & -- \\
\cutinhead{\itshape (GP $T$--$P$)}
$T_{0,\,m}$ & K & -- & -- & -- & $1165^{+82}_{-143}$ \\
$\alpha_{m}$ & -- & -- & -- & -- & $0.104^{+0.023}_{-0.044}$ \\
$\log_{10}a_{u}$ & -- & -- & -- & -- & $-0.36^{+0.11}_{-0.19}$ \\
$\log_{10}\tau_{u}$ & -- & -- & -- & -- & $-0.37^{+0.08}_{-0.13}$ \\
\cutinhead{\itshape (Negative log-likelihood)}
Normalized high-resolution spectra & -- & $-12239$ & $-12165$ & $-12251$ & $-12295$ \\
JWST medium-resolution flux density & -- & $-28$ & $-14$ & $-13$ & $-28$ \\
Total & -- & $-12268$ & $-12179$ & $-12265$ & $-12323$ \\
\enddata
\end{deluxetable*}

Using spectral models based on each CO line list, we performed NUTS sampling and obtained posterior distributions for all parameters. Posterior summaries are listed in Tables~\ref{table:result_Luhman16A} and \ref{table:result_Luhman16B}. 

Figure~\ref{fig:fit_PLTP_HITEMPH2} compares, for both components of Luhman 16, the data and the best-fitting spectral models computed with a power-law $T$–$P$ profile, a gray-cloud prescription, and $\hitemphco$. Overall, the models reproduce the high-resolution spectra well: the residual panels show no conspicuous absorption features that are clearly misfit. The largest residual structures may reflect limitations of the spectral model (e.g., line list incompleteness or pressure broadening uncertainties), although residual telluric contamination or detector-related systematics cannot be ruled out. Although not shown, the other two CO line lists yield comparably good fits.

To examine posterior structure, Figure~\ref{fig:corner_plot_line_lists} overlays selected parameters from the three line list runs in a corner plot. From Eq.~(\ref{eq:dtau_mol}), the molecular optical depth increment scales as $\Delta\tau_{\mathrm{mol},i} \propto \mathrm{VMR}_i/(\mu g)$ (at fixed $T$ and line shape), where the mean molecular weight $\mu$ is computed self-consistently from the adopted VMRs of all species at each sampling step. Since the variation in $\mu$ is negligibly small, its effect on the opacity is minor, and the spectra primarily constrain the ratio $\mathrm{VMR}_i/g$. Increases in $g$ can therefore be offset by proportional increases in $\mathrm{VMR}_i$, producing the observed positive covariance between $\log g$ and the molecular VMRs. In our spectra the key features of each species are largely separated in wavelength, so the relative line depths directly constrain abundance ratios; accordingly, the retrieved VMRs exhibit positive covariance.
These correlations between $\log g$ and the VMRs appear as diagonally elongated contours in the off-diagonal panels of Figure~\ref{fig:corner_plot_line_lists}. However, this correlation does not translate into a simple one-to-one trend in the one-dimensional summaries (Tables~\ref{table:result_Luhman16A} and \ref{table:result_Luhman16B}), because the correlation structures in the joint posteriors can also shift due to line-list–dependent systematics and the inferred cloud-deck pressure.


The retrieved $T$--$P$ profiles under the power-law + gray-cloud + $\hitemphco$ configuration are shown for each object in Figure~\ref{fig:est_PLTP_cloud_HITEMPH2}. Across all line lists, the inferred temperature ranges in the regions of largest contribution are consistent with the Sonora diamondback models \citep{Morley2024} at the literature effective temperatures. 
Here, $\tau_c$ denotes the cumulative optical depth due only to the gray-cloud opacity integrated downward from the top of the atmosphere, and the pressure level at $\tau_c = 1$ corresponds to the cloud deck in our model. The inferred cloud-deck pressures fall within the altitude ranges predicted by the condensation curves in the Sonora Diamondback models, indicating broad physical consistency with the cloud-base pressures of plausible condensates, although we do not attempt to identify a specific cloud species.

We use the VMR in the retrieval, so that the number density of each species is directly represented as $n_i = \mathrm{VMR}_i$. 
From these, we compute the elemental number densities as
\begin{eqnarray}
    &&n_ {\mathrm{C}}= n_{\mathrm{CO}}+n_{\mathrm{CH_4}}, \\
    &&n_{\mathrm{O}}= n_{\mathrm{CO}}+n_{\mathrm{H_2O}}, \\
    &&n_{\mathrm{F}}= n_{\mathrm{HF}}, \\
    &&n_{\mathrm{H}}= 2n_{\mathrm{H_2}}+4n_{\mathrm{CH_4}}+2n_{\mathrm{H_2O}}+n_{\mathrm{HF}},
\end{eqnarray}
and the carbon-to-oxygen ratio
\begin{equation}
    \label{eq:CO_ratio}
    \mathrm{C/O}=\frac{n_{\mathrm{C}}}{n_{\mathrm{O}}}.
\end{equation}
We also compute $\mathrm{[C/H]}$, $\mathrm{[O/H]}$, $\mathrm{[F/H]}$, and $\mathrm{[(C+O)/H]}$ as
\begin{eqnarray}
    &&\mathrm{[X/H]}=\log_{10}\left(\frac{n_{\mathrm{X}}}{n_{\mathrm{H}}}\right)-\log_{10}\left(\mathrm{X/H}\right)_{\odot},\ \mathrm{X}\in\{\mathrm{C,O,F}\}, \\
    &&\mathrm{[(C+O)/H]}=\log_{10}\left(\frac{n_{\mathrm{C}}+n_{\mathrm{O}}}{n_{\mathrm{H}}}\right)-\log_{10}\left(\mathrm{C+O \over H}\right)_{\odot},
\end{eqnarray}
where the solar ratios $\left(\mathrm{X/H}\right)_{\odot}$ follow \citet{Asplund2021}. For both Luhman 16A and 16B, the VMR-inferred C/O is $\sim 0.67$ --- slightly above solar (Figure~\ref{fig:corner_plot_line_lists}). We nevertheless find that the inferred C/O shifts by $\sim7\%$ depending on the adopted line list. Differences in the line shapes alter the inferred cloud-deck pressure and, in turn, the relative VMRs.

\begin{figure*}[h!]
    \includegraphics[width=0.5\textwidth]{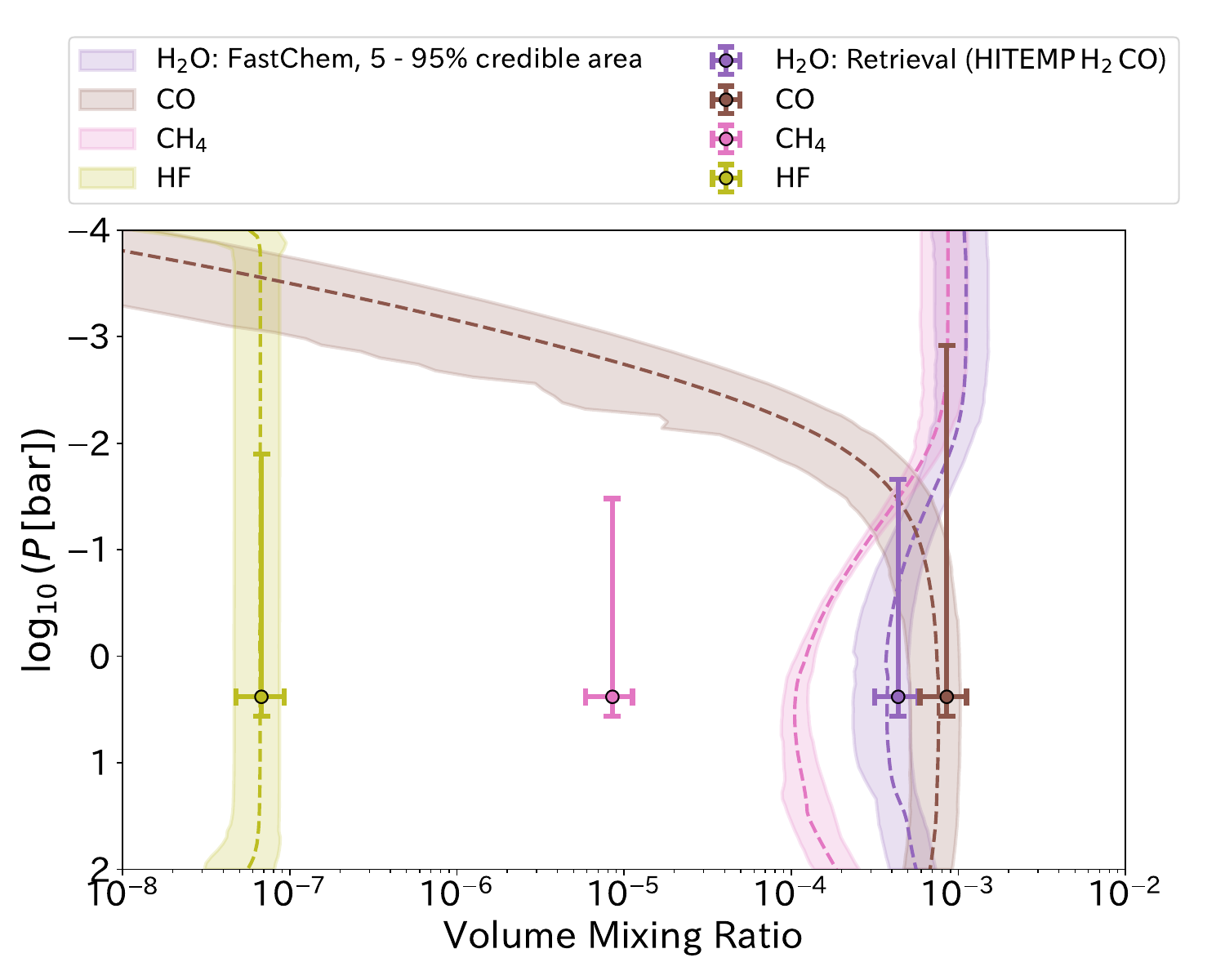}
    \includegraphics[width=0.5\textwidth]{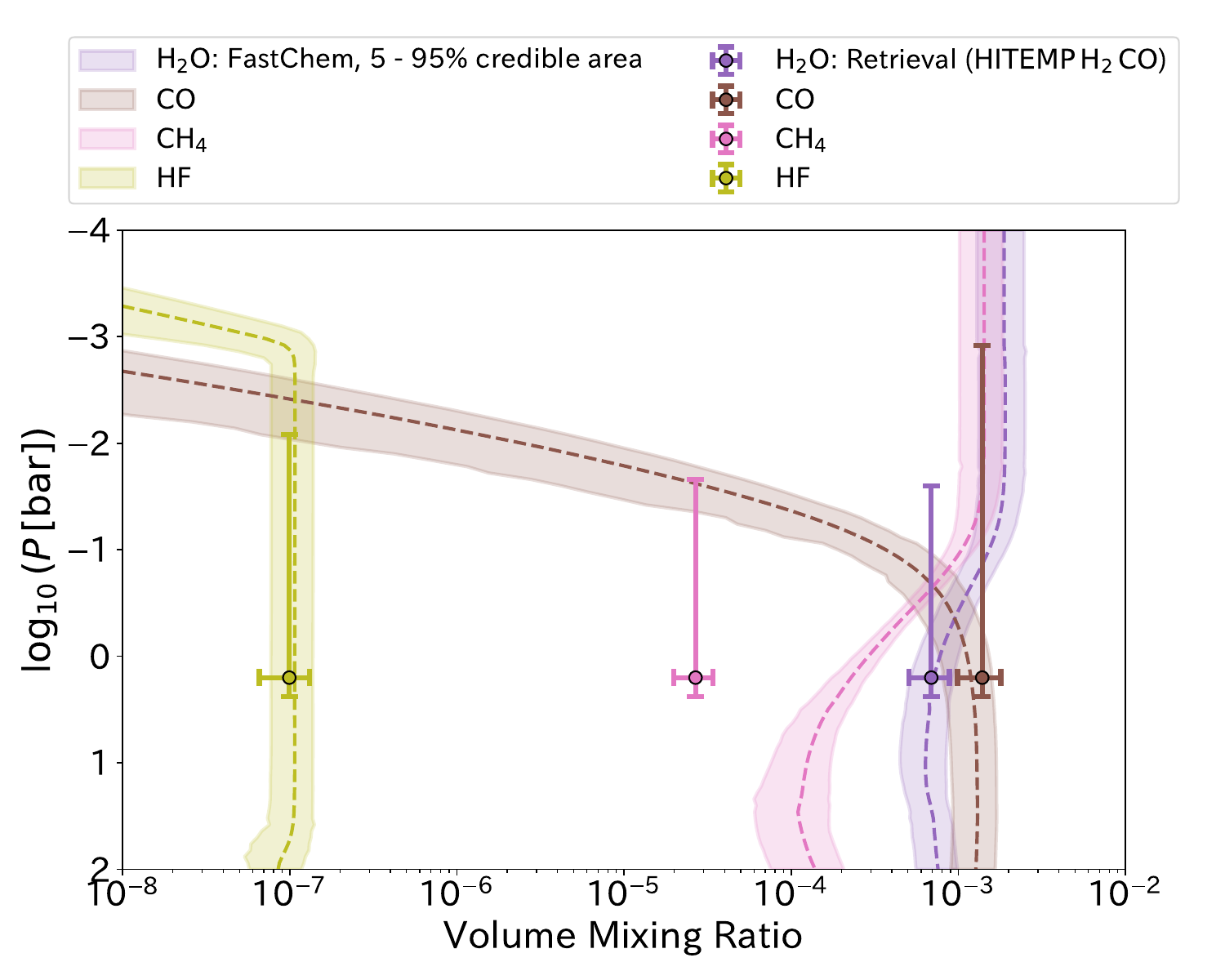}
    \caption{Retrieved VMRs of $\mathrm{H_2O}$ (purple), $\mathrm{CO}$ (brown), $\mathrm{CH_4}$ (pink), and $\mathrm{HF}$ (yellow) for Luhman 16A (left) and Luhman 16B (right). Points are plotted at the pressure of the contribution function peak; the vertical extent indicates the pressure range that contributes appreciably for each molecule, with the upper bound truncated at the cloud-deck pressure. For comparison, thermochemical-equilibrium VMRs computed with FastChem, including equilibrium condensation, are shown as 5–95\% credible bands.}
    \label{fig:volume_mixing_ratio_PLTP}
\end{figure*}

To interpret the retrieved abundances, we computed thermochemical-equilibrium VMRs with FastChem including equilibrium condensation \citep{fastchem, fastchem_cond}. In these calculations, we adjusted the solar C, O, and F abundances of \citet{Asplund2021} by our retrieved $\mathrm{[X/H]}$ values (all other elements fixed to solar). \citet{Asplund2021} tabulate $A(\mathrm{X})\equiv\log_{10}(n_{\mathrm{X}}/n_{\mathrm{H}})+12$ with $A(\mathrm{H})=12$; accordingly we adopt
\begin{equation}
    A(\mathrm{X}) = A_{\odot}(\mathrm{X}) + \mathrm{[X/H]},\quad \mathrm{X}\in\{\mathrm{C,O,F}\},
\end{equation}
evaluated separately for Luhman 16A and 16B using their respective $\mathrm{[X/H]}$. Numerically, $A_{\odot}(\mathrm{C})=8.46$, $A_{\odot}(\mathrm{O})=8.69$, and $A_{\odot}(\mathrm{F})=4.40$ \citep{Asplund2021}. The equilibrium VMRs computed with these inputs and the retrieved $T$–$P$ profiles (Figure~\ref{fig:volume_mixing_ratio_PLTP}) agree well for $\mathrm{CO}$, $\mathrm{H_2O}$, and $\mathrm{HF}$, indicating compatibility with chemical equilibrium under the inferred thermal structures. By contrast, the retrieved $\mathrm{CH_4}$ VMRs for both objects are about an order of magnitude lower than the FastChem predictions; we revisit this discrepancy in Section~\ref{subsec:GPTP_luh16}.

\subsubsection{Comparison with Medium-Resolution Flux Density}

\begin{figure*}[h!]
    \includegraphics[width=0.5\textwidth]{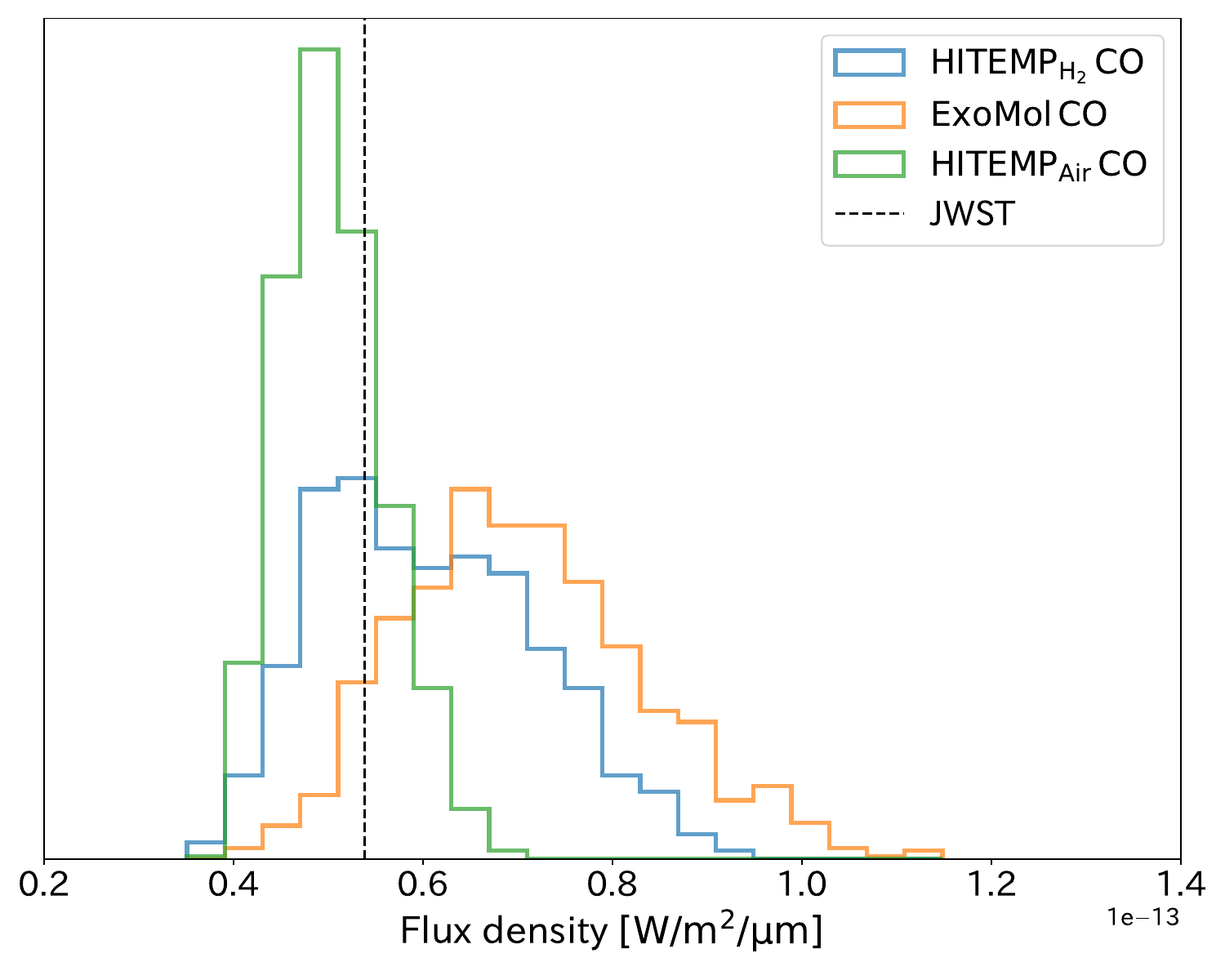}
    \includegraphics[width=0.5\textwidth]{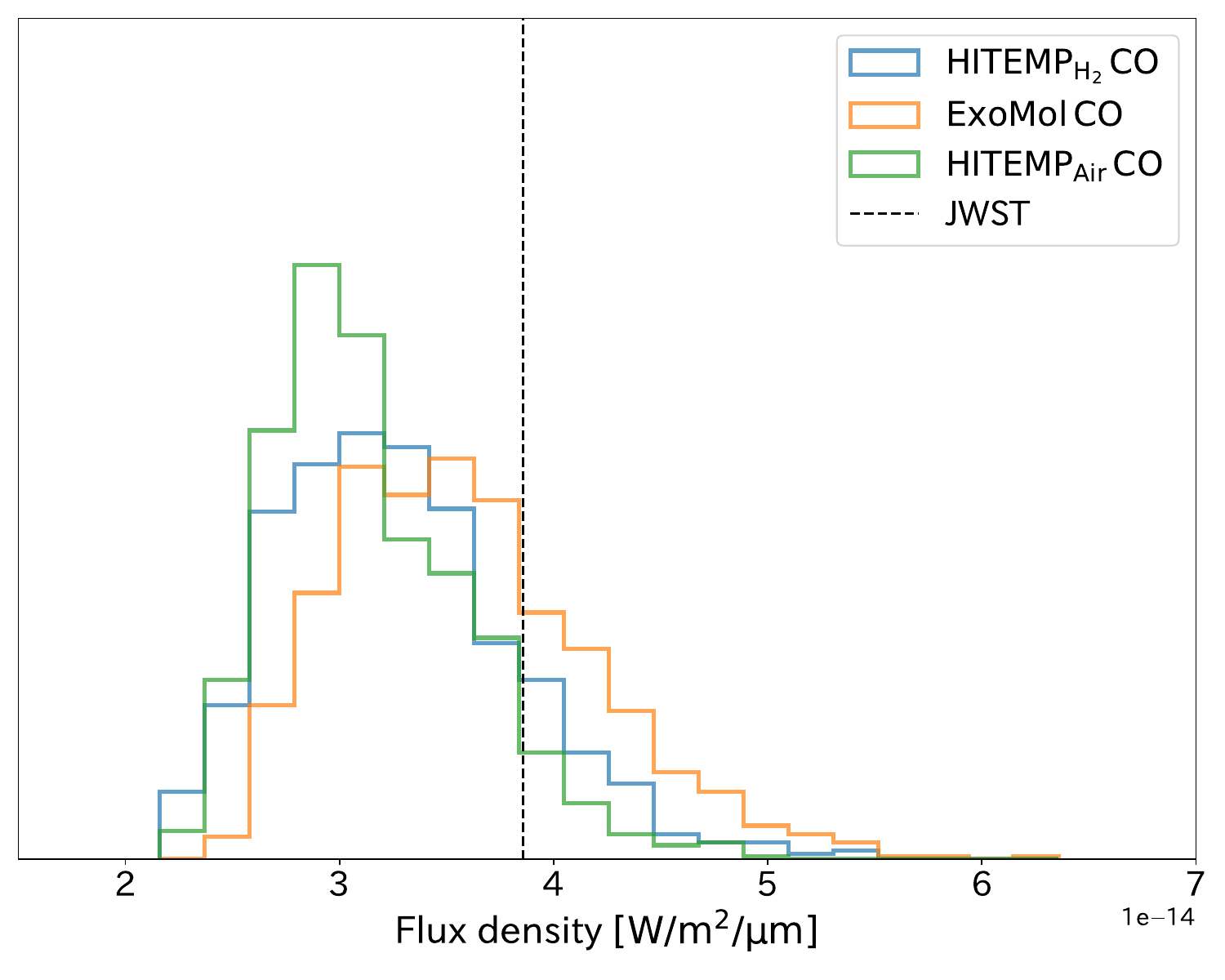}
    \caption{Comparison between the observed JWST medium-resolution flux density (vertical dashed lines) at the central wavelength of chip~1 ($2.30\,\mathrm{\mu m}$) and the model predictions based on different CO line lists (histograms). The model flux densities are converted from ExoJAX emergent fluxes to JWST units using Eq.~\ref{eq:wn_to_wl_fixed}. (Left) Luhman 16A. (Right) Luhman 16B.}
    \label{fig:flux_density_vs_JWST}
\end{figure*}

Although the retrievals above were performed on normalized spectra, the forward models also provide absolute flux densities. Since observational measurements of the absolute flux are available, checking their consistency with the model is also important for validating the retrieval. We therefore compared the JWST medium-resolution flux density with the model flux densities computed using each CO line list at the chip1 central wavelength $\lambda_1=2.30\,\mu\mathrm{m}$ (Figure~\ref{fig:flux_density_vs_JWST}).
To enable a direct comparison, we converted the mean emergent flux density per wavenumber from ExoJAX, $\boldsymbol{\bar{F}}_\mathrm{model}\,[\mathrm{erg/cm^2/s/cm^{-1}}]$, to the observed flux density in JWST units $F_{\mathrm{JWST}}[\mathrm{W/m^{2}/\mu m}]$ via
\begin{equation}
    F_{\mathrm{JWST}}(\lambda_1)
    = \left(\frac{R_\mathrm{BD}}{D}\right)^2 \boldsymbol{\bar{F}}_\mathrm{model}\;
      \frac{10}{\lambda_1^{2}},
    \label{eq:wn_to_wl_fixed}
\end{equation}
where $\lambda_1$ is in $\mu$m and the prefactor $10/\lambda_1^{2}$ results from $1\,\mathrm{erg}=10^{-7}\,\mathrm{J}$, $1\,\mathrm{cm^{-2}}=10^{4}\,\mathrm{m^{-2}}$,
and $d\nu/d\lambda_\mu=10^{4}/\lambda_\mu^{2}$ for wavenumber $\tilde{\nu}$ in $\mathrm{cm^{-1}}$ and wavelength in $\mu$m.
The brown dwarf radius $R_\mathrm{BD}=\sqrt{GM/g}$ is sampled from the retrieved surface gravity $g$ and mass prior,
and the distance $D$ is computed from the parallax $\varpi=500.5\pm0.11$\,mas measured by \citet{Sahlmann2015}. We also include a $3\%$ photometric uncertainty to account for systematic errors in JWST spectroscopy \citep{Biller2024}.
We find that the observations and predictions show reasonable agreement for both Luhman 16A and 16B across all line lists. 
Thus, our models are consistent not only in spectral morphology but also in the absolute flux density. 
While the strong prior on $\log g$ and the well-constrained mass and distance contribute to the agreement in the absolute flux density, the retrieved temperature structure and cloud properties also play a significant role. Therefore, the flux density agreement is not trivial; rather, it provides a valuable cross-check between the observed spectra and the model predictions. Because the prior on $\log g$ was informed by evolutionary models, this comparison also provides an external consistency check with those evolutionary constraints.


To assess the agreement across line lists, we compared the total negative log-likelihoods from the combined CRIRES spectra and JWST flux density; the values are shown in the bottom of Table~\ref{table:result_Luhman16A} and \ref{table:result_Luhman16B}. Among the three CO line lists, $\mathrm{HITEMP_{Air}}$ consistently yielded the poorest fit for both Luhman 16A and B. The remaining two, $\exomolco$ and $\hitemphco$, performed similarly, with $\hitemphco$ providing the best agreement for Luhman 16A and nearly matching $\exomolco$ for Luhman 16B. We therefore focus on the $\hitemphco$ results hereafter, noting that our conclusions are insensitive to the choice of line list.

\subsubsection{Impact of Photometric Variability in Luhman 16B}\label{ssec:PLTP_variability}

Luhman 16A shows $>2\%$ variability over a 7-hr baseline \citep{Biller2024}, and Luhman 16B exhibits $\sim$3\% variability over 5 hr \citep{Crossfield2014} and up to $\sim$6\% over 7 hr \citep{Biller2024}. Such variability, common among L/T transition brown dwarfs including Luhman 16AB, is often attributed to heterogeneous silicate/iron clouds whose evolving rotational modulation produces temporal brightness changes \citep{Radigan2014,Crossfield2014,Chen2024,Saumon_Marley2008}. The CRIRES time series used here comprise 14 spectra ($\sim$5 hr) for each component \citep{Crossfield2014} and show comparable normalized flux variability: 2.3\% for A and 3.2\% for B.

\begin{figure*}[h!]
    \includegraphics[width=0.5\textwidth]{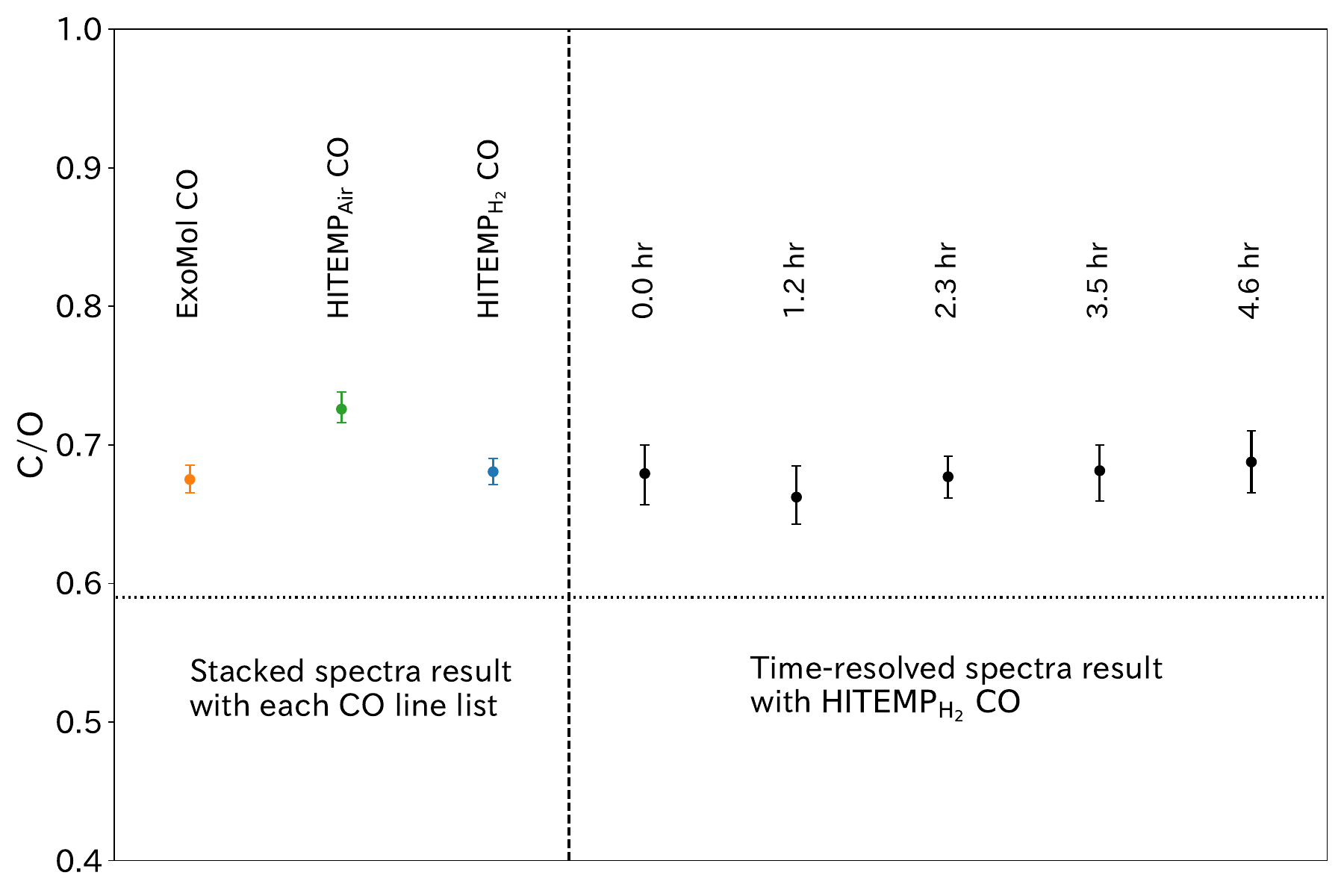}
    \includegraphics[width=0.5\textwidth]{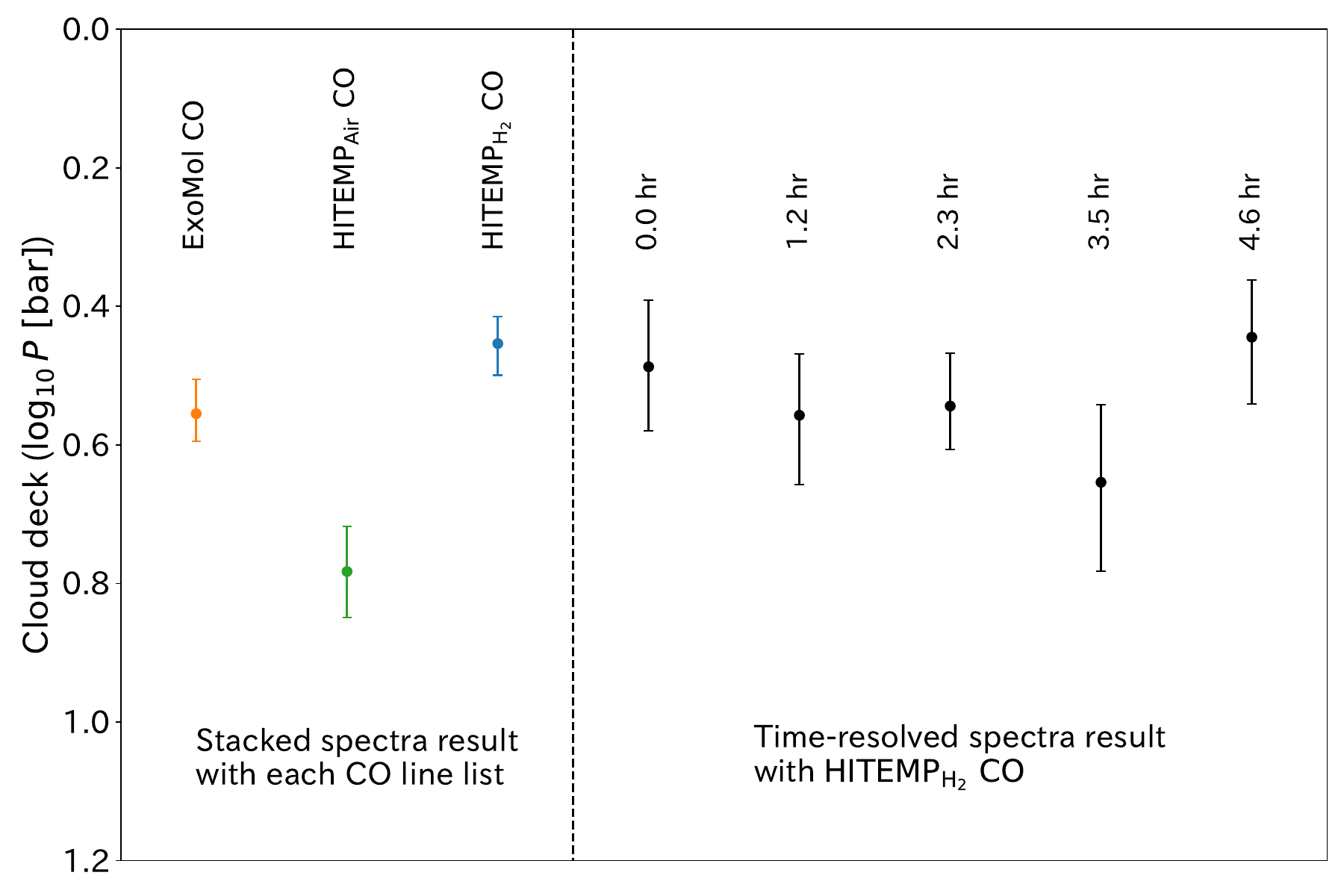}
    \caption{C/O and cloud deck pressure for Luhman 16B from stacked and time-resolved spectra. Left: C/O. Right: cloud deck pressure. In each panel, the region left of the vertical dashed line shows results from stacked spectra using three CO line lists (orange: $\mathrm{ExoMol}$; green: $\mathrm{HITEMP_{Air}}$; blue: $\mathrm{HITEMP_{H_2}}$). The region right of the dashed line shows epoch-by-epoch (time-resolved) retrievals using $\hitemphco$; from the 14 spectra spanning 0.0–5.0 hr we use five representative epochs. Error bars denote 90\% credible region. In the left panel, the dotted horizontal line marks the solar C/O \citep[$= 0.59$; ][]{Asplund2021}.}
    \label{fig:result_each_epoch}
\end{figure*}

Our baseline retrieval assumes a horizontally homogeneous (i.e., effectively averaged) atmosphere, neglecting cloud patchiness. To assess potential systematics on C/O and cloud‐top pressure from this assumption, we performed independent retrievals for each epoch of Luhman 16B using $\hitemphco$. We find no statistically significant differences relative to the stacked‐spectrum results in C/O or cloud deck pressure (Figure~\ref{fig:result_each_epoch}); quantitatively, the epoch-by-epoch C/O values agree with the stacked value ($\hitemphco$) to within at most $1.3\sigma$, and the cloud deck pressures to within at most $2.6\sigma$. Hence, for the present data, the impact of ignoring cloud inhomogeneity appears subdominant to systematics associated with the line lists.
This conclusion is consistent with the analysis of the $J$-band VLT/CRIRES$^+$ spectrum by \citet{deRegt2025}, who found that a patchy-cloud model is only weakly preferred for Luhman 16B and that the retrieval outcomes are largely insensitive to the inclusion of patchiness.

\section{Retrievals Using a Flexible Gaussian Process $T$--$P$ Profile}
\label{sec:GPTP}

Some high-resolution retrieval studies of brown-dwarf and exoplanet atmospheres adopt simple parameterizations of the $T$–$P$ profile, such as a power-law form \citep[e.g.,][]{Kawahara2022,Kawashima2025,Kasagi2025}. Although efficient, such profiles can be overly restrictive. More flexible parameterizations have therefore been proposed, including schemes that parameterize the temperature gradient in discrete pressure layers \citep{Zhang2023} and $P$-splines \citep[e.g.,][]{Line2015, deRegt2025}. However, these approaches still remain tied to specific functional forms and may fail to capture complex structures. 

As a potential future extension that takes advantage of the computational efficiency of HMC/NUTS in high-dimensional settings,\footnote{The convergence time of conventional random Metropolis–Hastings scales as $D^2$ with $D$ being the number of parameters, whereas HMC scales as $D^{5/4}$ \citep{HMC}. There are also studies suggesting that NUTS scales as $D^{1/4}$ when initialized in the concentration region of the canonical Gaussian measure \citep{Bou-Rabee2024}.} \citet{Kawahara2022} proposed a nonparametric $T$--$P$ profile model based on Gaussian processes \citep[GP,][]{2023ARA&A..61..329A}. That work, however, did not explore its practical applicability in real retrievals, its sensitivity to other parameters, or its validation through injection–recovery tests. Here we refine the GP-based $T$--$P$ profile model into a more practical framework and demonstrate its performance through application to the retrieval of Luhman 16AB data.

\subsection{Formulation of the GP $T$--$P$ Profile}
\label{subsec:GPTP}

To flexibly represent temperature--pressure ($T$--$P$) structures while maintaining numerical stability, 
we define the temperature profile as a GP in a whitened and logistically mapped latent space.
This approach ensures physically bounded temperatures, reduces parameter correlations during NUTS sampling,
and avoids the need for explicit box priors.

We model the temperature at 21 pressure layers, denoted $\boldsymbol{T}_{\mathrm{GP}}$,
as a logistic transformation of a underlying latent variable $\boldsymbol{u}$:
\begin{equation}
\boldsymbol{T}_{\mathrm{GP}} = T_{\min} + (T_{\max}-T_{\min})\,\sigma(\boldsymbol{u}),
\quad \sigma(\boldsymbol{u})=\frac{1}{1+e^{-\boldsymbol{u}}},
\end{equation}
where $T_{\min}$ and $T_{\max}$ are the physically allowed temperature limits,
set to 210\,K and 3500\,K, respectively.
This transformation guarantees $T_{\mathrm{GP},i} \in [T_{\min},T_{\max}]$ by construction.
The latent vector $\boldsymbol{u}$ follows a GP prior
with mean function $\boldsymbol{m}_u$ and covariance matrix $\boldsymbol{\Sigma}_u$:
\begin{equation}
\boldsymbol{u} \sim \mathcal{GP}\!\left(\boldsymbol{m}_u,\,\boldsymbol{\Sigma}_u\right).
\end{equation}
The covariance matrix is defined as
\begin{equation}
\Sigma_{u,\,ij}
= K_{u,\,ij}+\sigma_{u}^2\,\delta_{ij},
\end{equation}
where we fixed a small jitter term at $\sigma_{u}=10^{-6}$ for numerical stability. This negligible value ensures that the term has no physical impact on the retrieved $T$--$P$ profiles.
We adopt the Matérn 3/2 kernel,
\begin{equation}
K_{u,\,ij}
= a_{u}^2
\left(1+\frac{\sqrt{3}\,|\log_{10}P_i-\log_{10}P_j|}{\tau_{u}}\right)
\exp\!\left[-\frac{\sqrt{3}\,|\log_{10}P_i-\log_{10}P_j|}{\tau_{u}}\right],
\end{equation}
with amplitude $a_{u}$ and correlation length $\tau_{u}$
treated as free hyperparameters with uniform priors on $\log_{10}a_{u}$
and $\log_{10}\tau_{u}$.
The mean function $\boldsymbol{m}_u$ is defined as the logit transform of a baseline temperature profile
$\boldsymbol{T}_{\mathrm{mean}}$:
\begin{equation}
\boldsymbol{m}_u =
\log\!\left(\frac{\boldsymbol{T}_{\mathrm{mean}} - T_{\min}}{T_{\max}-T_{\min}}\right)-\log\!\left(1-\frac{\boldsymbol{T}_{\mathrm{mean}} - T_{\min}}{T_{\max}-T_{\min}}\right),
\end{equation}
where $\boldsymbol{T}_{\mathrm{mean}}$ follow a power-law form
\begin{equation}
T_{\mathrm{mean}}(P)
= T_{0,\,m}\!\left(\frac{P}{1~\mathrm{bar}}\right)^{\alpha_m},
\end{equation}
with the parameters $T_{0,\,m}$ and $\alpha_m$ inferred simultaneously with the GP hyperparameters.
For efficient sampling, we adopt the whitened representation:
\begin{equation}
\boldsymbol{u} = \boldsymbol{m}_u + \mathbf{L}\,\boldsymbol{z},
\quad \boldsymbol{z}\sim \mathcal{N}(\mathbf{0},\mathbf{I}),
\end{equation}
where $\mathbf{L}$ is the Cholesky factor of the covariance matrix $\boldsymbol{\Sigma}_u$, 
and $\mathbf{I}$ denotes the identity matrix. The whitened variable $\boldsymbol{z}$ is directly sampled in the NUTS inference, 
which improves convergence and decorrelates the GP components.
Finally, the temperature vector $\boldsymbol{T}_{\mathrm{GP}}$ at the 21 layers is
interpolated to the 101 layer grid via
$\mathrm{IP}(\boldsymbol{T}_{\mathrm{GP}})$.
The joint posterior is
\begin{eqnarray}
p\!\left(\boldsymbol{\theta},\,\boldsymbol{u},\,T_{0,m},\,\alpha_m,\,a_u,\,\tau_u \mid \boldsymbol{d}\right)
&\propto&
\mathcal{L}\!\left(\boldsymbol{d}\mid \boldsymbol{\theta},\,\mathrm{IP}\!\big(\boldsymbol{T}_{\rm GP}(\boldsymbol{u})\big)\right)\,
p\!\left(\boldsymbol{\theta}\right)\nonumber\\
&&\times\; p\!\left(\boldsymbol{u}\mid \boldsymbol{m}_u(T_{0,\,m},\alpha_m),\,a_u,\,\tau_u\right)\,
p\!\left(T_{0,\,m},\,\alpha_m,\,a_u,\,\tau_u\right),
\end{eqnarray}
which we sample using NUTS.

Relative to the GP formulation of \citet{Kawahara2022}, our implementation introduces several changes aimed at practical convergence and physical regularization. 
First, while \citet{Kawahara2022} adopted a RBF kernel, we instead use a Matérn~3/2 kernel. The Matérn kernel is less smooth than the RBF kernel, avoiding the overly smooth behavior inherent to the RBF assumption.
We also infer the kernel hyperparameters (the amplitude $a_{u}$ and correlation length $\tau_{u}$) jointly with the atmospheric parameters, thereby removing the need to set them manually.
In addition, instead of adopting a constant mean function as in \citet{Kawahara2022}, we use a power-law form characterized by $(T_{0,\,m},\,\alpha_m)$, whose parameters are inferred jointly with the GP. We find that a constant mean tends to force the retrieved profile to flatten toward the average value in the upper and lower layers, as is also hinted in figure 11 \citet{Kawahara2022}. The power-law mean model, on the other hand, better reflects the physically expected structure of self-luminous brown dwarfs while allowing the GP to capture deviations from it.

\subsection{Demonstration}
\label{subsec:Demonstration}

\begin{figure*}[h!]
    \includegraphics[width=0.5\textwidth]{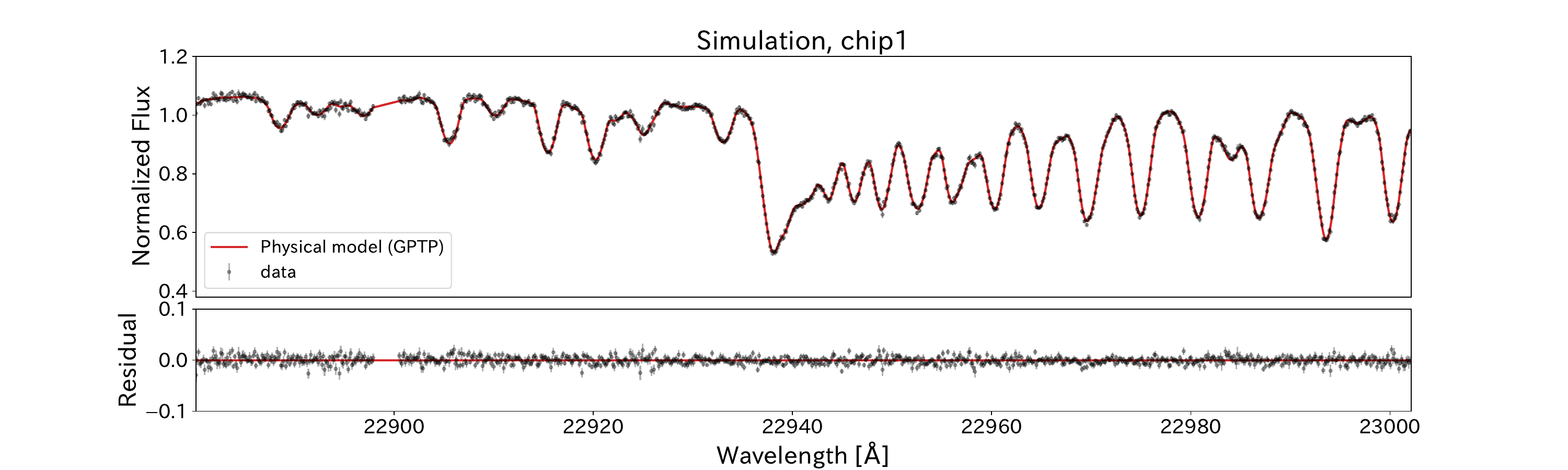}
    \includegraphics[width=0.5\textwidth]{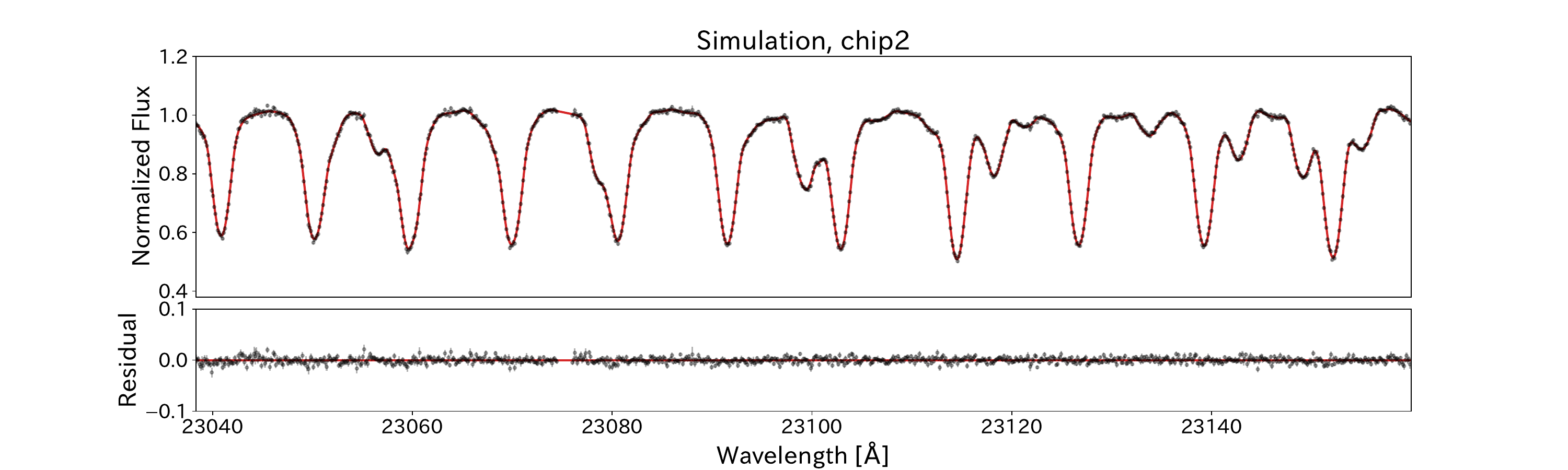}

    \includegraphics[width=0.5\textwidth]{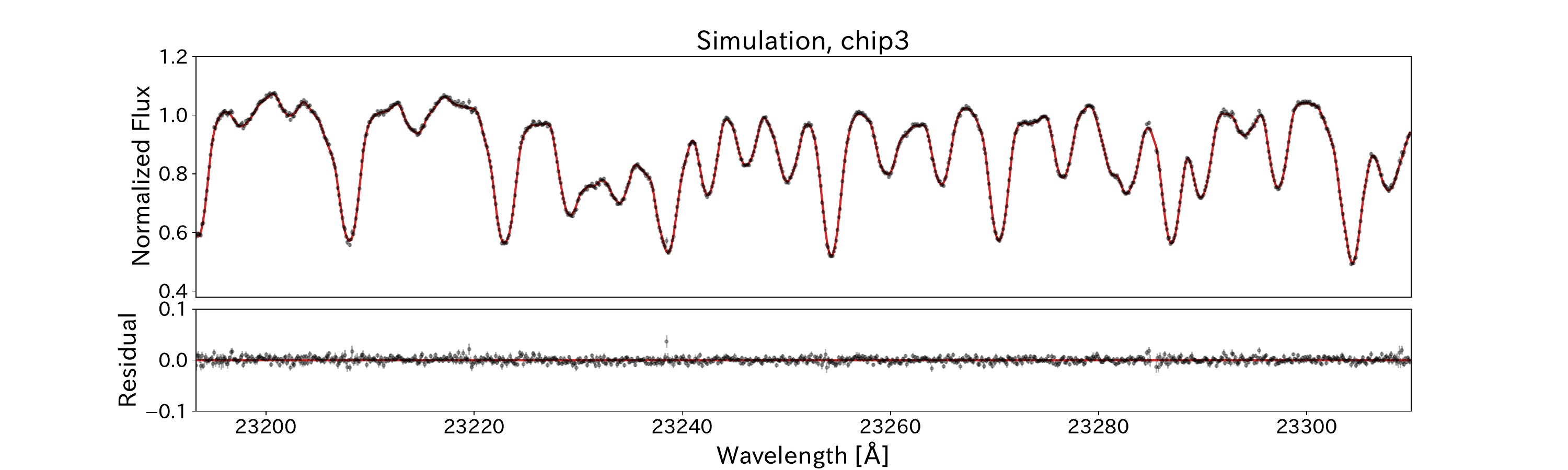}
    \includegraphics[width=0.5\textwidth]{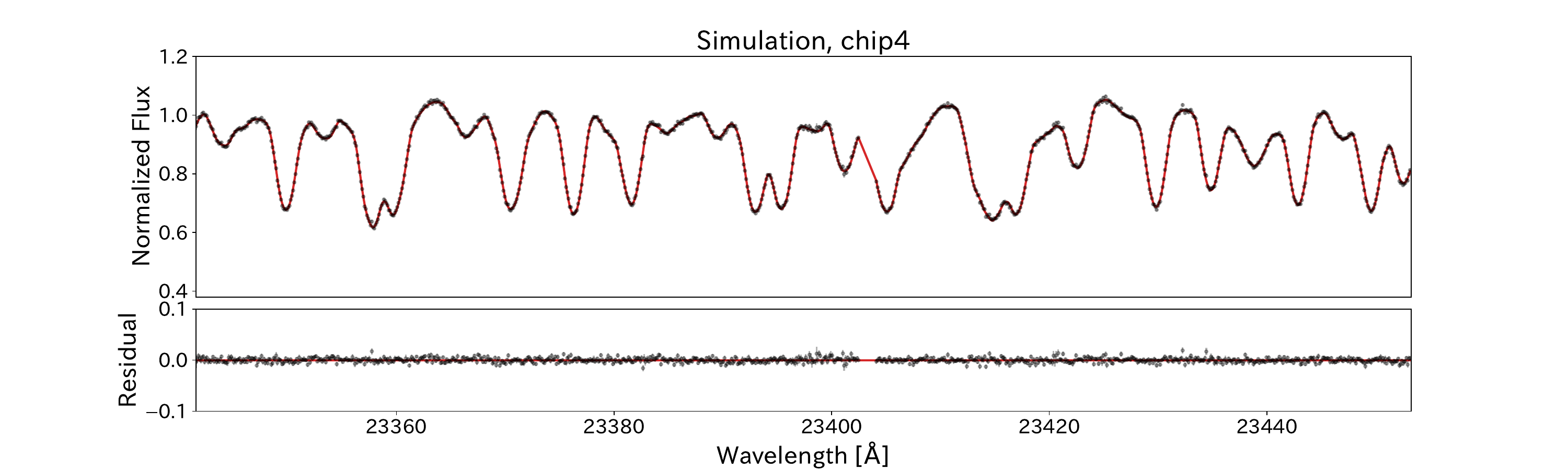}
    
    \caption{Retrieval results for the simulation spectra, modeled with the $\hitemphco$ line list and a Gaussian process $T$--$P$ profile. The wavelength coverage is split into four detector segments (chips 1–4) like CRIRES. For each chip, the top panel shows the simulated normalized flux data (black points) and the best-fit physical model (red line). The bottom panel shows the residuals (data minus the best-fit model). Each chip is normalized independently.}
    \label{fig:fit_sim_GPTP_HITEMPH2}
\end{figure*}

\begin{figure*}[h!]
    \includegraphics[width=0.5\textwidth]{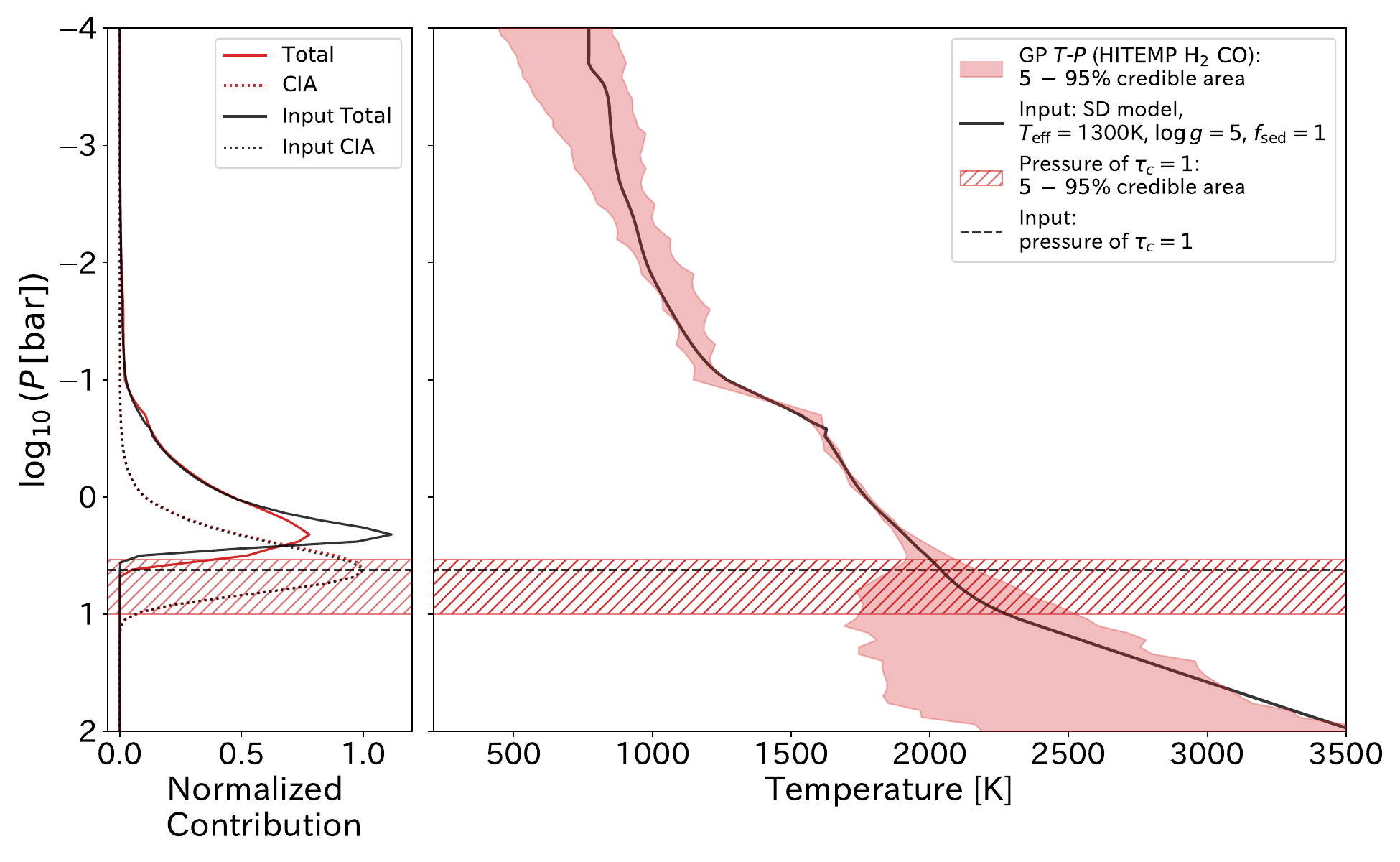}
    \includegraphics[width=0.5\textwidth]{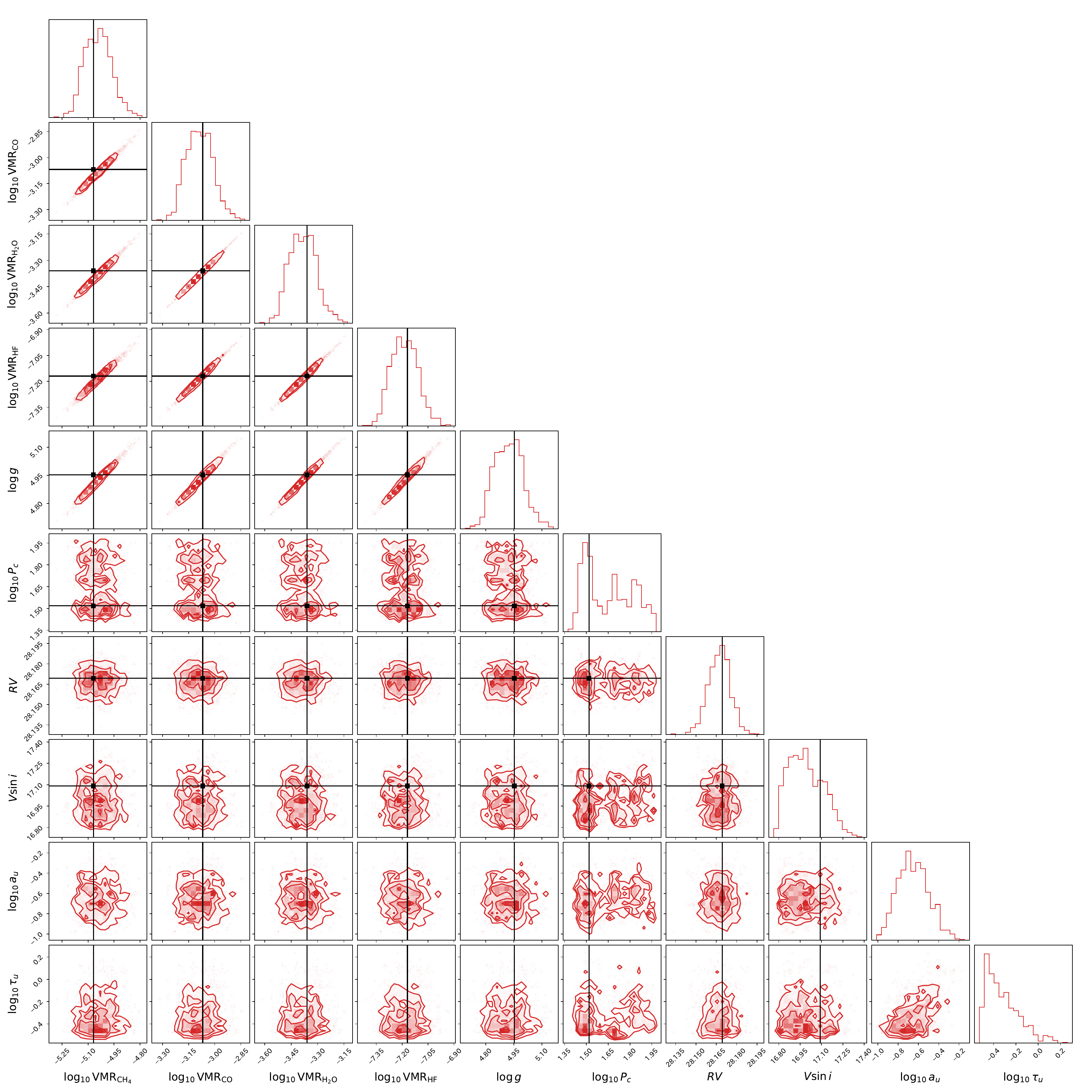}
    \caption{(Left) The Sonora diamondback (SD) $T$--$P$ profile model (black solid line) used to generate the simulation data, and the $T$--$P$ profile inferred with the GP $T$--$P$ model (red shaded region). (Right) The corresponding corner plot.}
    \label{fig:sim_GPTP_result}
\end{figure*}

We validate the GP $T$--$P$ framework using simulated spectra. All physical parameters except those describing the $T$--$P$ profile are fixed to the posterior medians from our retrieval of the Luhman 16A spectrum with a power-law $T$--$P$ profile and the $\hitemphco$ line list (Table~\ref{table:result_Luhman16A}). As the input $T$--$P$ profile for the simulations, we adopt a Sonora Diamondback model with $T_{\mathrm{eff}}=1300\,\mathrm{K}$, $f_{\mathrm{sed}}=1$, $\log g=5.0$, $[\mathrm{M/H}]=1.0$, and solar C/O. The synthetic data $\boldsymbol{d}_{\mathrm{sim}}$ are created by evaluating the spectral model $\boldsymbol{f}(\boldsymbol{\theta})$ at the simulation inputs $\boldsymbol{\theta}_{\mathrm{sim}}$ and adding Gaussian noise with a flux-density uncertainty $\boldsymbol{\sigma}_{\mathrm{sim}}$ matched to the S/N of the Luhman~16A high-resolution data:
\begin{equation}
    \label{eq:sim_data}
    \boldsymbol{d}_{\mathrm{sim}}=\boldsymbol{f}\!\left(\boldsymbol{\theta_{\mathrm{sim}}}\right)+\mathcal{N}\!\left(0, \boldsymbol{\sigma}_{\mathrm{sim}}\right).
\end{equation}
Because noise is drawn independently at each pixel, the likelihood used in the retrieval is
\begin{equation}
    \label{eq:sim_likelihood}
    \mathcal{L}\!\left(\boldsymbol{\theta}\right)=\prod_{i=1}^{n}\frac{1}{\sqrt{2\pi\sigma_{\mathrm{sim},\,i}^2}}
    \exp\!\left[-\frac{\left(d_{\mathrm{sim},\,i}-f_{i}\!\left(\boldsymbol{\theta}\right)\right)^2}{2\sigma_{\mathrm{sim},\,i}^2}\right].
\end{equation}
Priors for the parameters follow Table~\ref{tab:prior_Luhman16AB} (shared with Luhman~16A), except for $T_{0,\,m}$, the power-law $T$--$P$ parameters, and the GP for the spectral likelihood. We set the prior for $T_{0,\,m}$ to $\mathcal{U}\left(1000,\,2500\right)$ to ensure that the temperature range of the input Sonora Diamondback $T$--$P$ profile is fully covered by the allowed parameter space.

The simulation spectra and best fit are shown in Figure~\ref{fig:fit_sim_GPTP_HITEMPH2}. The retrieved $T$--$P$ profile and posteriors are shown in Figure~\ref{fig:sim_GPTP_result}. The GP $T$--$P$ model successfully reproduces input physical parameters as well as non–power-law shape in the input $T$--$P$ profile at pressures where the contribution function peaks. Conversely, in regions with little or no contribution (e.g., above the photosphere or deep beneath the cloud deck), the temperature uncertainty naturally broadens, reflecting the paucity of information more faithfully than a power-law model. This behavior demonstrates that the GP $T$--$P$ approach can capture complex thermal structures where the data are informative, while appropriately inflating uncertainties where they are not.

\subsection{Application to Luhman 16AB}
\label{subsec:GPTP_luh16}

\begin{figure*}[h!]
    \includegraphics[width=0.5\textwidth]{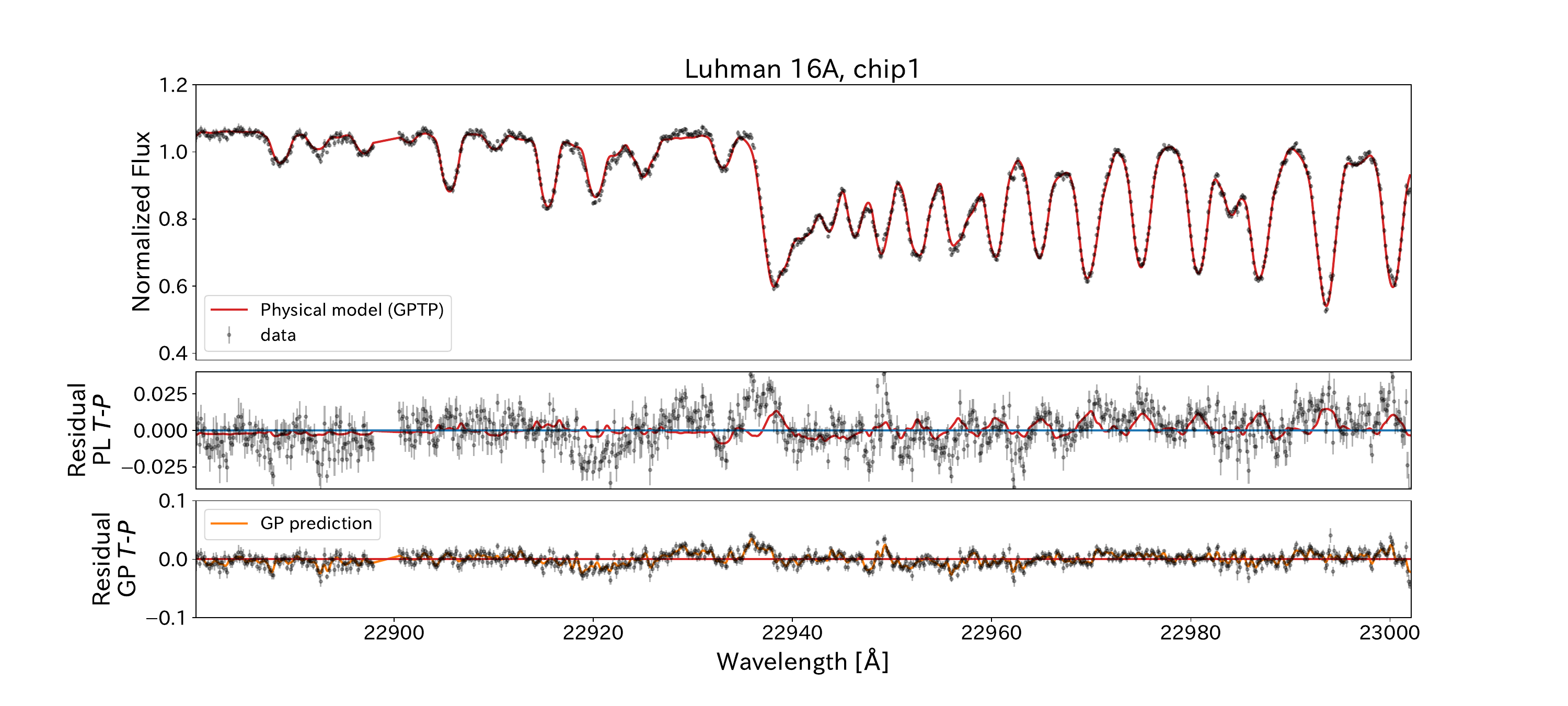}
    \includegraphics[width=0.5\textwidth]{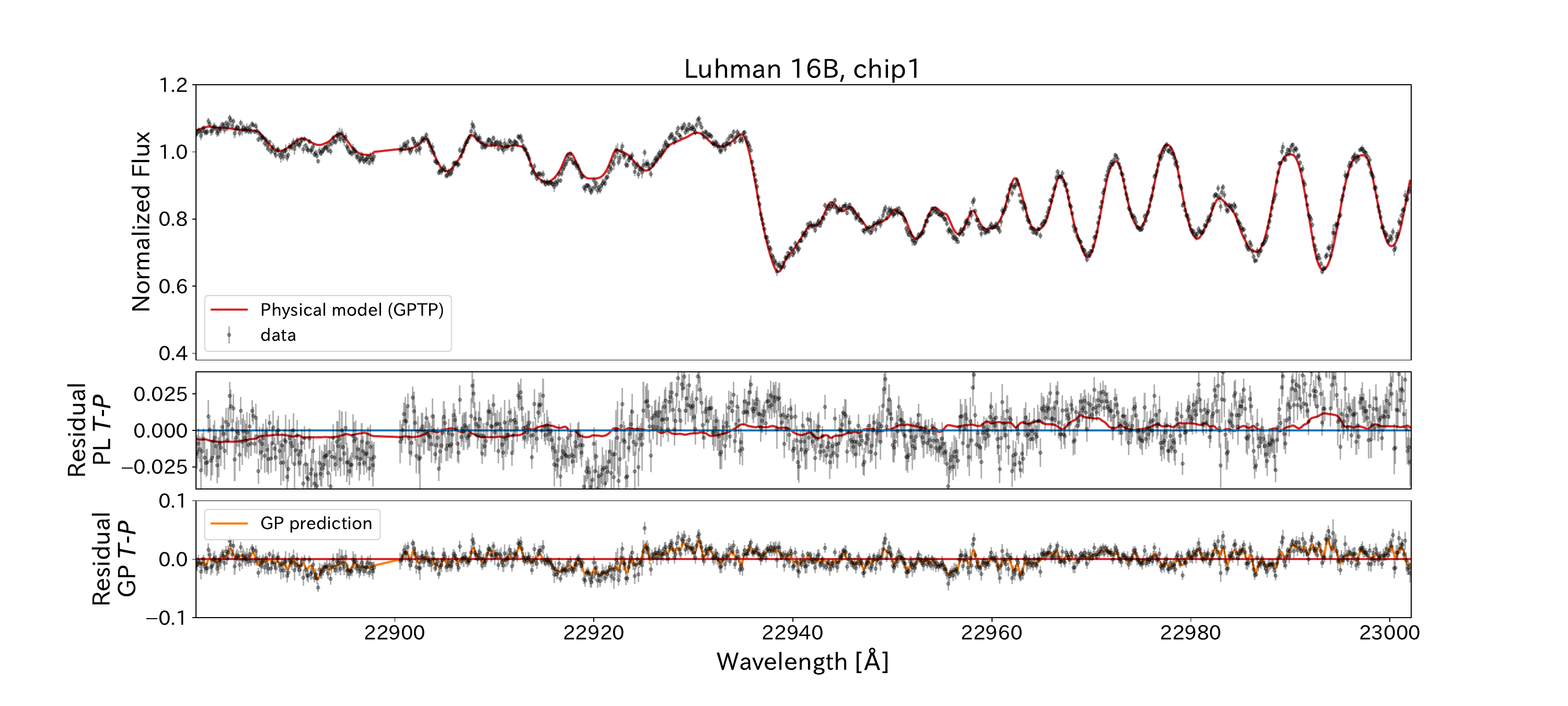}

    \includegraphics[width=0.5\textwidth]{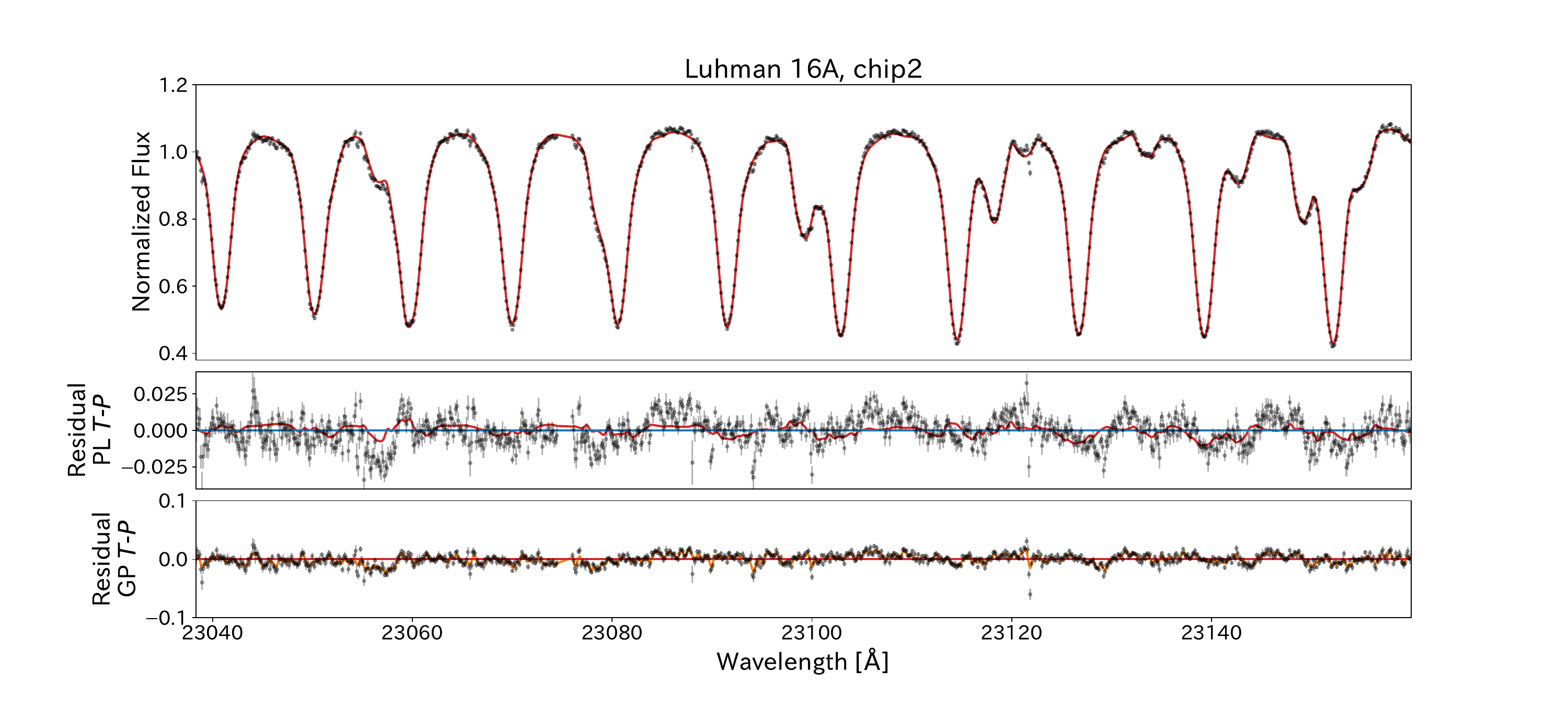}
    \includegraphics[width=0.5\textwidth]{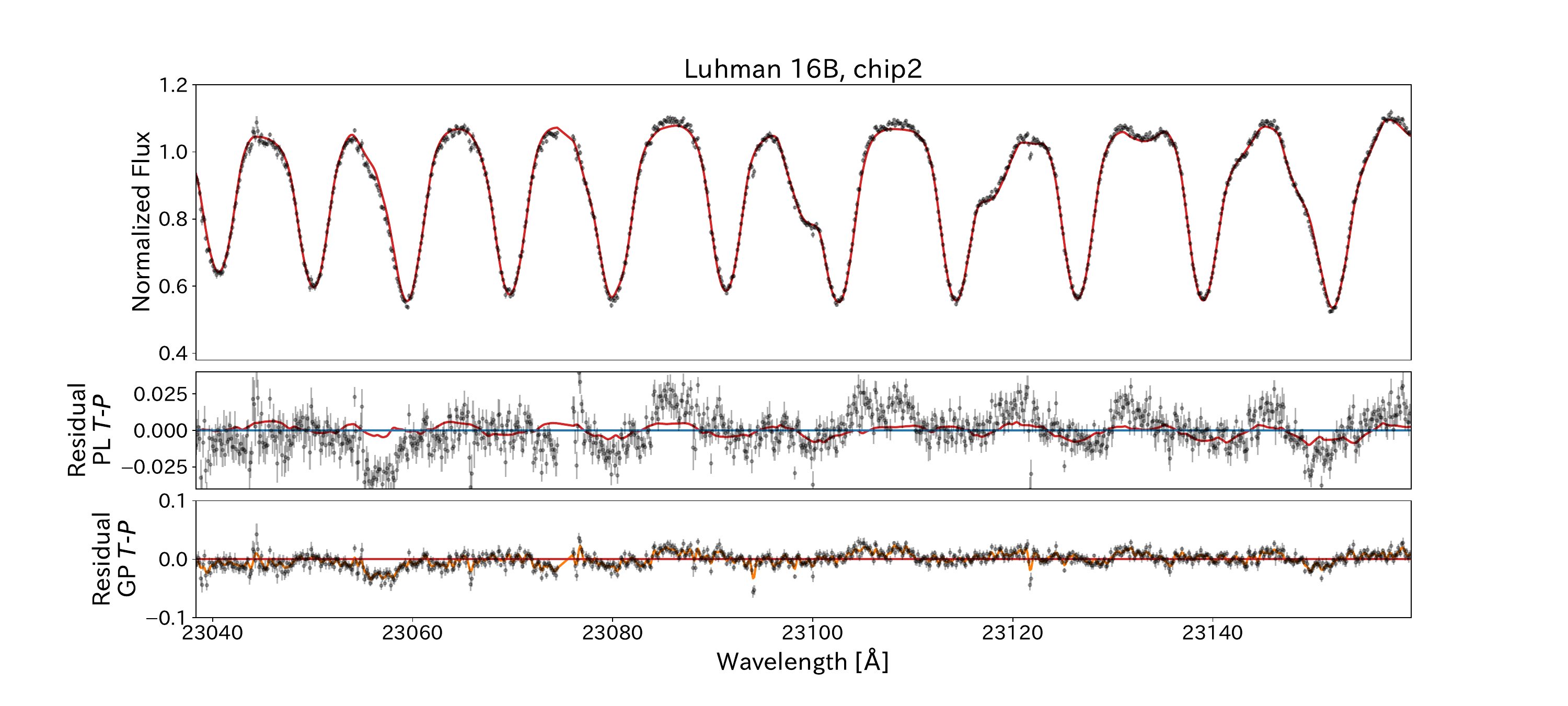}

    \includegraphics[width=0.5\textwidth]{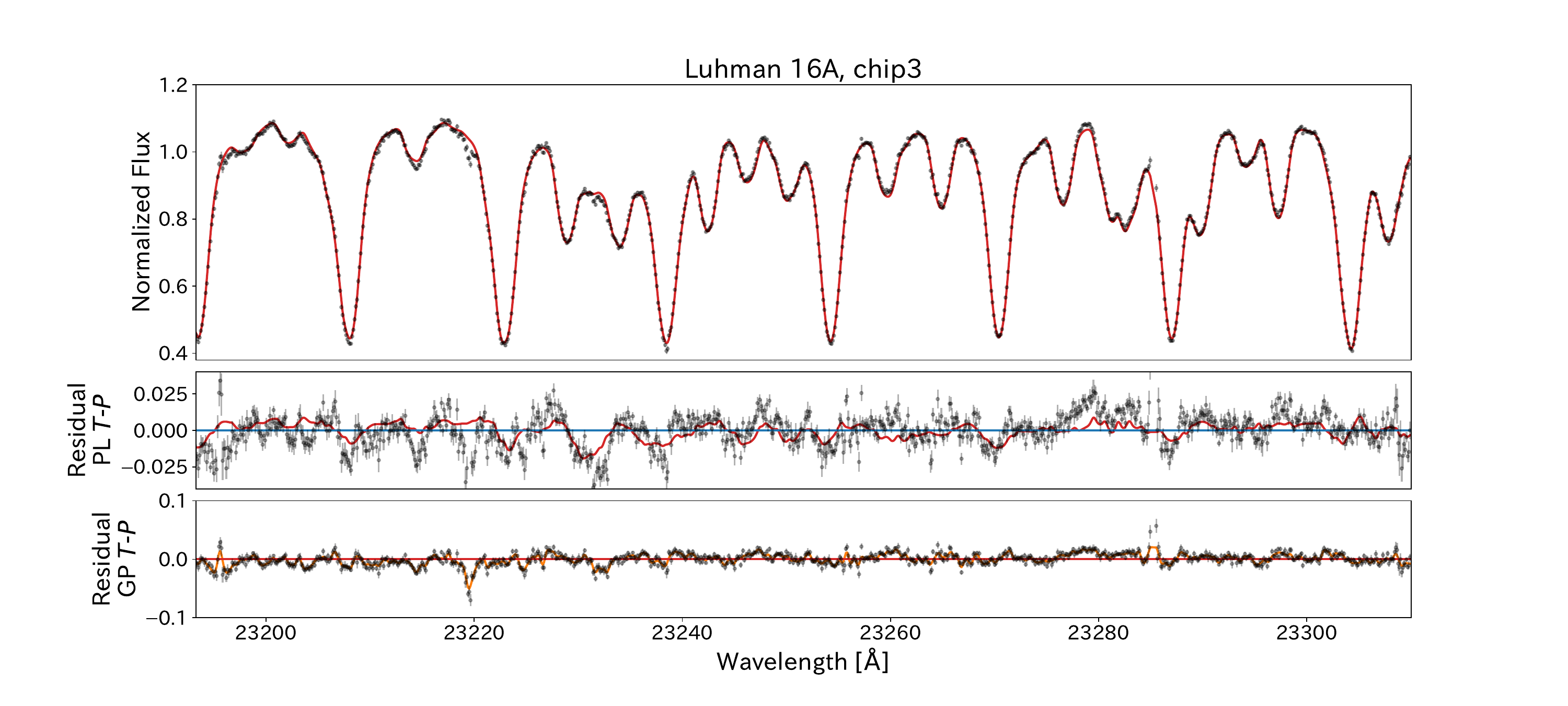}
    \includegraphics[width=0.5\textwidth]{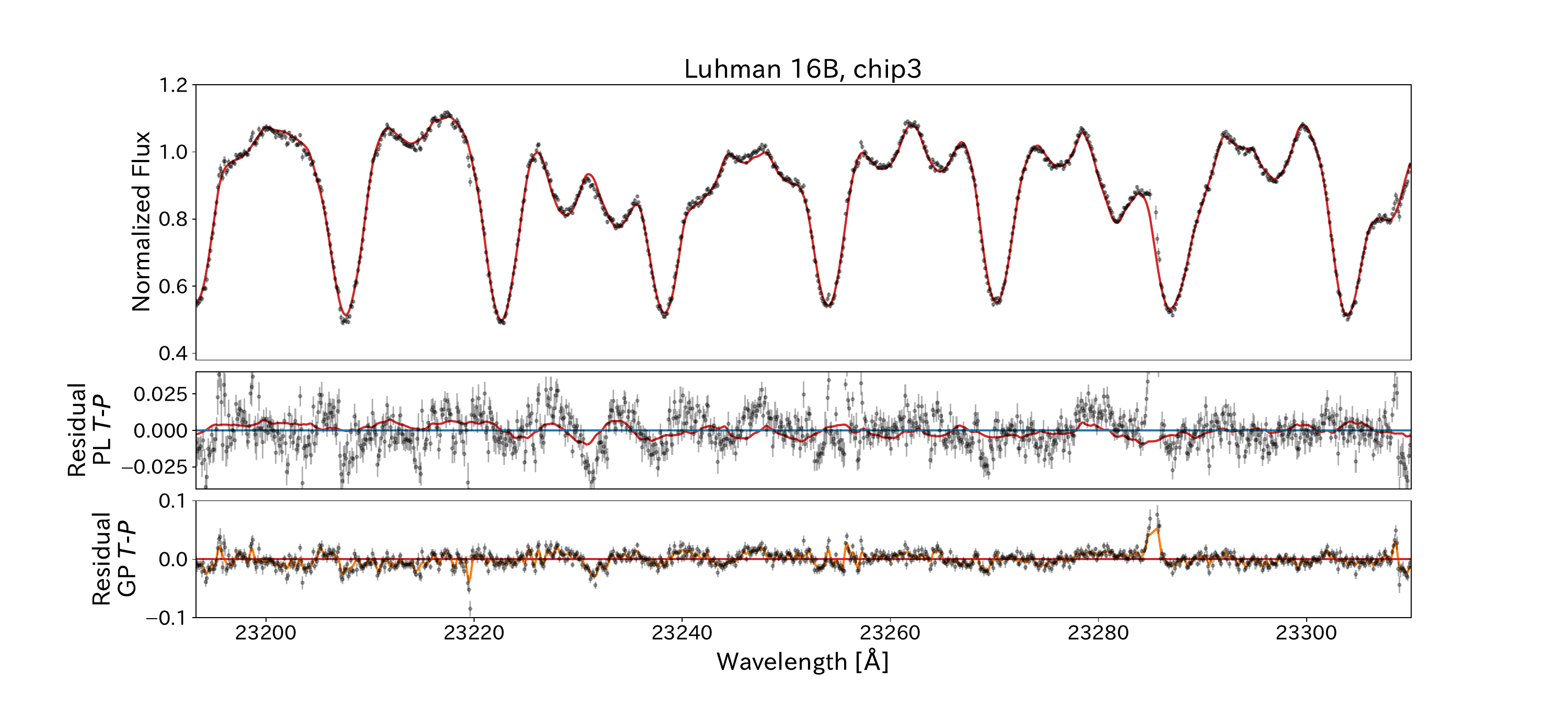}

    \includegraphics[width=0.5\textwidth]{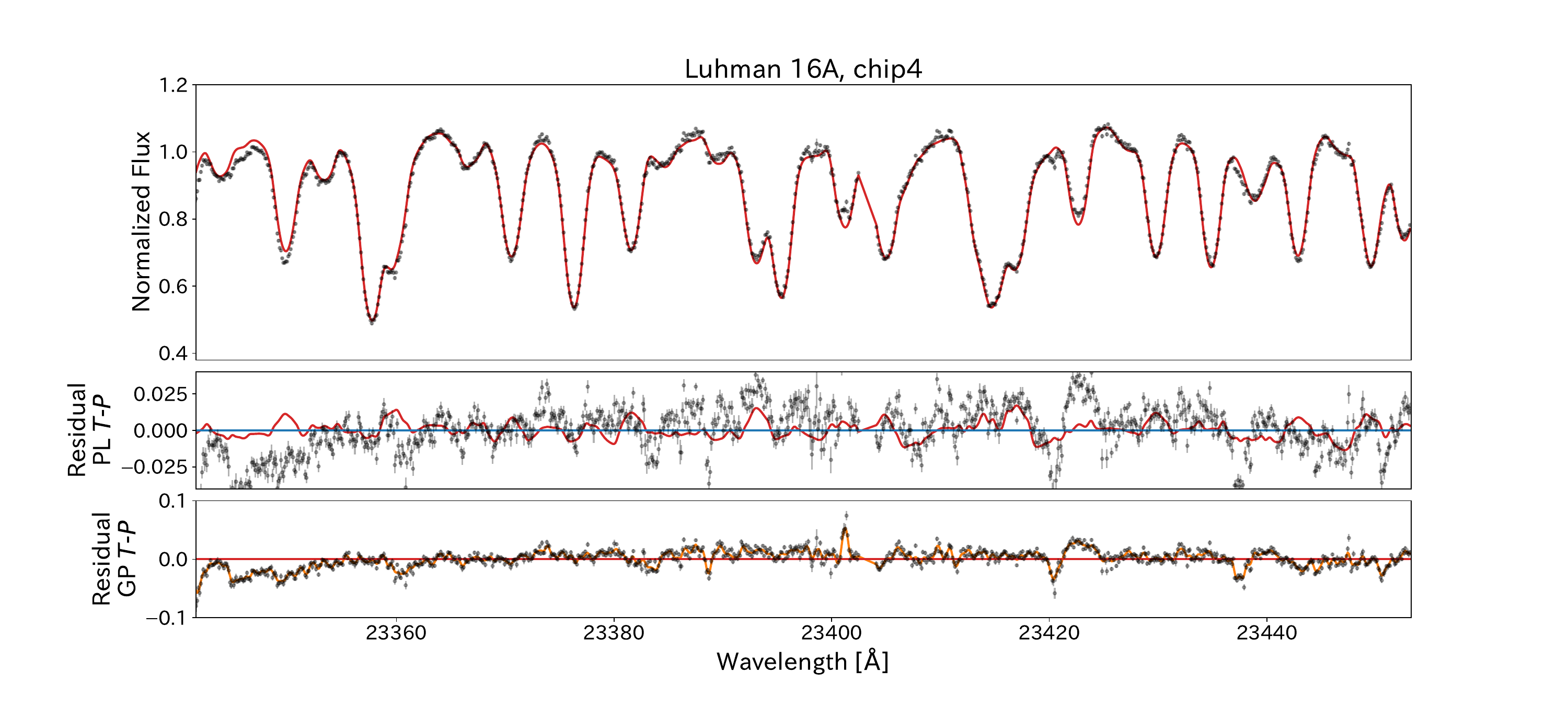}
    \includegraphics[width=0.5\textwidth]{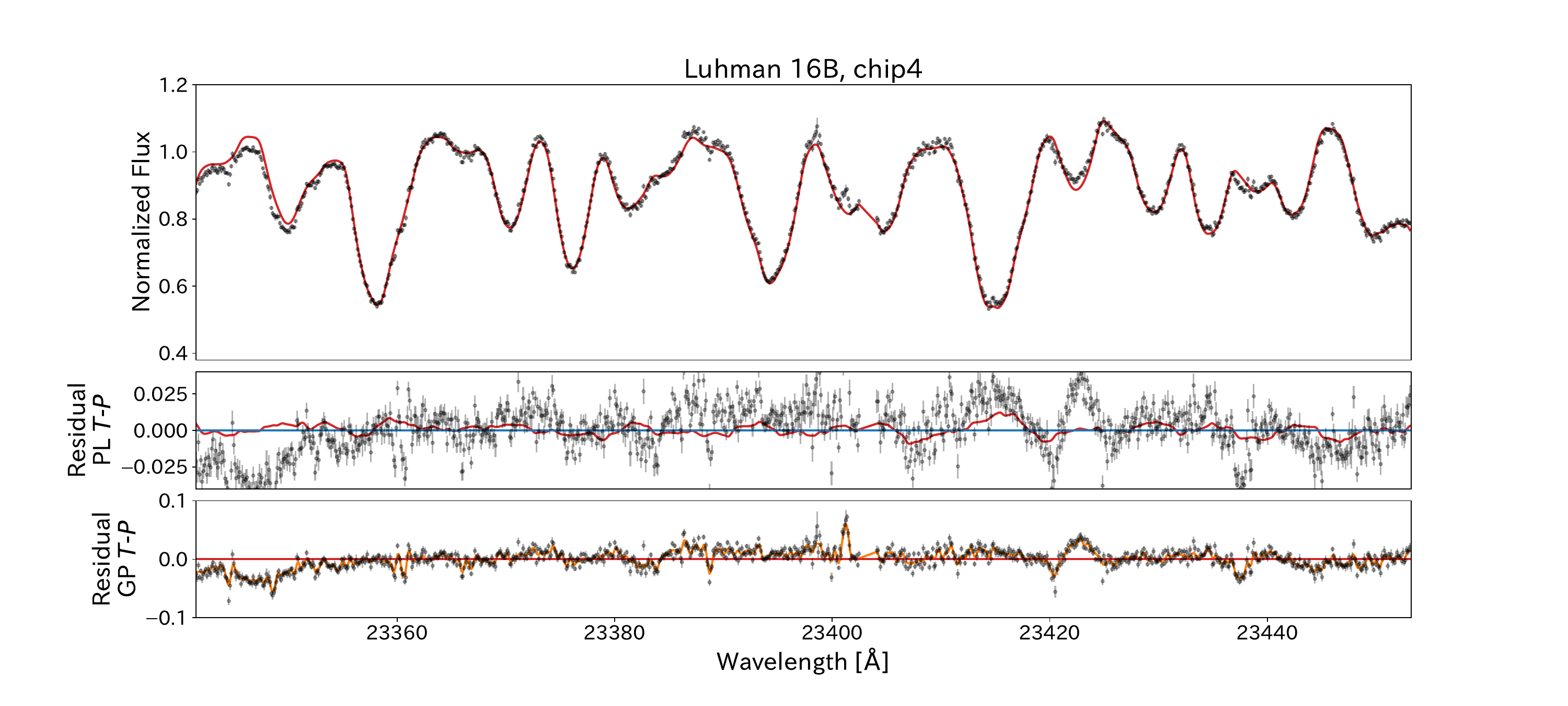}
    
    \caption{Retrieval results for the VLT/CRIRES high-resolution spectra of Luhman 16A (left) and Luhman 16B (right), modeled with the $\hitemphco$ line list and a Gaussian process $T$--$P$ profile. The wavelength coverage is split into four detector segments (chips 1–4). For each chip, the top panel shows the normalized flux data (black points) and the best-fit physical model (red line). The middle panel shows the residuals relative to the power-law $T$--$P$ model with the $\hitemphco$ line list (data minus the power-law $T$--$P$ best-fit model), along with the difference between the GP and power-law $T$--$P$ best-fit models (red line). The bottom panel shows the residuals relative to the GP $T$--$P$ model (data minus the GP $T$--$P$ best-fit model); orange lines indicate the correlated noise predicted by the GP. Each chip is normalized independently.}
    \label{fig:fit_GPTP_HITEMPH2}
\end{figure*}

\begin{figure*}[h!]
    \includegraphics[width=0.5\textwidth]{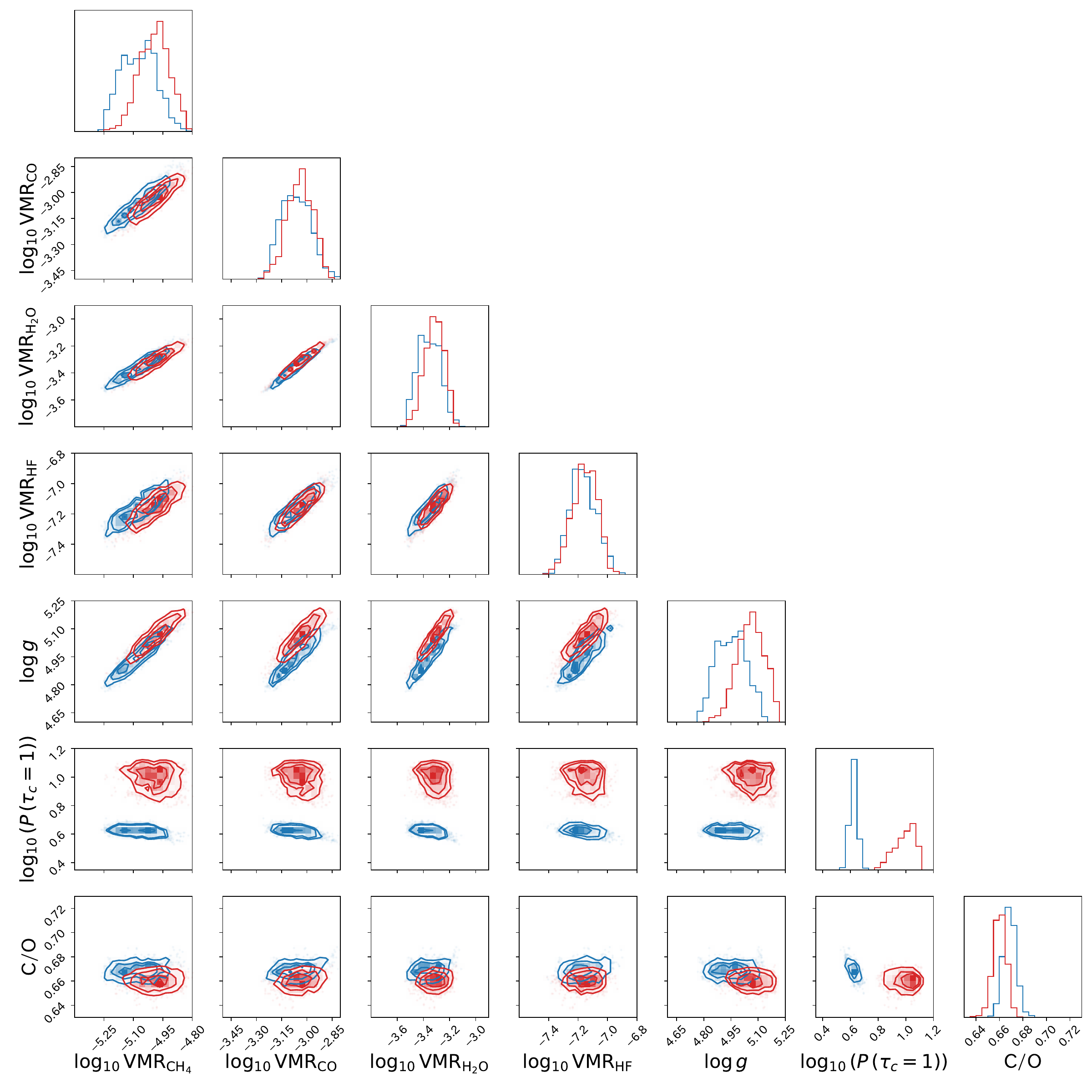}
    \includegraphics[width=0.5\textwidth]{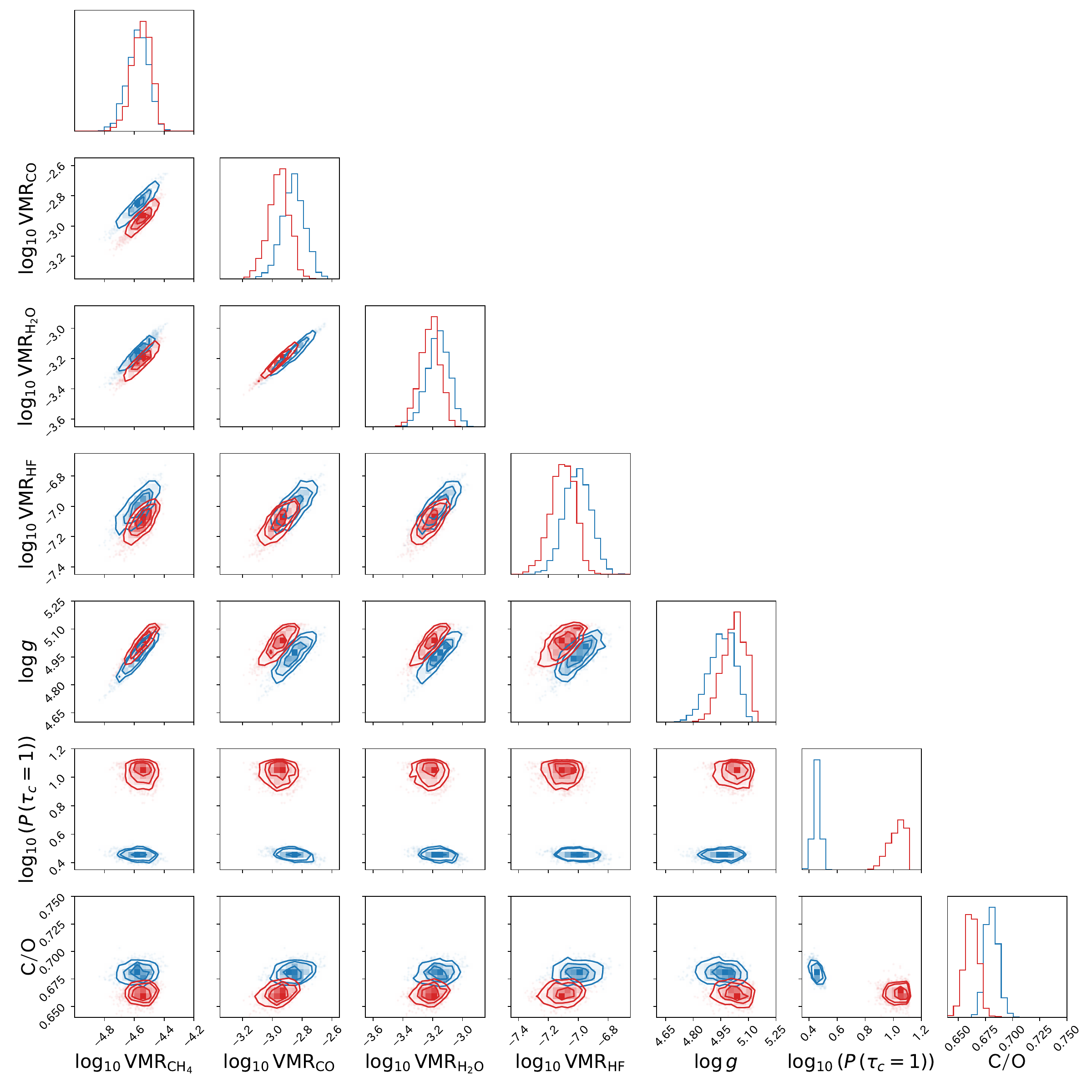}
    \caption{Comparison of posteriors obtained with a power-law $T$--$P$ profile (blue) and with a GP $T$--$P$ profile (red), both using the $\hitemphco$ line list. (Left) Luhman 16A. (Right) Luhman 16B.}
    \label{fig:comp_post_PL_GP}
\end{figure*}

\begin{figure*}[h!]
    \includegraphics[width=0.5\textwidth]{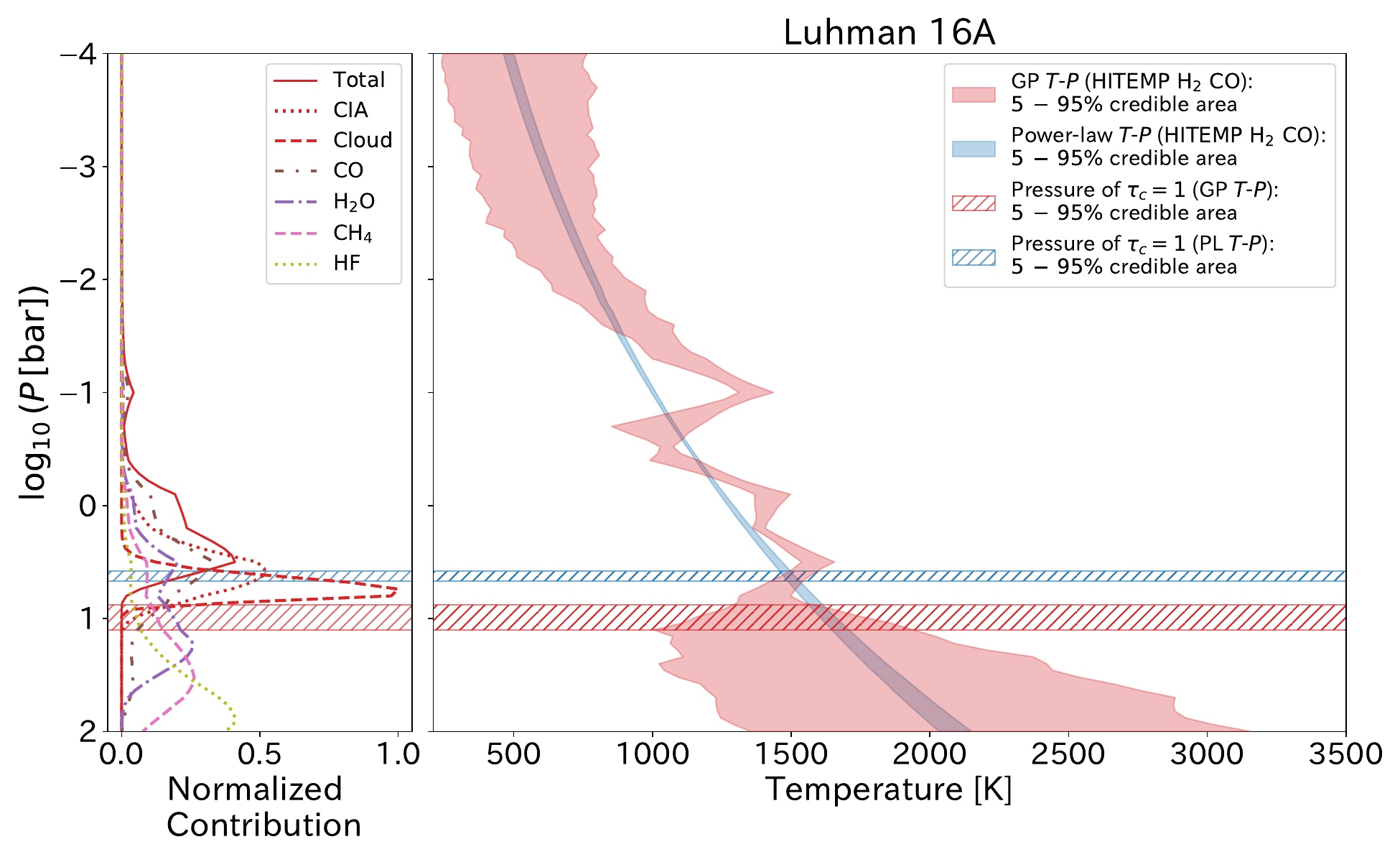}
    \includegraphics[width=0.5\textwidth]{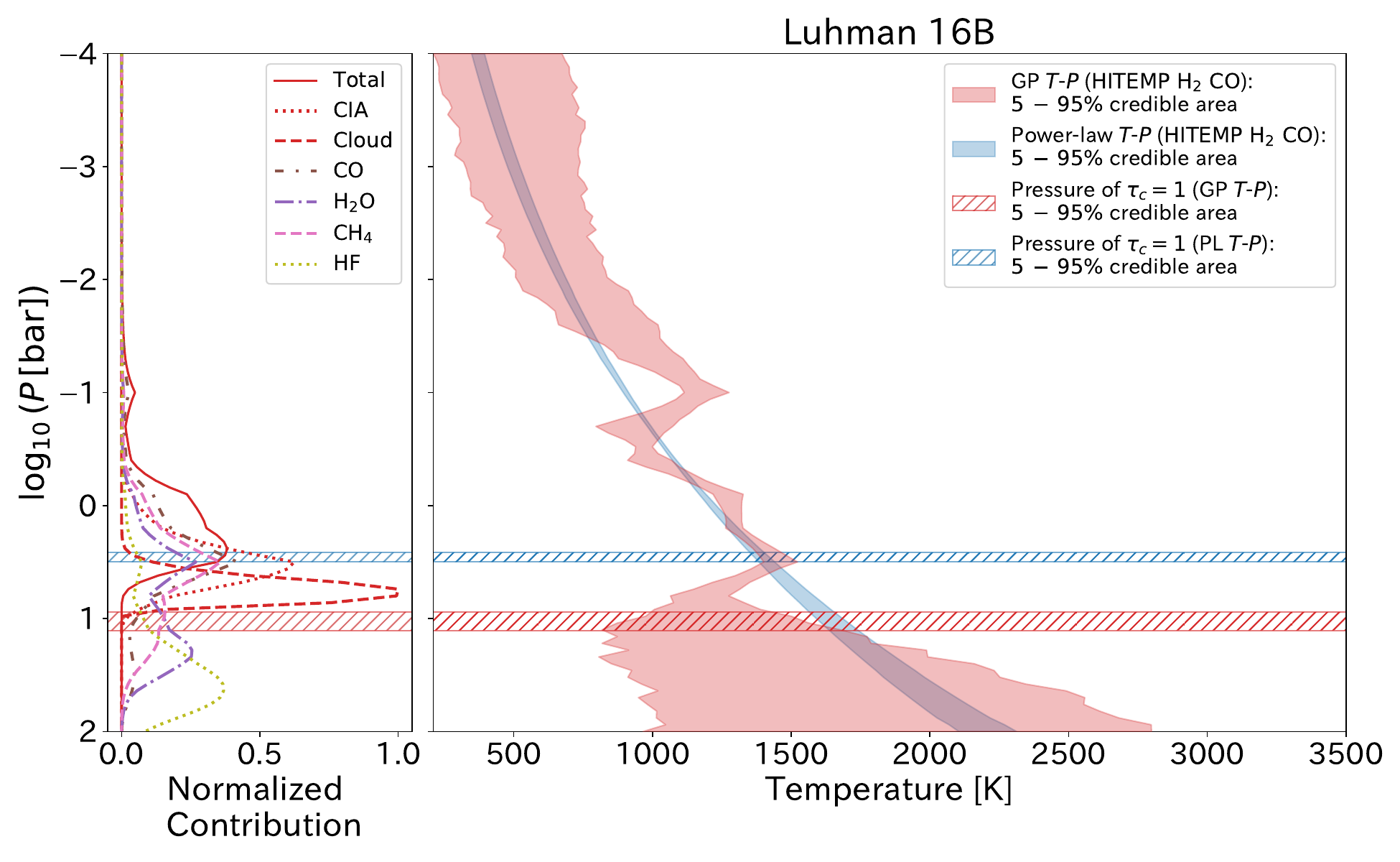}
    \caption{Retrieved $T$--$P$ profiles and contribution functions: GP (red shaded) versus power-law (blue) under the gray-cloud + $\hitemphco$ configuration. (Left) Luhman 16A. (Right) Luhman 16B.}
    \label{fig:TP_GP_vs_PL}
\end{figure*}

\begin{figure*}[h!]
    \includegraphics[width=0.5\textwidth]{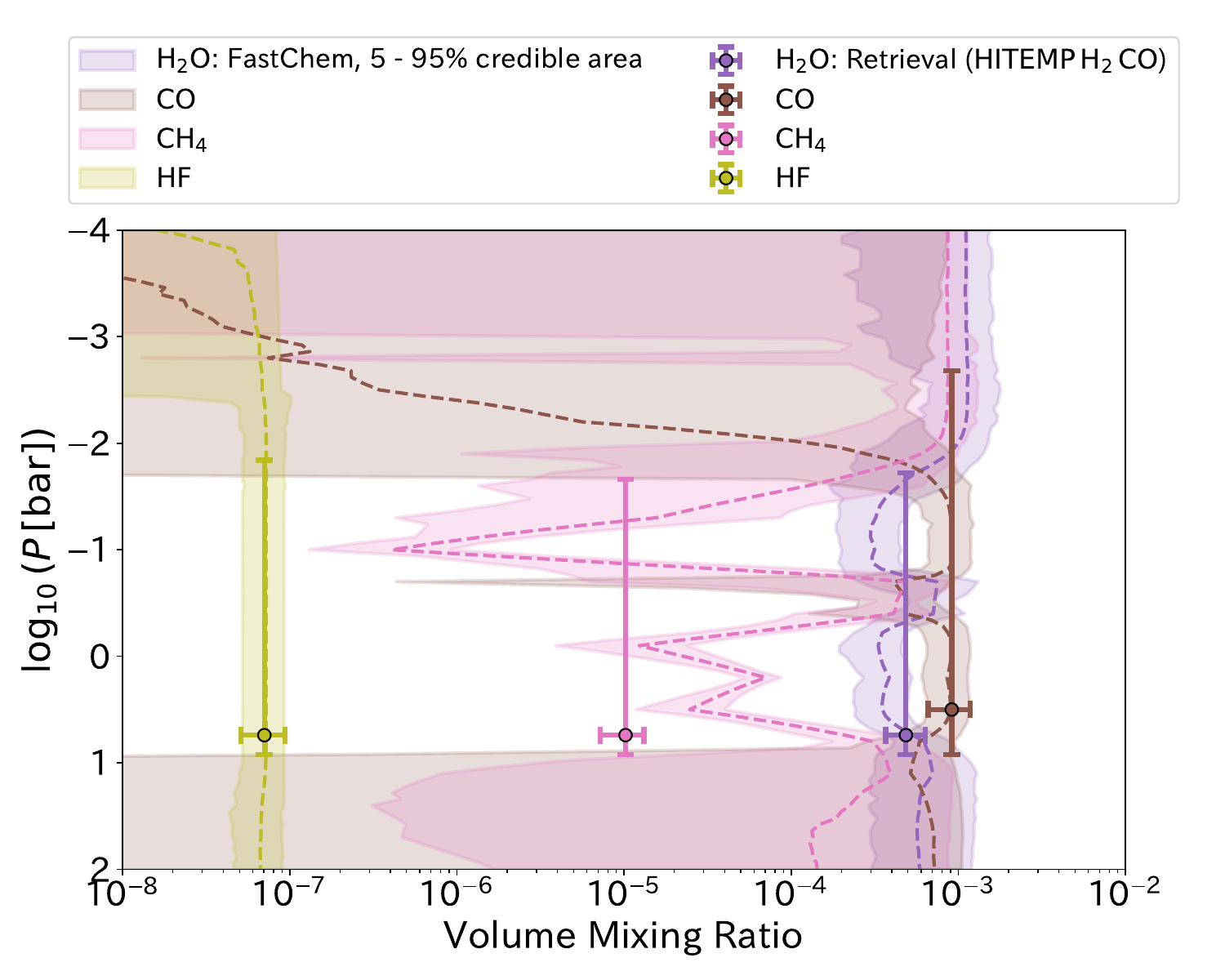}
    \includegraphics[width=0.5\textwidth]{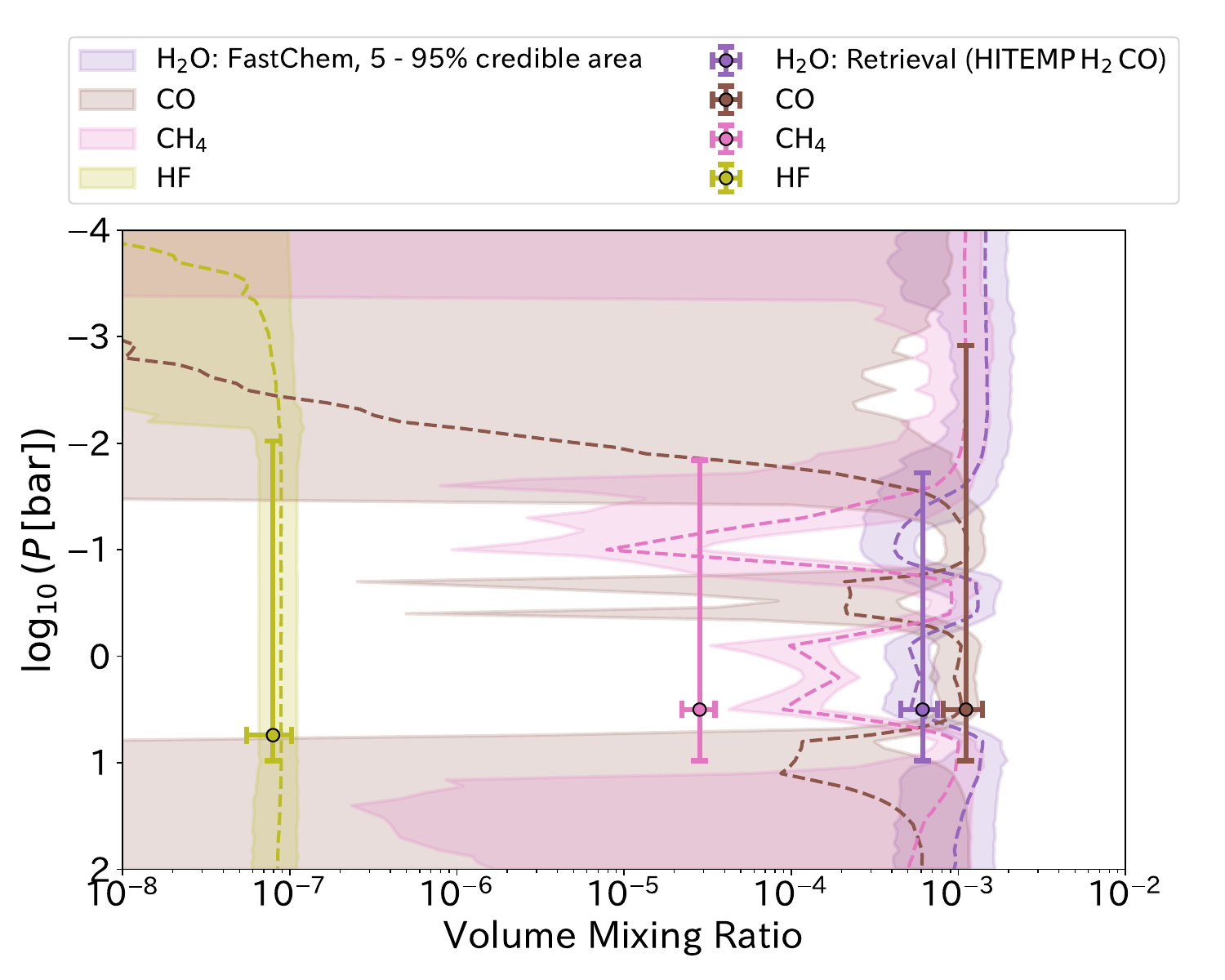}
    \caption{Comparison between the retrieved volume mixing ratios and FastChem equilibrium predictions (solar abundances from \citealt{Asplund2021}). (Left) Luhman 16A. (Right) Luhman 16B.}
    \label{fig:mixing_ratio}
\end{figure*}

We applied the GP $T$--$P$ retrieval to the CRIRES data of Luhman 16AB. Best-fit model spectra for both components are shown in Figure~\ref{fig:fit_GPTP_HITEMPH2}. Corner plots for the selected parameters are shown in Figure~\ref{fig:comp_post_PL_GP};
the retrieved $T$--$P$ profiles are shown in Figure~\ref{fig:TP_GP_vs_PL} along with the power-law profiles from Section~\ref{sec:PLTP}; the retrieved parameter values are summarized in the rightmost columns of Tables~\ref{table:result_Luhman16A} and \ref{table:result_Luhman16B}.

Figure~\ref{fig:TP_GP_vs_PL} shows that the broad trends of the inferred $T$--$P$ profiles are consistent with those obtained under the power-law parameterization. Consistently, the molecular abundances and C/O ratios agree with the power-law results within uncertainties, indicating that these bulk parameters are robust to the choice of $T$--$P$ profile. As in the simulation test, the temperature uncertainties increase at pressures where the spectral contribution is small, yielding a more conservative characterization than the power-law model. 

At the same time, the GP $T$--$P$ profiles infer departures from a power-law in the detailed structure. In both components, we find evidence for localized inversions near $\log_{10} P \sim 0.5$ and $\log_{10} P \sim -1$. The feature around $\log_{10} P \sim 0.5$ roughly coincides with the cloud-top pressure inferred in the power-law analysis; an approximately isothermal layer can mimic the behavior of a gray cloud, underscoring that cloud-top inferences are sensitive to the flexibility allowed in the $T$--$P$ profile. The higher-altitude feature near $\log_{10} P\sim -1$ corresponds to subtle differences in the depths of line cores arising from cooler layers. Whether this structure is physical remains uncertain; possibilities include residual line-list/systematic effects or additional weak condensate/haze opacity at those pressures. Indeed, Sonora models suggest some condensates may be expected in this regime depending on composition and thermal structure. Regardless, the information content from these levels is limited, so any model deficiency there has a negligible impact on global parameters such as C/O for the present data.

Figure~\ref{fig:mixing_ratio} shows the comparison with FastChem, analogous to the power-law case. While the overall trends remain similar, the model-predicted ranges broaden owing to the propagated $T$--$P$ uncertainty. The effect is most pronounced for $\mathrm{CH_4}$ in deeper layers, for which the apparent discrepancy seen under the power-law assumption is no longer compelling once the increased structural uncertainty is accounted for.
This highlights that accurately characterizing the uncertainty in regions with modest contribution is essential when assessing possible departures from chemical equilibrium. In this regard, the GP $T$--$P$ profile provides a useful and conservative framework.

\section{Summary and Discussion}
\label{sec:Summary}

Using \textsc{ExoJAX}, we performed high-resolution retrievals of Luhman~16AB under a gray-cloud assumption and inferred elemental abundance ratios. Taking the $\hitemphco$ line list and a power-law $T$--$P$ parameterization as a baseline, we then (i) repeated the analysis with alternative CO line lists and (ii) introduced a more flexible $T$--$P$ description based on GPs to assess the robustness of the inferred parameters.

\subsection{Comparison with Previous Work}

In addition to the present work, several other high-resolution retrieval studies of Luhman 16AB have recently been reported. Since the spectral wavelength ranges and model setups differ across these works, we focus here on comparing the retrieved physical parameters, namely, molecular abundances and cloud parameters.  

\paragraph{Abundance} The molecules included in our model, H$_2$O, CH$_4$, CO, and HF, have all been detected in the $H$- and $K$-band Gemini South/IGRINS high-resolution ($R \sim 28,000$) spectra of Luhman 16AB presented by \citet{Ishikawa2025}. Their study compared the spectra against forward models generated with \texttt{PICASO}, which incorporates radiative--convective equilibrium, Sonora Bobcat $T$--$P$ profiles, VIRGA cloud models, and nonequilibrium chemistry parameterized by a vertical eddy diffusion coefficient $K_{zz}$. The retrieved volume mixing ratios (VMRs) of the common molecules (their Figure~8) are broadly consistent with ours, but tend to be smaller, likely because they fixed the metallicity to solar whereas we inferred slightly super-solar values. Both H$_2$O and HF have also been detected in the CRIRES$^+$/VLT $J$-band high-resolution retrieval by \citet{deRegt2025}. Their modeling adopted a $T$--$P$ parameterization based on the temperature gradient at five knots (as in \citealp{Zhang2023}), and clouds were treated as gray opacity decreasing with a power-law of pressure from the cloud base upward. Despite these differences, the retrieved molecular abundances are consistent with ours. Neither study provides explicit estimates of the C/O ratio: \citet{Ishikawa2025} fixed the metallicity and C/O to solar, and in the spectral range analyzed by \citet{deRegt2025}, no carbon-bearing molecules were detected.  

\paragraph{Cloud} \citet{Ishikawa2025} also detected H$_2$ lines. By comparing the inferred H$_2$ column density with the interstellar gas-to-dust ratio, they suggested that dust clouds may reside deeper than the H$_2$ absorption layer, located at $\sim 3\,\mathrm{bar}$ (or $\log_{10} P \sim 0.5$) in both components A and B. Their self-consistent models likewise imply that clouds are present at similar or deeper pressures relative to the H$_2$ line-forming region. In the power-law cloud model of \citet{deRegt2025}, the cloud opacity is dominated by contributions around $\log_{10} P \sim 0.5$--1, and this estimate is robust to assumptions about patchiness. In our analysis as well, the inferred cloud top pressures consistently fall within $\log_{10} P \sim 0.5$--1, broadly agreeing with these results. On the other hand, the quantitative estimates depend on assumptions about the $T$--$P$ profile and line lists, which in turn propagate into the inferred C/O ratio as discussed in the next subsection.

\subsection{$\CO$ Ratios}

\begin{figure*}[h!]
    \centering
    \includegraphics[width=0.9\textwidth]{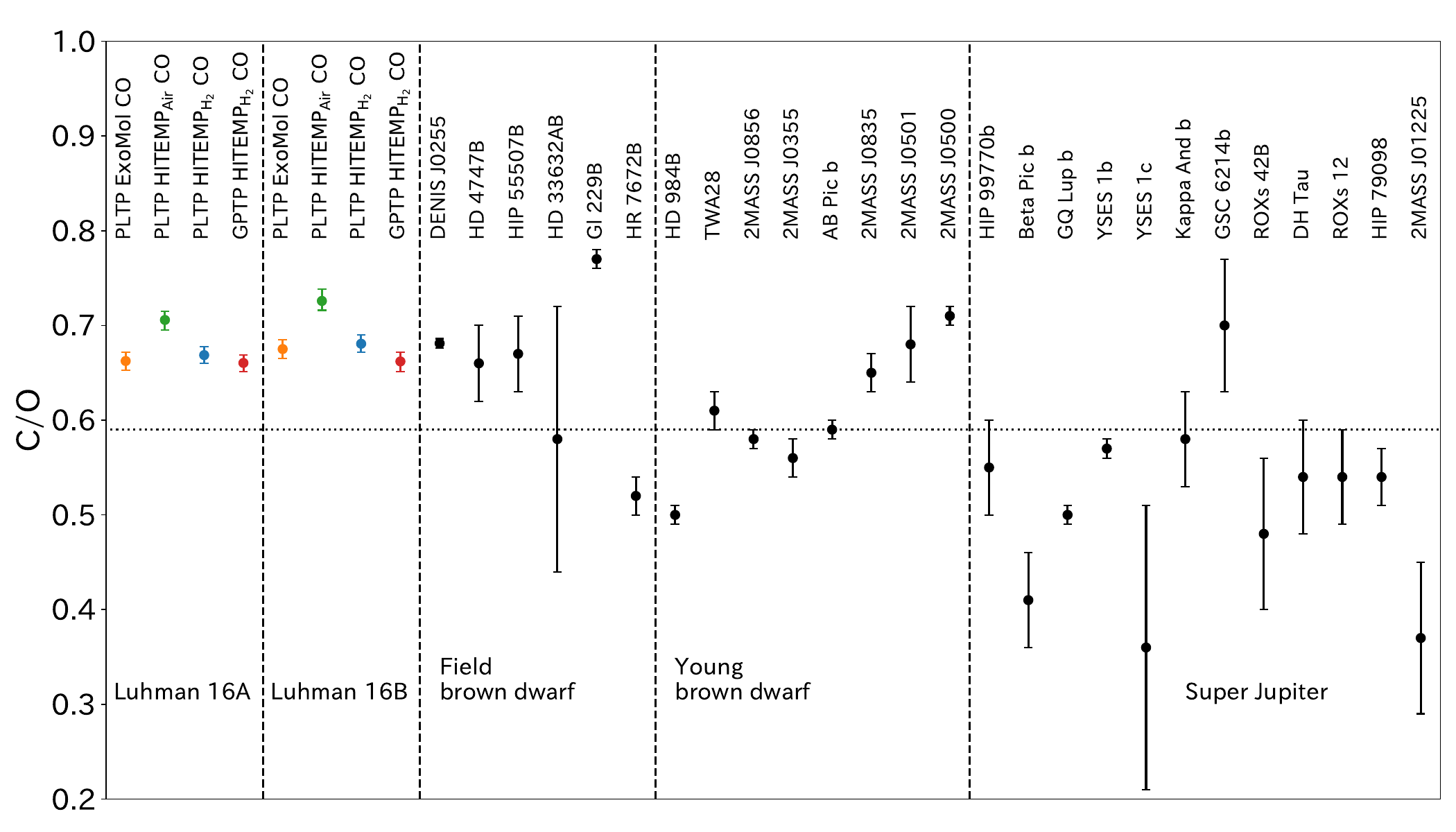}
    \caption{Comparison of $\CO$ for Luhman 16AB derived in this work with $\CO$ values reported for other brown dwarfs and directly-imaged super Jupiters from high-resolution spectroscopic analyses (Table~\ref{tab:co_lit}). Abbreviations: PLTP = power-law $T$--$P$ profile; GPTP = Gaussian-process $T$--$P$ profile.}
    \label{fig:CO_ratio}
\end{figure*}

Figure~\ref{fig:CO_ratio} compares $\CO$ inferred with each CO line list configuration and with the GP $T$--$P$ model (using $\hitemphco$) to literature values for other brown dwarfs and super Jupiters. For both A and B, the choice of CO line list introduces a systematic offset of $\sim$7\% in $\CO$. By contrast, the assumed $T$--$P$ profile (power-law versus GP) does not yield a significant difference in $\CO$. An epoch-by-epoch analysis of the Luhman~16B time series likewise reveals no statistically significant impact from cloud-cover variability on the retrieved $\CO$ or molecular VMRs (Section~\ref{ssec:PLTP_variability}). We therefore conclude that, in our framework, the principal systematic on $\CO$ arises from the CO line list, whereas the $T$--$P$ shape and photometric variability are not dominant. 
In all cases, the C/O values for components A and B agree within $\approx2\sigma$, consistent with the expectation that isolated brown dwarfs (and their binaries) form in a star-like manner.

At face value, Figure~\ref{fig:CO_ratio} hints that brown dwarfs may exhibit higher $\CO$ than super Jupiters. Although this is a collection of inhomogeneous inferences from the literature, our measurements for Luhman~16AB, together with the associated systematic-error budget, reinforce this possibility. While differences in formation pathways (i.e., gravitational instability vs. core accretion) remain a possible explanation, the contrast could also reflect differences in the efficiency of oxygen sequestration by silicate (and other) clouds, as emphasized in previous low-resolution studies \citep[e.g.,][]{Line2015,Line2017}: more efficient rainout in field brown dwarfs would raise the gas-phase C/O, whereas weaker sequestration in super Jupiters would leave C/O closer to the stellar value. A decisive test will require a larger, uniformly analyzed sample of both planets and brown dwarfs, so that inter-study systematics are minimized and C/O can be compared on equal footing. 
Building on such a uniform baseline, joint analyses of high-resolution spectra (to constrain $T$--$P$) and JWST coverage (to probe silicate or other cloud composition and vertical structure) can further mitigate cloud--$T$--$P$--chemistry degeneracies, thereby providing a more direct test of the efficiency of oxygen sequestration by rainout. We return to the prospects and implementation of such joint retrievals in Section~\ref{subsec:joint_JWST}.

\subsection{Usefulness and Caveats of the GP $T$--$P$ Model}

For both simulations and real data, the GP $T$--$P$ model captures structures that are more complex than a simple power-law and yields a more faithful assessment of uncertainties. In the present application, these differences have only a modest impact on bulk parameters such as $\CO$. However, as discussed in Section~\ref{subsec:GPTP_luh16}, comparisons between retrieved molecular abundances and chemical-equilibrium predictions are sensitive to how accurately $T$--$P$ uncertainties are estimated; in this context, the GP framework is important because it enables predictive uncertainties to faithfully reflect the spectra's information content, remaining conservative where the data are weakly informative. The GP approach is also sensitive to residual model--data mismatches and thus shows promise for diagnosing deficiencies in current models, potentially pointing to additional physics.

On the other hand, as indicated in Section~\ref{subsec:GPTP_luh16}, degeneracies between the $T$--$P$ shape and cloud opacity can also allow nonphysical temperature structures if left unconstrained. Mitigations include imposing weak physical regularization (e.g., monotonicity or stability constraints) on flexible $T$--$P$ profiles and/or extending the analysis to broader wavelength coverage where cloud signatures are stronger. The latter strategy can help balance flexibility with physical plausibility; see also Section~\ref{subsec:joint_JWST}.

\begin{figure*}[h!]
    \includegraphics[width=\textwidth]{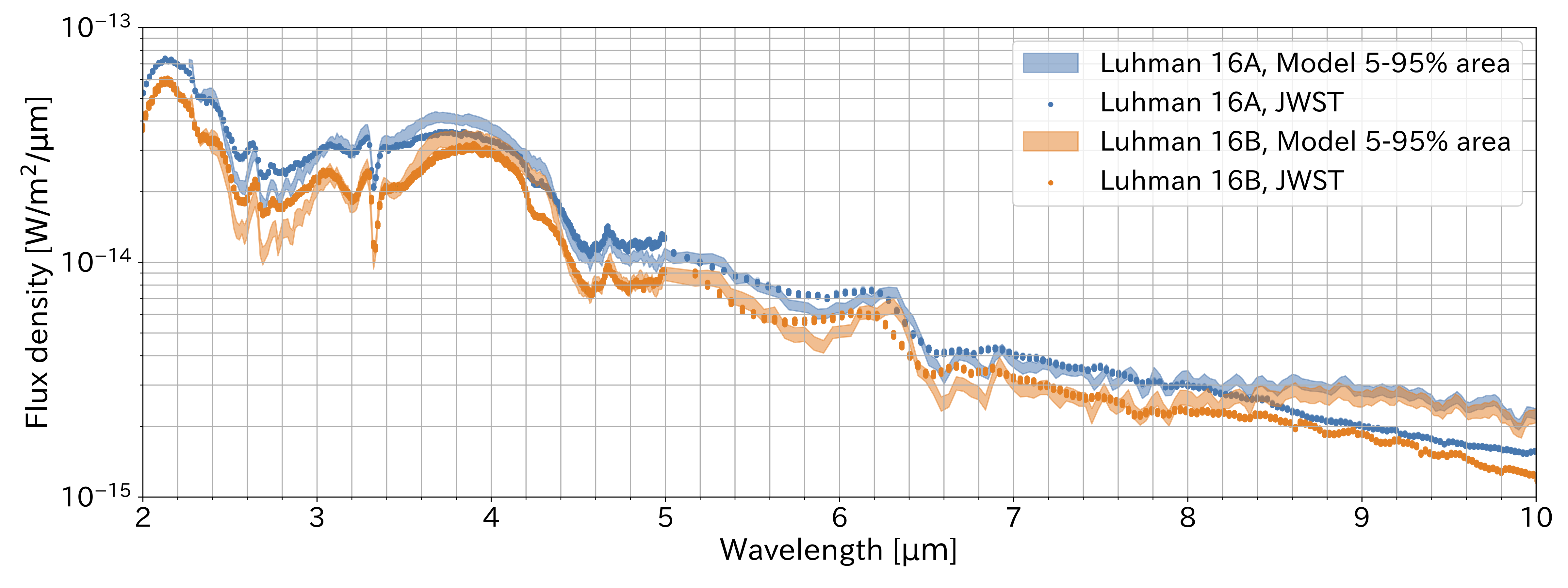}
    \caption{Comparison between the medium-resolution spectra predicted by extrapolating our high-resolution retrievals and the JWST observations. Shaded bands show the 90\% credible intervals of the predicted spectra computed with the $\hitemphco$ line list and a power-law $T$--$P$ profile (blue: Luhman 16A; orange: Luhman 16B). Points show the corresponding JWST medium-resolution spectra from \citet{Biller2024, Chen2025} (blue: A; orange: B).}
    \label{fig:model_vs_jwst}
\end{figure*}

\subsection{Cloud Degeneracies and Joint Analysis with JWST}
\label{subsec:joint_JWST}

Throughout this work, we find that the inferred cloud-top pressure is sensitive to assumptions about the $T$--$P$ profile and the adopted line list. These systematics propagate into the retrieved VMRs and, via absolute flux, into radius inferences. This underscores the value of combining broad wavelength information with high spectral resolution. 

In the current era, JWST provides such leverage. In Figure~\ref{fig:model_vs_jwst}, we compare the model spectra extrapolated to $10\,\mathrm{\mu m}$ from our high-resolution analysis (power-law $T$--$P$, gray cloud, $\hitemphco$) against JWST medium-resolution spectra \citep{Biller2024, Chen2025}. Beyond $\sim8.5\,\mathrm{\mu m}$ the JWST flux densities fall below the extrapolated models, consistent with absorption by silicate particles. Similar features have also been reported in medium-resolution spectra of L dwarfs with the \textit{Spitzer} Space Telescope, linked to the emergence and sedimentation of silicate clouds \citep{Cushing2006,Suarez2022}. More recently, JWST has revealed mid-infrared silicate absorption in directly imaged planets: \citet{Hoch2025} report a clear $8.5$--$11\,\mathrm{\mu m}$ feature in YSES-1\,c, demonstrating that particle size, composition (e.g., Mg silicates and Fe-bearing pyroxenes), vertical distribution, and the $T$--$P$ profile all shape the mid-infrared continuum and feature morphology. These findings highlight the opportunity to gain a more complete understanding of cloud properties by constraining clouds and the $T$--$P$ structure simultaneously. Together with the mid-infrared flux deficit relative to our extrapolated models, these comparisons motivate joint retrievals that combine high-resolution ground-based spectra with JWST data.
High-resolution spectra exploit the height dependence of individual line formation to tightly constrain the $T$--$P$ profile, while JWST provides broad coverage to probe cloud composition and vertical structure. Thus, joint retrievals offer a powerful approach to advancing our understanding of cloud physics and composition, fully leveraging the complementary strengths of broad wavelength coverage and high spectral resolution. 

Building upon these insights, future work will incorporate more detailed cloud modeling that accounts for realistic particle size distributions, compositions, and vertical structures. While the present analysis adopted a simple gray cloud prescription (Eq.~\ref{eq:dtau_cloud}) to focus on the line list and $T$--$P$ effects, these developments --- particularly when combined with joint retrievals using JWST and high-resolution ground-based spectra --- will enable a more comprehensive understanding of cloud formation and evolution in substellar atmospheres.

\begin{acknowledgments}
We gratefully acknowledge Ian J.~M. Crossfield for providing the reduced VLT/CRIRES spectra of Luhman 16AB.
Based on observations collected at the European Southern Observatory under ESO programme(s) 291.C-5006(A). We also thank the anonymous reviewer for providing comments that helped improve the manuscript.
This study was supported by JSPS KAKENHI Grant Numbers 24KJ1555 (H.Y.), 21H04998 (K.M., Y.K., and H.K.), 22H05150 (Y.K.), 25K17429 (Y.K.), 23H01224 (Y.K. and H.K.), and 23H00133 (H.K.).
We used ChatGPT (OpenAI, GPT-5, Sept 2025) for language editing. No part of the scientific analysis was performed by ChatGPT.
The authors are responsible for the content and interpretation.
 
\end{acknowledgments}





%
\facilities{VLT (CRIRES)}

\software{
corner \citep{corner},
ExoJAX \citep{Kawahara2022, Kawahara2024}, FastChem \citep{fastchem, fastchem_cond}, JAX \citep{jax2018github}, NumPyro \citep{bingham2018pyro, phan2019composable}, tinygp \citep{tinygp}
          }


\appendix

Table~\ref{tab:co_lit} provides the literature values, together with the corresponding references, for the objects shown in Figure~\ref{fig:co_mass}.
\begin{deluxetable*}{lcccccc}
\tablecaption{Literature C/O Ratios and Masses for Brown Dwarfs and Super Jupiters\label{tab:co_lit}}
\tablehead{
\colhead{Object} & \colhead{Class\tablenotemark{a}} & \colhead{Comp.\tablenotemark{b}} &
\colhead{C/O} & \colhead{$M$ [$\mathrm{M_J}$]} & \colhead{$T_{\mathrm{eff}}$ [K]} & \colhead{Ref.} \\[-4pt]
\colhead{} & \colhead{} & \colhead{} & \colhead{} & \colhead{} & \colhead{}
}
\startdata
DENIS J0255  & FBD  & N & $0.681\pm0.005$ & $45.7\pm4.4$ & $1554^{+263}_{-53}$ & (1) \\
HD 4747 B     & FBD  & Y & $0.66\pm0.04$   & $67.2\pm1.8$ & $1554^{+128}_{-218}$ & (2) \\
HIP 55507 B   & FBD  & Y & $0.67\pm0.04$   & $88.0^{+3.4}_{-3.2}$ & $2350\pm50$ & (3) \\
HD 33632 Ab   & FBD  & Y & $0.58\pm0.14$   & $46\pm8$     & $1628^{+109}_{-97}$ & (4), (5) \\
Gl 229 B      & FBD  & Y & $0.77\pm0.01$   & $71.47^{+0.57}_{-0.61}$\tablenotemark{c} & $850\pm50$ & (6), (7)  \\
HR 7672 B     & FBD  & Y & $0.52\pm0.02$   & $72.84^{+0.71}_{-0.72}$ & $1806\pm77$ & (8) \\
2MASS J0835  & FBD  & N & $0.65\pm0.02$   & $62.47\pm15.86$ & $1754\pm112$ & (9), (10) \\
2MASS J0501  & FBD  & N & $0.68\pm0.04$   & $21.45\pm13.71$ & $1720\pm55$ & (9), (11) \\
2MASS J0500  & FBD  & N & $0.71\pm0.01$   & $63.68\pm14.44$ & $1793\pm72$ & (9), (12) \\
HD 984 B      & YBD  & Y & $0.50\pm0.01$   & $61\pm4$     & $2730^{+120}_{-180}$ & (13), (14), (15) \\
TWA 28       & YBD  & N & $0.61\pm0.02$   & $20.9\pm5.0$ & $2382\pm42$ & (16), (17), (18)\\
2MASS J0856  & YBD  & N & $0.58\pm0.01$   & $14.4\pm1.4$ & $2380\pm32$ & (16), (18), (19) \\
2MASS J0355  & YBD  & N & $0.56\pm0.02$   & $19^{+7}_{-5}$ & $1527\pm14$ & (20), (21) \\
HIP 99770 b   & SJ & Y & $0.55\pm0.05$   & $16.1^{+5.4}_{-5.0}$ & $1400^{+200}_{-150}$ & (22), (23) \\
$\beta$ Pic b & SJ & Y & $0.41\pm0.05$   & $11.9\pm3.0$ & $1724\pm15$ & (24), (25), (26) \\
GQ Lup b     & SJ & Y & $0.50\pm0.01$   & $25^{+11}_{-15}$ & $2650\pm100$ & (27), (28) \\
YSES 1 b      & SJ & Y & $0.57\pm0.01$   & $21.8\pm3.0$ & $1893\pm10$ & (29), (30) \\
YSES 1 c      & SJ & Y & $0.36\pm0.15$   & $7.2\pm0.7$  & $958^{+163}_{-76}$ & (29), (30) \\
$\kappa$ And b & SJ & Y & $0.58\pm0.05$ & $22\pm9$     & $1680^{+60}_{-50}$ & (31) \\
GSC 6214 b    & SJ & Y & $0.70\pm0.07$   & $21\pm6$     & $1860^{+170}_{-110}$ & (31) \\
ROXs 42 B     & SJ & Y & $0.48\pm0.08$   & $13\pm5$     & $2270^{+170}_{-140}$ & (31) \\
DH Tau b     & SJ & Y & $0.54\pm0.06$   & $12\pm4$     & $2250^{+120}_{-100}$ & (31) \\
ROXs 12 b    & SJ & Y & $0.54\pm0.05$   & $19\pm5$     & $2500\pm140$ & (31) \\
HIP 79098 b  & SJ & Y & $0.54\pm0.03$   & $28\pm13$    & $2360^{+70}_{-80}$ & (31) \\
2MASS J01225 & SJ & Y & $0.37\pm0.08$   & $25\pm12$    & $1710^{+170}_{-160}$ & (31) \\
AB Pic b     & SJ & Y & $0.59\pm0.01$   & $30^{+20}_{-18}$ & $2000^{+100}_{-300}$ & (32), (33) \\
\enddata
\tablenotetext{a}{Class: FBD = Field Brown Dwarf; YBD = Young Brown Dwarf ($\lesssim \text{few}\times 10\,\mathrm{Myr}$); SJ = Super Jupiter}
\tablenotetext{b}{Companion flag: Y = bound to a host star (long-period); N = isolated.}
\tablenotetext{c}{Gl 229 B has been reported to be a binary, based on high–angular-resolution observations with GRAVITY/VLTI together with radial-velocity monitoring with CRIRES+/VLT \citep{Xuan2024b}.}
\tablecomments{References—(1)~\citet{deRegt2024}; (2)~\citet{Xuan2022}; (3)~\citet{Xuan2024c}; (4)~\citet{Hsu2024}; (5)~\citet{Currie2020}; (6)~\citet{Kawashima2025}; (7)~\citet{Calamari2022}; (8)~\citet{Wang2022}; (9)~\citet{Mulder2025}; (10)~\citet{Schlawin2017}; (11)~\citet{Zapatero2014}; (12)~\citet{Filippazzo2015}; (13)~\citet{Costes2024}; (14)~\citet{Franson2022}; (15)~\citet{Johnson-Groh2017}; (16)~\citet{Picos2024}; (17)~\citet{Scholz2005}; (18)~\citet{Cooper2024}; (19)~\citet{Morales2021}; (20)~\citet{Zhang2021}; (21)~\citet{Vos2022}; (22)~\citet{Zhang2024b}; (23)~\citet{Currie2023}; (24)~\citet{Landman2024}; (25)~\citet{Lacour2021}; (26)~\citet{Chilcote2017}; (27)~\citet{Picos2025}; (28)~\citet{Schwarz2016}; (29)~\citet{Zhang2024a}; (30)~\citet{Wood2023}; (31)~\citet{Xuan2024a}; (32)~\citet{Gandhi2025}; (33)~\citet{Bonnefoy2010}}
\end{deluxetable*}


\bibliography{sample701}{}



\end{document}